\documentclass[oldversion, a4paper]{aa}

\usepackage{graphicx}
\usepackage[varg]{txfonts}
\usepackage{natbib}

\newcommand{\ngc}{NGC~}
\newcommand{\azv}{AzV~}

\newcommand{\Mv}{\mbox{$M_{V}$}}
\newcommand{\Av}{\mbox{$A_{V}$}}
\newcommand{\Rv}{\mbox{$R_{V}$}}

\newcommand{\fastwind}{{\sc fastwind}}
\newcommand{\pikaia}{{\sc pikaia}}
\newcommand{\cmfgen}{{\sc cmfgen}}
\newcommand{\tlusty}{{\sc tlusty}}

\newcommand{\ha}{H$\alpha$}
\newcommand{\hb}{H$\beta$}
\newcommand{\hg}{H$\gamma$}
\newcommand{\hd}{H$\delta$}

\newcommand{\redchi}{\mbox{$\chi_{\rm red}^2$}}
\newcommand{\magn}{\mbox{$^{\rm m}$}}

\newcommand{\teff}{\mbox{$T_{\rm eff}$}}
\newcommand{\logg}{\mbox{$\log{{g}}$}}
\newcommand{\loggc}{\mbox{$\log{{g}}_{\rm c}$}}
\newcommand{\mdot}{\mbox{$\dot{M}$}}
\newcommand{\yhe}{\mbox{$Y_{\rm He}$}}
\newcommand{\rstar}{\mbox{$R_{\star}$}}
\newcommand{\rsun}{\mbox{$R_{\sun}$}}
\newcommand{\lstar}{\mbox{$L_{\star}$}}
\newcommand{\lsun}{\mbox{$L_{\sun}$}}
\newcommand{\Ms}{\mbox{$M_{\rm s}$}}
\newcommand{\Mev}{\mbox{$M_{\rm ev}$}}
\newcommand{\msun}{\mbox{$M_{\sun}$}}
\newcommand{\zsun}{\mbox{$Z_{\sun}$}}
\newcommand{\kmsec}{\mbox{km\,s$^{-1}$}}
\newcommand{\cmsecsec}{\mbox{cm\,s$^{-2}$}}
\newcommand{\msunyr}{\mbox{$M_{\sun}{\rm yr}^{-1}$}}
\newcommand{\Dmom}{\mbox{$D_{\rm mom}$}}
\newcommand{\Qnull}{\mbox{$Q_{0}$}}

\newcommand{\vturb}{\mbox{$v_{\rm turb}$}}
\newcommand{\vmac}{\mbox{$v_{\rm mac}$}}
\newcommand{\vr}{\mbox{$v_{\rm r}$}}
\newcommand{\vrot}{\mbox{$v_{\rm rot}$}}
\newcommand{\veq}{\mbox{$v_{\rm eq}$}}
\newcommand{\vmin}{\mbox{$v_{\rm min}$}}
\newcommand{\vmax}{\mbox{$v_{\rm max}$}}
\newcommand{\vsini}{\mbox{$v_{\rm r}\sin i$}}
\newcommand{\vinf}{\mbox{$v_{\infty}$}}
\newcommand{\vesc}{\mbox{$v_{\rm esc}$}}
\newcommand{\vrgauss}{\mbox{$v^{\rm G}_{r}$}}
\newcommand{\vrmaxwell}{\mbox{$v^{\rm M}_{r}$}}

\newcommand{\hei}{\ion{He}{i}}
\newcommand{\heii}{\ion{He}{ii}}

\defcitealias{mokiem05}{Paper~I}

\begin{document}

\title{The VLT-FLAMES survey of massive stars: \\Mass loss and rotation
of early-type stars in the SMC}

\titlerunning{Mass loss and rotation of early-type stars in the SMC}

\author{M.\,R.~Mokiem\inst{1}
   \and A.~de~Koter\inst{1}
   \and C.\,J.~Evans\inst{2}
   \and J.~Puls\inst{3}
   \and S.\,J.~Smartt\inst{4}
   \and P.~A.~Crowther\inst{5}
   \and A.~Herrero\inst{6,7}
   \and N.~Langer\inst{8}
   \and D.\,J.~Lennon\inst{9,6}
   \and F.~Najarro\inst{10}
   \and M.\,R.~Villamariz\inst{11,6}
   \and S.-C.~Yoon\inst{1}
}


\institute{
  Astronomical Institute Anton Pannekoek, University of Amsterdam,
  Kruislaan 403, 1098~SJ Amsterdam, The Netherlands
  \and
  UK Astronomy Technology Centre, Royal Observatory, Blackford Hill,
  Edinburgh, EH9 3HJ, UK
  \and
  Universit\"ats-Sternwarte M\"unchen, Scheinerstr. 1,
  D-81679 M\"unchen, Germany
  \and
  The Department of Pure and Applied Physics,
  The Queen's University of Belfast,
  Belfast BT7 1NN, Northern Ireland, UK
  \and
  Department of Physics and
  Astronomy, University of Sheffield, Hicks Building, Hounsfield Rd,
  Shefffield, S3 7RH,  UK
  \and
  Instituto de Astrof\'{\i}sica de Canarias, E-38200, La Laguna,
  Tenerife, Spain
  \and
  Departamento de Astrof\'{\i}sica, Universidad de La Laguna,
  Avda.\ Astrof\'{\i}sico Francisco S\'anchez, s/n, E-38071
  La Laguna, Spain
  \and
  Astronomical Institute, Utrecht University, Princetonplein 5,
  3584 CC Utrecht, The Netherlands
  \and
  The Isaac Newton Group of Telescopes,
  Apartado de Correos 321, E-38700,
  Santa Cruz de La Palma, Canary Islands, Spain
  \and 
  Instituto de Estructura de la Materia, Consejo Superior de
  Investigaciones Cient\'{\i}ficas, CSIC, Serrano 121, E-28006
  Madrid, Spain
  \and
  Grantecan S.A., E-38200, La Laguna,
  Tenerife, Spain
}

\date{DRAFT, \today}

\abstract{ We have studied the optical spectra of a sample of 31 O-
and early B-type stars in the Small Magellanic Cloud, 21 of which are
associated with the young massive cluster \ngc346. Stellar parameters
are determined using an automated fitting method (Mokiem et al.\
2005), which combines the stellar atmosphere code \fastwind\ (Puls et
al.\ 2005) with the genetic algorithm based optimisation routine
\pikaia\ (Charbonneau 1995). Comparison with predictions of stellar
evolution that account for stellar rotation does not result in a
unique age, though most stars are best represented by an age of
1--3~Myr. The automated method allows for a detailed determination of
the projected rotational velocities. The present day \vsini\
distribution of the 21 dwarf stars in our sample is consistent with an
underlying rotational velocity (\vr) distribution that can be
characterised by a mean velocity of about $160 - 190~\kmsec$ and an
effective half width of $100 - 150~\kmsec$. The \vr\ distribution must
include a small percentage of slowly rotating stars. If predictions of
the time evolution of the equatorial velocity for massive stars within
the environment of the SMC are correct (Maeder \& Meynet 2001), the
young age of the cluster implies that this underlying distribution is
representative for the initial rotational velocity distribution. The
location in the Hertzsprung-Russell diagram of the stars showing
helium enrichment is in qualitative agreement with evolutionary tracks
accounting for rotation, but not for those ignoring \vr. The mass loss
rates of the SMC objects having luminosities of $\log \lstar/\lsun
\gtrsim 5.4$ are in excellent agreement with predictions by Vink, de
Koter \& Lamers (2001). However, for lower luminosity stars the winds
are too weak to determine \mdot\ accurately from the optical
spectrum. Two of three spectroscopically classified Vz stars from our
sample are located close to the theoretical zero age main sequence, as
expected.  Three additional objects of lower luminosity, which are not
given this classification, are also found to lie near the ZAMS. We
argue that this is related to a temperature effect inhibiting
relatively cool stars from displaying the spectral features
characteristic for the Vz luminosity class.}


\maketitle

\section{Introduction}
Mass loss and rotation play a crucial role in the evolution of the
most massive stars. These two processes are linked. To illustrate
this, fast rotation may lead to an enhanced mass loss, as is, for
instance, observed in the case of Be stars \citep{lamers91} and
$\eta$\,Carinae \citep{smith03, vanboekel03}. Conversely, the stellar
winds of these stars will cause loss of angular momentum, leading to
spin down. In a Galactic environment the spin down of massive stars is
expected to be relatively rapid, essentially ``erasing'' their initial
rotational velocity properties within only a few million years
\citep{meynet00}. For an environment characteristic of the Small
Magellanic Cloud (SMC) it is predicted that the initial conditions of
rotational velocity remain preserved during the main sequence life of
O-type stars \citep{maeder01}, as the strength of stellar winds
diminishes with decreasing metal content \citep[e.g.][]{abbott82,
kudritzki87, vink01}.

The study of a young SMC cluster containing a substantial population
of O-type main sequence stars would allow the role of metallicity on
mass loss and angular momentum loss during the early evolution of
massive stars to be better constrained. This provides a better insight
into the current and perhaps even initial rotational conditions from
which this evolution unfolds. Among others, this would provide
important constraints on the physics of the formation of massive
stars, for instance on the role of magnetic fields and the lifetime of
accretion disks \citep[see e.g.][]{porter03}.

The distribution of projected rotational velocities \vsini\ of
Galactic early-type stars has been studied by \cite{conti77b},
\cite{penny96} and \cite{howarth97}. The first extragalactic studies
of rotational velocities have been presented by \cite{keller04} for a
sample of Large Magellanic Cloud (LMC) B-type stars in the vicinity of
the main sequence turnoff and by \cite{penny04} for predominately
giant and supergiant SMC and LMC O-type stars. So far, no systematic
study for early main sequence O-type stars in a low metallicity
environment has been attempted.

The young cluster \ngc346 in the Small Magellanic Cloud (SMC) contains
a substantial population of O-type main sequence stars. In the context
of the {\em VLT-FLAMES Survey of Massive Stars} \citep{evans05}, we
have used the {\em Fibre Large Array Multi-Element Spectrograph} at
the {\em ESO Very Large Telescope} to obtain spectra of several dozen
dwarf O-type stars in this cluster. Our genetic algorithm based
fitting method \citep{mokiem05} is used to perform a homogeneous
spectroscopic analysis of this set of stars, including the projected
rotational velocity as one of the fitting parameters. The \vsini\
information obtained from this analysis may be compared to models of
the present and possibly initial (see above) \vsini\
distribution. This is a main aim of this paper.

The strength of radiatively driven winds is predicted to be reduced
for decreasing metal content. In the last two decades this prediction
has been quantified by different groups who found a metallicity
dependence of the mass loss rate of $\mdot(Z)\propto Z^{0.5 - 0.7}$
\citep[e.g.][]{kudritzki87, puls00, vink01}. Qualitatively this
metallicity dependence has been confirmed by several authors
\citep[e.g.][]{puls96, bouret03, evans04b, massey05}. However, until
now a quantitative comparison of the theoretically predicted and the
empirically determined $Z$ dependence is still lacking. The analysis
of our SMC targets will provide insight in the wind characteristics of
objects with a metal content approximately five times lower than in
Galactic objects. Consequently, given the large number of objects that
we have analysed in a homogeneous way, the current study will for the
first time be able to provide such a quantitative comparison.

While comparing mass loss rate predictions with observed values it is
important to realise that recent studies of the wind strengths of not
too luminous ($\log \lstar/\lsun \lesssim 5.3$) Galactic and SMC dwarf
O-type stars \citep{bouret03, martins04, martins05b} seem to indicate
that the mass loss rates are significantly (up to two orders of
magnitude) lower than expected from theory \citep*{vink01}. If this is
indeed the case it would imply that our chances of actually observing
a \vr\ pattern that is representative of the initial distribution will
increase, as less angular momentum will be carried away in the stellar
wind. However, from the viewpoint of our understanding of the physics
of relatively weak stellar winds ($\sim$$10^{-8}$ -- $10^{-6}
\msunyr$) the situation is obviously less desirable.

The mass-loss rates reported by \citeauthor{bouret03} and
\citeauthor{martins05b} are based on an analysis of unsaturated
ultraviolet resonance lines. The \mdot\ sensitivity that can be
obtained using such lines is $\sim$$10^{-9}$ -- $10^{-8}
\msunyr$. This is, in principle, much better than is possible using
\ha\ as the mass loss diagnostic. Using our automated fitting method
and spectra having a signal-to-noise of 50--200 (typical for those
secured within the context of our VLT-FLAMES programme) we can push
the \ha\ method as low as $\sim$$2\times 10^{-7}\,\msunyr$. This
provides some overlap with the weak wind regime so far reserved for
the UV line method. It thus allows for a first investigation of the
question whether the so-called ``weak wind problem'' points to missing
physics in the theoretical predictions of mass loss or that there may
be problems with the mass loss diagnostic, i.e.\ that we do not fully
understand the formation of UV resonance lines.

The structure of this paper is as follows: in Sect.~\ref{sec:data} we
describe the data set that has been analysed using our genetic
algorithm based fitting method, which is discussed in
Sect.~\ref{sec:method}. The stellar properties of our sample
determined in this analysis are presented in Sect.~\ref{sec:fun_par}.
In sections~\ref{sec:par_vr_dist}, \ref{sec:weak_winds} and
\ref{sec:age} we investigate and discuss, respectively, the underlying
rotational velocity distribution of our sample, the mass loss rates we
have determined in the context of weak winds and the evolutionary
status of the cluster \ngc346. Sect.~\ref{sec:sum} summarises and
lists the conclusions of our study. Finally, in the appendix the fits
and comments on the individual objects are provided.

\section{Data description}
\label{sec:data}

The sample considered here is mainly drawn from the targets observed
in the SMC as part of the VLT-FLAMES survey of massive stars
\citep[see][]{evans05}. The survey observed two fields in the SMC,
centered on the clusters \ngc346 and \ngc330.  Here we analyse each of
the O-type stars and three luminous B-type stars observed in the
\ngc346 field, and two O-type stars from the older \ngc330 cluster.

To improve our sampling of luminosity and temperature within the
O-type domain the FLAMES targets were supplemented by eight field
stars from the catalogue of \citet[hereafter AzV, note that in other
studies these are also often denoted by AV]{azzopardi75, azzopardi82}.
These stars were observed using the Ultraviolet and Visual Echelle
Spectrograph (UVES) at the VLT as part of the programs 67.D-0238,
70.D-0164 and 074.D-0109 (P.I. Crowther).

\begin{table*}[p]
  \caption{Basic parameters. Identifications: FLAMES targets
  (``NGC330'' and ``NGC346'') from \cite{evans06}, ``MPG'' from
  \cite{massey89}, ``AzV'' from \cite{azzopardi75, azzopardi82} and
  ``Sk'' from \cite{sanduleak68}. Photometric data for the FLAMES
  targets is from \cite{momany01}; for objects with a primary \azv
  identification these are from \cite{massey02} and
  \cite{massey04}. Spectral types of the FLAMES targets are from
  \cite{evans06}. For convenience previously published spectral types
  are also given. Sources of these classifications are W77
  \citep{walborn77}; NMC \citep{niemela86}; G87 \citep{garmany87}; MPG
  \citep{massey89}; W00 \citep{walborn00}; EH04
  \citep{evans04c}. Spectral types for the field O-type stars were
  taken from the catalogues of \cite{garmany87} and
  \cite{walborn02b}. Wind velocities given without brackets are from
  \cite{garmany88, haser98, prinja98, crowther02, bouret03, evans04b,
  evans04a, massey04}. Values between brackets are calculated from the
  escape velocity at the stellar surface.}
  \label{tab:data}
  \begin{center}
  \begin{tabular}{llllllcccc}
  \hline\\[-9pt] \hline \\[-7pt]
  Primary ID & \multicolumn{3}{c}{Cross-IDs} & Spectral & Published & $V$ & \Av   & \Mv & \vinf \\[2pt]
     & \multicolumn{1}{c}{MPG} & \multicolumn{1}{c}{\azv} & \multicolumn{1}{c}{Sk} & Type & ST & & & & [\kmsec]\\[1pt]
	\hline\\[-9pt]
\object{\ngc346-001}$^{a)}$ & MPG~789 & \azv232 & Sk~80   & O7\,Iaf+      & O7\,Iaf+ [W77] & 12.31 & 0.40 & $ -6.99 $ &  1330 \\
                            &         &         &         &               &  O7\,If [MPG] \\[3.5pt]
\object{\ngc346-007}$^{a)}$ & MPG~324 & ...     & ...     & O4\,V((f+))   & O4-5\,V [NMC] & 14.07 & 0.12 & $ -4.95 $ &  2300  \\
                            &         &         &         &               & O4\,V((f)) [MPG]\\
                            &         &         &         &               & O4\,((f)) [W00]\\[3.5pt]
\object{\ngc346-010}        & ...     & \azv226 & ...     & O7\,IIIn((f)) & O7\,III [G87] & 14.37 & 0.31 & $ -4.84 $ & [1832] \\[3.5pt]
\object{\ngc346-012}        & ...     & \azv202 & ...     & B1\,Ib        & B1\,III [G87] & 14.39 & 0.09 & $ -4.60 $ & [1568] \\[3.5pt]
\object{\ngc346-018}        & MPG~217 & ...     & ...     & O9.5\,IIIe    & ... & 14.78 & 0.56 & $ -4.68 $ & [1481] \\[3.5pt]
\object{\ngc346-022}        & MPG~682 & ...     & ...     & O9\,V         & O8\,V [MPG] & 14.91 & 0.16 & $ -4.14 $ & [3305] \\[3.5pt]
\object{\ngc346-025}$^{a)}$ & MPG~848 & ...     & ...     & O9\,V         & O8.5\,V [MPG] & 14.95 & 0.12 & $ -4.07 $ & [2816] \\[3.5pt]
\object{\ngc346-026}        & MPG~12  & ...     & ...     & B0\,IV        & O9.5\,V [MPG] & 14.98 & 0.50 & $ -4.42 $ & [2210]\\
                            &         &         &         &               & O9.5-B0\,V\,(N str) [W00]\\
                            &         &         &         &               & O9.5\,III [EH04] \\[3.5pt]
\object{\ngc346-028}        & MPG~113 & ...     & ...     & OC6\,Vz       & O6\,V [MPG] & 15.01 & 0.19 & $ -4.08 $ & [2369] \\
                            &         &         &         &               & OC6\,Vz [W00]$^{b)}$\\[3.5pt]
\object{\ngc346-031}        & ...     & ...     & ...     & O8\,Vz        & ... & 15.02 & 0.16 & $ -4.04 $ & [2461] \\[3.5pt]
\object{\ngc346-033}        & MPG~593 & ...     & ...     & O8\,V         & ... & 15.07 & 0.19 & $ -4.02 $ & [4102] \\[3.5pt]
\object{\ngc346-046}        & ...     & ...     & ...     & O7\,Vn        & ... & 15.44 & 0.12 & $ -3.58 $ & [2723] \\[3.5pt]
\object{\ngc346-050}        & MPG~299 & ...     & ...     & O8\,Vn        & O9\,V [MPG] & 15.50 & 0.03 & $ -3.43 $ & [2661] \\[3.5pt]
\object{\ngc346-051}        & MPG~523 & ...     & ...     & O7\,Vz        & O7\,V+neb [MPG] & 15.51 & 0.16 & $ -3.54 $ & [3233] \\[3.5pt]
\object{\ngc346-066}        & MPG~213 & ...     & ...     & O9.5\,V       & ... & 15.75 & 0.16 & $ -3.30 $ & [2931] \\[3.5pt]
\object{\ngc346-077}        & MPG~238 & ...     & ...     & O9\,V         & B0: [MPG] & 15.88 & 0.40 & $ -3.42 $ & [2205] \\[3.5pt]
\object{\ngc346-090}        & ...     & ...     & ...     & O9.5\,V       & ... & 15.96 & 0.37 & $ -3.31 $ & [2978] \\[3.5pt]
\object{\ngc346-093}        & MPG~304 & ...     & ...     & B0\,V         & ... & 16.01 & 0.34 & $ -3.23 $ & [3516] \\[3.5pt]
\object{\ngc346-097}        & MPG~519 & ...     & ...     & O9\,V         & ... & 16.06 & 0.74 & $ -3.58 $ & [4004] \\[3.5pt]
\object{\ngc346-107}        & MPG~559 & ...     & ...     & O9.5\,V       & ... & 16.20 & 0.09 & $ -2.79 $ & [2547] \\[3.5pt]
\object{\ngc346-112}        & MPG~327 & ...     & ...     & O9.5\,V       & ... & 16.24 & 0.19 & $ -2.85 $ & [2368] \\[1pt]
 \hline\\[-9pt]              
\object{\ngc330-013}        & ...     & \azv186 & ...     & O8.5\,II-III((f)) & O7\,III [G87] & 14.00 & 0.56 & $ -5.46 $ &  1600 \\
                            &         &         &         &                   & O8\,III((f)) [EH04] \\[3.5pt]
\object{\ngc330-052}        & ...     & ...     & ...     & O8.5\,Vn      & O8\,V [EH04]& 15.69 & 0.16 & $ -3.36 $ & [2005] \\[1pt]
 \hline\\[-9pt]              
\object{AzV 14}             & ...     & ...     & Sk~9    & O5\,V         & ... & 13.55 & 0.47 & $ -5.81 $ &  2000  \\[3.5pt]
\object{AzV 15}             & ...     & ...     & Sk~10   & O7\,II        & ... & 13.12 & 0.40 & $ -6.18 $ &  2125  \\[3.5pt]
\object{AzV 26}             & ...     & ...     & Sk~18   & O7\,III       & ... & 12.46 & 0.47 & $ -6.90 $ &  2150  \\[3.5pt]
\object{AzV 95}             & ...     & ...     & ...     & O7\,III       & ... & 13.78 & 0.43 & $ -5.55 $ &  1700  \\[1pt]
\object{AzV 243}            & ...     & ...     & Sk~84   & O6\,V         & ... & 13.84 & 0.47 & $ -5.53 $ &  2125  \\[3.5pt]
\object{AzV 372}            & ...     & ...     & Sk~116  & O9\,Iabw      & ... & 12.59 & 0.50 & $ -6.81 $ &  1550  \\[3.5pt]
\object{AzV 388}            & ...     & ...     & ...     & O4\,V         & ... & 14.09 & 0.34 & $ -5.15 $ &  1935  \\[3.5pt]
\object{AzV 469}            & ...     & ...     & Sk~148  & O8.5\,II((f)) & ... & 13.12 & 0.47 & $ -6.25 $ &  1550  \\[1pt]
 \hline           
  \end{tabular}
  \end{center}
 $^{a)}$ binary;
 $^{b)}$ the source of the type adopted here
\end{table*}

Table~\ref{tab:data} lists some of the basic observational properties
of the programme stars analysed together with common aliases used in
other studies. For a full description of the observational properties
of the FLAMES targets see \cite{evans06}. Note that five of these
objects and six of the field stars were recently also analysed using
line blanketed stellar atmosphere models \citep{crowther02, bouret03,
evans04b, massey04, heap06}. A comparison with our findings is
presented in the appendix. Each of the FLAMES targets were observed
with the Giraffe spectrograph at least six times, at each of six
wavelength settings. The effective resolving power of the observations
is $R \simeq 20\,000$, full details are given by \cite{evans06}. The
multiple exposures, often at different epochs, allowed for the
detection of variable radial velocities, with a number of binaries
detected \citep[see][]{evans06}. Three of the stars analysed here are
detected as binaries (\ngc346-001 \& \ngc346-025) or are found to have
variable velocities suggestive of binarity (\ngc346-007), as noted in
Table~\ref{tab:data}.  Note that none of these appear to have massive
companions, which could have a significant effect on the derived
parameters if they were present.

A general description of the reduction of the FLAMES data was given by
\cite{evans05}.  The most pertinent part of the reductions for the
current study is that of sky subtraction.  A master sky-spectrum was
created from combining the sky fibres (typically 15), individually
scaled by their relative fibre throughput.  In general the sky
background is low for our (relatively bright) targets, and this
approach successfully removes the background contribution.  However,
in regions such as \ngc346 which has strong nebular emission, accurate
subtraction of nebular features in multi-fibre data is notoriously
difficult. This does not hamper our analysis, except in the core of
the \ha\ Balmer line. For most stars the core nebular emission is
well-resolved and we simply only consider the wings of the profile in
the automated line fits. Tests assessing the impact of possible
residual nebular contamination in the line wings or over-subtraction
of sky components are discussed in Sect.~\ref{sec:neb-em}.

For each wavelength range the individual sky-subtracted spectra were
co-added, then normalised using a cubic-spline fit to the
continuum. These normalised data were finally merged to obtain a final
spectrum covering 3850--4750 and 6300--6700~\AA.  The combined spectra
have a typical signal-to-noise ratio of 50--200, depending on the
magnitude of the target.

Spectral types for the field O-type stars were taken from the
catalogues of \cite{garmany87} and \cite{walborn02b}. The FLAMES
spectra were classified by visual inspection, using published
standards, in particular the atlas of \cite{walborn90}, with
consideration to the lower metallicity environment of the SMC
\citep[e.g.][]{walborn00}. The classifications quoted in
Table~\ref{tab:data} are from \cite{evans06}.

The majority of the field stars were observed with the VLT-UVES
spectrograph during a visitor run on 27-28 September 2001 under
program 67.D-0238. UVES is a two arm cross-dispersed echelle
spectrograph, the red arm of which contains a mosaic of two
detectors. A standard blue setting, with central wavelength 437~nm
provided continuous coverage between 3730--5000~\AA, recorded on a
single 2$\times 4$K EEV CCD. A non-standard red setting with central
wavelength 830~nm included an identical EEV CCD covered
6370--8313~\AA, whilst a 2$\times$4K MIT-LL CCD covered
8290--10250~\AA. A 1$''$ slit was used in variable seeing conditions
to provide a spectral resolution of 0.09~\AA\ at \ha. Exposure times
ranged from 1200 to 3000 seconds.

Subsequently, \azv388 was observed with UVES on 5 Dec.\ 2002 under
service program 70.D-0164, using a standard setup with central
wavelength 390~nm (blue, 3300--4500~\AA) and 564~nm (red, 4620--5600
and 5675--6650~\AA), plus a second red setup with central wavelength
520~nm (4170--5160 and 5230--6210~\AA). The exposure time was 2700~s
for each setup, with a 1.2$''$ slit. Finally, \azv95 was observed with
UVES under service program 074.D-0109 on Nov.\ 27 2004, using standard
setups with central wavelengths 390/564~nm and 437/860~nm, with
individual 2200~s exposures, again with a 1.2$''$ slit. In all cases,
the two-dimensional CCD frames were transformed to extracted
one-dimensional spectra using the UVES pipeline software. The S/N in
individual blue spectra ranged from $\sim$50 to $\sim$120.

The quoted photometry for the FLAMES targets in the \ngc346 and
\ngc330 fields is taken from the ESO Imaging Survey (EIS) pre-FLAMES
release by \cite{momany01}. These data compare well to the values
obtained previously by \cite{massey89} and \cite{azzopardi75,
azzopardi82}, with an average difference of 0.04 magn for both $V$ and
$(B-V)$ and incidental (two cases) maximum differences of
approximately 0.1 magn. Photometric data for the field stars was
taken from the UBVR CCD survey of \cite{massey02} and \cite{massey04}.

The interstellar extinction (\Av) listed in Table~\ref{tab:data} was
calculated using intrinsic colours from \citet[and references
therein]{johnson66} and by assuming a ratio of total to selective
extinction of $\Rv=3.1$.  Using these \Av\ values, extinction
corrected visual magnitudes ($V_0$) were calculated from the observed
$V$-band magnitudes.  Finally, the absolute visual magnitude \Mv, was
calculated taking a distance modulus of 18.9~magn \citep{westerlund97,
harries03,hilditch05}.

\section{Analysis method}
\label{sec:method}

To analyse the large number of spectra we use an automated fitting
method. This method was developed by \citet[hereafter
\citetalias{mokiem05}]{mokiem05} for the fitting of the profiles of
hydrogen and helium lines and provides a means for an unbiased and
homogeneous analysis of large samples of early-type stars. For a
detailed description we refer to this paper. Here we will only give a
brief description of the method and modifications of its
implementation.

The automated fitting method consists of two main components. The
first component is the stellar atmosphere code \fastwind\ of
\cite{puls05}. This fast performance code incorporates non-LTE and
line blanketing following the concept of a unified model atmosphere to
synthesise hydrogen and helium line profiles. To optimise the input
parameters for \fastwind, i.e.\ to determine the fit parameters for a
certain spectrum, the genetic algorithm based optimisation routine
\pikaia\ \citep{charbonneau95} is used. This second component is
capable of global optimisation and in combination with \fastwind\
provides, as was shown in \citetalias{mokiem05}, a robust method for
the fitting of the hydrogen and helium spectra for a broad range of O-
and early B-type stars.

In short the automated method determines the best fit by calculating
consecutive generations of \fastwind\ models. In every generation the
models which fit the observed spectrum the best are selected and their
parameters are crossbred and mutated to create new sets of
parameters. Using these sets a new generation of models is
calculated. The procedure is repeated until the fit quality of the
best-fitting model is maximised. This fit quality is defined as the
inverted sum of the reduced chi squared, \redchi, values of the
observed hydrogen and helium line profiles and the synthetic line
profiles. In this sum each spectral line has a weight assigned. These
weights are used to express the accuracy with which the model
atmosphere code is believed to be able to reproduce certain lines. For
instance, the \hei\ line at 4471~\AA\ is given a low weight for late
type supergiants because of complications due to the so-called
``generalised dilution effect'' \citep{voels89} from which this line
suffers. A lower weight is assigned to the neutral helium singlet
lines for early and mid-type because the codes \fastwind\ and \cmfgen\
\citep{hillier98} show a discrepancy in the predictions of these
diagnostics \citep{puls05}. In \citetalias{mokiem05} the full
weighting scheme is described and discussed.

\begin{table*}[!ht]
  \caption{Results of the formal tests. Input parameters of the formal
  test models are given in the ``In'' column and parameters obtained
  with the automated fitting method by fitting synthetic data created
  from these models are listed in the ``Out'' column. The results were
  obtained by evolving a population of 72 \fastwind\ models over a
  course of 200 generations.}
  \label{tab:form-tests}
  \begin{center}
  \begin{tabular}{lrrrrrrrrr}
  \hline\\[-9pt] \hline \\[-7pt]
                            & Set A& Search       &      & Set B & Search       &      & Set C  & Search              \\[2pt]
                            & In   & range        & Out  & In    & range        & Out  & In     & range        & Out  \\[1pt]
  \hline \\[-9pt]
  Spectral type             & O3\,I&              &      & O5.5\,I&             &      & O9.5\,V&                     \\[3.5pt]
  \teff\ [kK]               & 47.0 & [44, 50]     & 46.9 & 40.0  & [37, 43]     & 39.6 & 33.0   & [29, 35]     & 33.0 \\[3.5pt]
  \logg\ [\cmsecsec]        & 3.80 & [3.4, 4.1]   & 3.82 & 3.6   & [3.3, 4.0]   & 3.57 & 4.00   & [3.6, 4.3]   & 3.97 \\[3.5pt]
  \rstar\ [\rsun]           & 18.0 &              & $-$  & 20.0  &              & $-$  & 8.0    &              & $-$  \\[3.5pt]
  $\log \lstar$ [\lsun]     & 6.17 &              & $-$  & 5.96  &              & $-$  & 4.83   &              & $-$  \\[3.5pt]
  \vturb\ [\kmsec]          & 5.0  & [0, 20]      & 8.01 & 15.0  & [0, 20]      & 12.1 & 10.0   & [0, 20]      & 14.5 \\[3.5pt]
  \yhe                      & 0.15 & [0.05, 0.30] & 0.16 & 0.10  & [0.05, 0.30] & 0.10 & 0.10   & [0.05, 0.30] & 0.09 \\[3.5pt]
  \vsini\ [\kmsec]          & 200  & [100, 300]   & 194  & 250   & [100, 300]   & 243  & 300    & [200, 400]   & 302  \\[3.5pt]
  \mdot\ [$10^{-6}$\msunyr] & 2.9  & [0.2, 8.0]   & 3.2  & 1.9   & [0.1, 5.0]   & 1.5  & 0.022  & [0.001, 0.1] & 0.031\\[3.5pt]
  $\beta$                   & 1.20 & [0.5, 1.5]   & 1.15 & 1.0   & [0.5, 1.5]   & 1.04 & 0.80   & [0.5, 1.5]   & 0.98 \\[3.5pt]
  \vinf\ [\kmsec]           & 3000 &              & $-$  & 2200  &              & $-$  & 2000   &              & $-$  \\[1pt]
  \hline
  \end{tabular}
  \end{center}
\end{table*}

\subsection{Fit parameters}
For the fitting of the spectra we allow for seven free
parameters. These parameters are the effective temperature \teff, the
surface gravity $g$, the helium over hydrogen number density \yhe, the
microturbulent velocity \vturb, the projected rotational velocity
\vsini, the mass loss rate \mdot\ and the exponent of the beta-type
velocity law describing the supersonic wind regime of the stellar
atmosphere. In contrast to \citetalias{mokiem05}, \vsini\ is now
treated as a free parameter. This could be done because the FLAMES
spectra are of sufficiently high resolution to allow for a self
consistent determination of this parameter (see
Sect.~\ref{sec:form_tests}). Note that this parameter can also be
determined relatively accurately using alternative methods (see
Sect.~\ref{sec:vsini}). The reason that we still incorporate it as a
free parameter is that it allows for a meaningful estimate of the
error in this parameter. More importantly, it allows for the
propagation of this uncertainty in the error estimates of the other
fit parameters.

An important parameter which cannot be determined from the optical
spectra of O stars is the terminal velocity of the wind
\vinf. Therefore, if a value obtained from the analysis of ultraviolet
(UV) wind lines is available for a certain object, we keep \vinf\
fixed at that value. If no \vinf\ determination is available, the
scaling relation of \vinf\ with the escape velocity at the stellar
surface (\vesc) is used throughout the fitting process. For early-type
stars this scaling implies that the ratio $\vinf/\vesc = 2.6$ is
adopted \citep{lamers95}. In some cases this produced relatively large
wind velocities (see Tab.~\ref{tab:data}). However, these individual
cases do not have a significant impact on our results, as for most of
these objects we could only derive upper limits for the mass loss rate
(also see Sect.~\ref{sec:wind_param}).

As the current implementation of the fitting method only analyses the
hydrogen and helium lines, no explicit abundance values other than the
ratio of these two elements can be determined. Therefore, we adopted
fixed values for the atmospheric abundances of the background metals.
For these values we use the Solar abundances from \citet[][and
references therein]{grevesse98} scaled proportionally with respect to
mass ratios by the same factor for all elements heavier than helium.
Iron, due to its strong line blanketing effect on the stellar
atmosphere and emergent spectrum, can be considered to be the most
important metal element. Large differences between the abundances of
other metals, such as nitrogen, have been reported for the Galaxy and
SMC \citep[e.g.][]{trundle04}, although their effect upon the
spectroscopic analysis of the hydrogen and helium lines are
negligible.  Consequently, we set the metallicity scaling factor equal
to the iron abundance ratio of the SMC. From the analysis of early
type SMC stars this ratio is found to be 1/5 times Solar
\citep{rolleston03}.

\subsection{Formal tests}
\label{sec:form_tests}

As argued in \citetalias{mokiem05} so-called formal tests, i.e.\
fitting of synthetic data, are an integral part of the automated
fitting method. First of all such tests are necessary to assess
whether the data quality is sufficient to secure a successful
determination of the global optimum in parameter space, i.e.\ whether
it will recover the global best fit. Secondly, they are necessary to
estimate the minimum number of generations that have to be calculated
in order to find this best fit. Consequently, by fitting synthetic
data with a similar quality as our observed spectra, we can establish
the minimum number of generations that have to be calculated to
safeguard that the best possible fit will be obtained when fitting the
real spectra.

Similar to \citetalias{mokiem05} three data sets were created based on
\fastwind\ models with parameters representing different types of
early-type stars. For these three data sets, which are denoted by A, B
and C, the input parameters of the \fastwind\ models are listed in
Tab.~\ref{tab:form-tests}. Set A and set B correspond to bright hot
supergiants. The parameters in set C represent a cooler dwarf O-type
star. For the mass loss rates we adopted values based on the
prediction of \cite{vink01} assuming $Z_{\rm SMC}=0.2\,\zsun$.

Using the line profiles from the \fastwind\ models the synthetic data
was created by first convolving the profiles with a rotational
broadening profile. To the broadened profiles Gaussian distributed
noise corresponding to a signal-to-noise ratio of 50, was added. This
value approximately corresponds to the lowest S/N in our
sample. Finally, as nebular emission requires us to ignore the cores
of hydrogen and neutral helium lines in the fitting of our target
stars, we also removed these cores from our test data set. In case of
the \hei\ lines 2~\AA\ from the central core was cut out. From the
hydrogen lines, with exception of \ha, 3~\AA\ was removed. Of all
observed line profiles \ha\ exhibits the strongest nebular
contamination. Therefore, a larger region of 5~\AA\ was removed from
its core.

The lines from the FLAMES data that will be fitted are the hydrogen
Balmer lines \ha\, \hg\ and \hd; the \hei\ singlet line at 4387~\AA;
the \hei\ triplet lines at 4026, 4471 and 4713~\AA, where the first
line is actually a blend with \heii; and the \heii\ lines at 4200,
4541 and 4686~\AA. In the formal test we also fit this set of
lines. Table~\ref{tab:form-tests} lists, for the above set of lines
for the different data sets the final fit parameters obtained by
evolving a population of 72 \fastwind\ models over a course of 200
generations. Also listed in Tab.~\ref{tab:form-tests} are the
parameter ranges for each parameter in which the fitting method was
allowed to search for the best fit.

For all three tests the automated method was able to find the global
optimum. There are some differences between the final fit parameters
and the input parameters. However, these can be explained by the low
quality of the data simulated and the fact that the sensitivity of
some parameters becomes less in certain parts of parameter space. The
latter explains the difference between the wind parameters found in
case of data set C. As the wind of this object is very weak (order
$10^{-8}$~\msunyr) very little information about it is available in
the spectrum. Consequently, the fit parameters describing the wind
($\beta$, \mdot) are relatively poorly constrained, with the errors in
\mdot\ becoming as large as the actual \mdot\ value.

The low S/N also decreases the sensitivity of the fit parameters. For
decreasing S/N the global optimum in parameter space becomes
shallower. Consequently, the error on the fit parameters increases, as
these are a measure of the width of this optimum (see
\citetalias{mokiem05}). This explains the differences in the
microturbulent velocities ($\sim$45\%), as well as the mass loss rate
found for data set~B.

To locate the best fit, respectively, approximately 100, 50 and 60
generations were needed in case of data set A, B and C. To obtain
robust results we adopt 150 generations as the minimum value in
fitting our programme stars.

\subsection{Nebular emission}
\label{sec:neb-em}

The subtraction of nebular emission features in multi-fibre data can
be problematic. The use of a combined sky-spectrum, for instance, does
not always result in a complete removal of the nebular component and
in other cases may result in an over-subtraction of nebular
features. To best cope with these potential problems in the automated
fitting method we only consider the wings of the line profiles,
ignoring any core nebular contamination. As in most cases the core
nebular feature is well resolved \citep[also see][]{evans05} and as
the removal of the line cores does not hamper an accurate
determination of the fit parameters (see Sect.~\ref{sec:form_tests}),
this approach seems justified. It is much more difficult to assess the
extent to which the determination of the fundamental parameters might
be influenced by residual nebular contamination or subtraction effects
in the line wings. Here we can only perform limited tests. To assess
the effect of too much sky-subtraction we looked in more detail to the
non-sky-subtracted data for \ngc346-010 and \ngc346-077.

To determine if the derived mass loss rates are affected by
over-subtraction of nebular features in the \ha\ Balmer line, we
refitted the non-sky-subtracted spectrum of the O7 giant
\ngc346-010. From the targets observed with FLAMES this object has the
smallest mass loss rate that we could determine accurately, i.e.\ with
error estimates smaller than 0.2~dex. Therefore, if incorrect
sky-subtractions would be an issue, the mass loss rate estimate of
this object would be affected the most. In comparison with the fit
parameters determined from the sky-subtracted spectrum, no significant
differences are found for any of the parameters obtained from the
non-sky-subtracted spectrum. In more detail, the mass loss derived
from the non-sky-subtracted spectrum is found to be larger by the
small amount of $\sim$0.07~dex, and can be attributed to the slightly
larger \logg\ ($\sim$0.08~dex). As no UV spectrum is available for
this object, \vinf\ was scaled with \vesc, resulting in an
approximately ten percent higher terminal flow velocity. As \mdot\ is
connected to \vinf\ through the continuity equation, this explains the
0.07~dex increase in the mass loss rate.

The second test we performed was on the spectrum of the O9 dwarf
\ngc346-077. This relatively faint object suffers from quite severe
nebular contamination in its line profiles. We find that the fit
parameters obtained from the non-sky-subtracted spectrum compare well
with the values determined from the sky-subtracted spectrum. Within
the error bars the two parameter sets, again, are in agreement. Only a
relatively large difference is found for the microturbulent
velocity. This parameter was found to be reduced by $\sim$14~\kmsec\
in the fit of the non-sky-subtracted spectrum. However, we do not
attribute this relatively large change to issues due to nebular
contamination. Instead, as the formal tests have shown, an accurate
determination of this parameter is notoriously difficult for low
signal to noise spectra (see also Sect.~\ref{sec:vturb}).

Now, what if our combined sky-spectrum underestimates the real sky
background in the line wings? To assess this, ideally one would like
to compare the current fitting results to results obtained from data
with a local sky-subtraction. For \ngc346-001 we had the opportunity
to perform such a test, as also the spectrum analysed by
\cite{crowther02} was available. A fit of this spectrum resulted in
parameters nearly identical to those obtained from the VLT-FLAMES
spectrum. In particular the wind parameters $\beta$ and $\mdot$, which
are expected to be the most sensitive to nebular contamination, were
found to agree within, respectively, 10 percent and 0.03~dex. In
itself, this agreement is reassuring, though, we note that \ngc346-001
is located away from the cluster centre \citep[see][]{evans06}.
Consequently, other objects might still suffer more from nebular
contamination. With respect to the mass loss rate determinations, we
note that from the objects in \ngc346 only four have a reliable mass
loss rate. Out of these only one object (\ngc346-033) lies close to
the core of the cluster. Therefore, its mass loss may be more
uncertain than is suggested by the formal errors (also see
Sect.~\ref{sec:wind_param}). The other three objects (\ngc346-001,
-010 and -012) lie at relatively large distances from the core, where
we anticipate the background contribution to be relatively small.

\begin{table*}[ht]
  \caption{Fundamental parameters determined using GA optimised
  spectral fits, with \teff\ in kK, \logg\ and \loggc\ in \cmsecsec,
  \rstar\ in \rsun, \lstar\ in \lsun, \vturb\ and \vsini\ in \kmsec,
  \mdot\ in \msunyr, \Ms\ and \Mev\ in \msun\ and \Qnull\ in number of
  photons per second. Results were obtained using a population of 72
  \fastwind\ models evolved over a minimum of 150
  generations. Gravities corrected for centrifugal acceleration
  (\loggc) were used to calculate the spectroscopic masses
  (\Ms). Evolutionary masses (\Mev) were derived from the tracks of
  \cite{charbonnel93}. Note that ``NGC'' is omitted from the
  identifications of the FLAMES targets.}
  \label{tab:results}
  \begin{center}
  \begin{tabular}{llcccrccrrclrrc}
  \hline\\[-9pt] \hline \\[-7pt]
  ID  & ST & \teff & \logg & \loggc & \multicolumn{1}{c}{\rstar} & $\log \lstar$ & \yhe
  & \multicolumn{1}{c}{\vturb} & \multicolumn{1}{c}{\vsini} & \mdot & \multicolumn{1}{c}{$\beta$}
  & \multicolumn{1}{c}{\Ms} & \multicolumn{1}{c}{\Mev} & \multicolumn{1}{c}{$\log \Qnull$} \\[1pt]
 \hline \\[-9pt]
346-001 & O7\,Iaf+ 	&  34.1 & 3.35 & 3.36 & 29.3 & 6.02 & 0.24 & 20.0 & 74 & 6.04$\cdot 10^{-6}$ & 1.15 & 71.5 & 65.5  & 49.60 \\[3.5pt]
346-007 & O4\,V((f+)) 	&  42.8 & 3.95 & 3.95 & 9.7 & 5.45 & 0.08 & 12.0 & 105 & 2.30$\cdot 10^{-7}$ & 0.80 & 30.9 & 39.2  & 49.16 \\[3.5pt]
346-010 & O7\,IIIn((f)) &  35.9 & 3.54 & 3.69 & 10.2 & 5.20 & 0.12 & 19.6 & 313 & 6.02$\cdot 10^{-7}$ & 0.80 & 18.6 & 27.4 & 48.76 \\[3.5pt]
346-012 & B1\,Ib 	&  26.3 & 3.35 & 3.35 & 12.1 & 4.80 & 0.07 & 11.1 & 29 & 1.24$\cdot 10^{-8}$ & 0.74 & 12.0 & 16.6  & 46.75 \\[3.5pt]
346-018 & O9.5\,IIIe 	&  32.7 & 3.33 & 3.37 & 11.1 & 5.10 & 0.10 & 0.0 & 138 & 9.65$\cdot 10^{-8}$ & 0.80 & 10.6 & 23.6  & 48.55 \\[3.5pt]
346-022 & O9\,V 	&  36.8 & 4.20 & 4.20 & 7.3 & 4.95 & 0.09 & 8.6 & 55 & 1.06$\cdot 10^{-7}$ & 0.80 & 31.3 & 23.5    & 48.37 \\[3.5pt]
346-025 & O9\,V 	&  36.2 & 4.07 & 4.08 & 7.2 & 4.90 & 0.10 & 6.3 & 138 & 1.25$\cdot 10^{-7}$ & 0.80 & 23.0 & 22.6   & 48.31 \\[3.5pt]
346-026 & B0\,IV 	&  32.6 & 3.76 & 3.76 & 9.2 & 4.93 & 0.11 & 10.6 & 67 & 5.25$\cdot 10^{-8}$ & 0.80 & 17.7 & 20.7   & 48.06 \\[3.5pt]
346-028 & OC6\,Vz 	&  42.9 & 3.97 & 3.97 & 6.5 & 5.10 & 0.16 & 10.3 & 27 & 1.00$\cdot 10^{-7}$ & 0.80 & 14.3 & 31.9   & 48.81 \\[3.5pt]
346-031 & O8\,Vz 	&  39.5 & 3.99 & 3.99 & 6.7 & 4.99 & 0.16 & 3.9 & 18 & 5.71$\cdot 10^{-8}$ & 0.80 & 15.8 & 26.7    & 48.60 \\[3.5pt]
346-033 & O8\,V 	&  39.9 & 4.44 & 4.45 & 6.6 & 4.99 & 0.07 & 17.5 & 188 & 7.42$\cdot 10^{-7}$ & 0.80 & 44.2 & 27.1  & 48.54 \\[3.5pt]
346-046 & O7\,Vn 	&  39.7 & 4.17 & 4.25 & 5.4 & 4.81 & 0.12 & 13.2 & 340 & 1.01$\cdot 10^{-7}$ & 0.80 & 18.7 & 24.0  & 48.39 \\[3.5pt]
346-050 & O8\,Vn 	&  37.2 & 4.16 & 4.25 & 5.2 & 4.67 & 0.13 & 10.2 & 357 & 7.30$\cdot 10^{-8}$ & 0.80 & 17.9 & 20.6  & 48.12 \\[3.5pt]
346-051 & O7\,Vz 	&  41.6 & 4.33 & 4.33 & 5.2 & 4.87 & 0.10 & 6.8 & 18 & 1.73$\cdot 10^{-7}$ & 0.80 & 21.5 & 26.5    & 48.50 \\[3.5pt]
346-066 & O9.5\,V 	&  35.6 & 4.25 & 4.26 & 5.2 & 4.59 & 0.09 & 16.4 & 129 & 9.75$\cdot 10^{-8}$ & 0.80 & 18.0 & 18.9  & 47.88 \\[3.5pt]
346-077 & O9\,V 	&  36.5 & 3.99 & 4.03 & 5.3 & 4.65 & 0.09 & 14.6 & 177 & 7.22$\cdot 10^{-8}$ & 0.80 & 10.8 & 19.9  & 48.09 \\[3.5pt]
346-090 & O9.5\,V 	&  34.9 & 4.26 & 4.28 & 5.3 & 4.56 & 0.09 & 11.3 & 188 & 9.82$\cdot 10^{-8}$ & 0.80 & 19.4 & 18.3  & 47.76 \\[3.5pt]
346-093 & B0\,V 	&  34.4 & 4.42 & 4.43 & 5.2 & 4.53 & 0.09 & 11.2 & 187 & 1.49$\cdot 10^{-7}$ & 0.80 & 26.3 & 17.8  & 47.60 \\[3.5pt]
346-097 & O9\,V 	&  37.5 & 4.49 & 4.49 & 5.6 & 4.75 & 0.08 & 8.5 & 22 & 2.03$\cdot 10^{-7}$ & 0.80 & 35.5 & 21.7    & 48.14 \\[3.5pt]
346-107 & O9.5\,V 	&  35.9 & 4.23 & 4.23 & 4.1 & 4.40 & 0.09 & 5.0 & 55 & 4.06$\cdot 10^{-8}$ & 0.80 & 10.4 & 17.9    & 47.73 \\[3.5pt]
346-112 & O9.5\,V 	&  34.4 & 4.15 & 4.17 & 4.3 & 4.36 & 0.10 & 15.6 & 143 & 2.44$\cdot 10^{-8}$ & 0.80 & 9.8 & 16.6   & 47.53 \\[1pt]
\hline\\[-9pt]														          
330-013 & O8.5\,II-III((f))\hspace{-8pt} &  34.5 & 3.40 & 3.41 & 14.1 & 5.40 & 0.18 & 19.1 & 73 & 2.96$\cdot 10^{-7}$ & 1.55 & 18.6 & 32.2 & 48.92 \\[3.5pt]
330-052 & O8.5\,Vn 	&  35.7 & 3.91 & 4.02 & 5.2 & 4.60 & 0.16 & 11.3 & 291 & 3.66$\cdot 10^{-8}$ & 0.80 & 10.5 & 19.0  & 48.00 \\[1pt]
\hline\\[-9pt]														          
\azv14  & O5\,V 	&  45.3 & 4.10 & 4.11 & 13.9 & 5.86 & 0.10 & 18.2 & 212 & 2.67$\cdot 10^{-7}$ & 0.80 & 90.9 & 61.7 & 49.60 \\[3.5pt]
\azv15  & O7\,II 	&  39.4 & 3.69 & 3.70 & 18.3 & 5.82 & 0.10 & 2.9 & 135 & 1.12$\cdot 10^{-6}$ & 1.12 & 60.9 & 53.9  & 49.53 \\[3.5pt]
\azv26  & O7\,III 	&  40.1 & 3.75 & 3.75 & 25.2 & 6.17 & 0.09 & 0.9 & 128 & 1.71$\cdot 10^{-6}$ & 1.17 & 132.0 & 85.7 & 49.86 \\[3.5pt]
\azv95  & O7\,III 	&  38.2 & 3.66 & 3.66 & 13.8 & 5.56 & 0.13 & 13.2 & 68 & 3.56$\cdot 10^{-7}$ & 1.16 & 32.1 & 39.3  & 49.19 \\[3.5pt]
\azv243 & O6\,V 	&  42.6 & 3.94 & 3.94 & 12.8 & 5.68 & 0.12 & 0.0 & 59 & 2.64$\cdot 10^{-7}$ & 1.37 & 52.4 & 49.0   & 49.39 \\[3.5pt]
\azv372 & O9\,Iabw 	&  31.0 & 3.19 & 3.22 & 28.7 & 5.83 & 0.11 & 20.0 & 135 & 2.04$\cdot 10^{-6}$ & 1.28 & 49.3 & 49.8 & 49.23 \\[3.5pt]
\azv388 & O4\,V 	&  43.3 & 3.95 & 3.96 & 10.6 & 5.55 & 0.09 & 13.2 & 163 & 3.34$\cdot 10^{-7}$ & 0.80 & 37.5 & 43.4 & 49.27 \\[3.5pt]
\azv469 & O8.5\,II((f)) &  34.0 & 3.41 & 3.42 & 20.6 & 5.70 & 0.17 & 19.8 & 81 & 1.10$\cdot 10^{-6}$ & 1.16 & 40.5 & 43.6  & 49.20 \\[1pt]
  \hline
  \end{tabular}
  \end{center}
A value of $\beta=0.80$ corresponds to an assumed fixed value
\end{table*}

\begin{table*}[ht]
  \caption{Optimum width based error estimates for the seven fit
  parameters. The ND entries correspond to error in \vturb\ that reach
  up to the maximum allowed value of \vturb\ and, therefore, are
  formally not defined. See text for details on the calculation of the
  uncertainties in the derived parameters. Units: \teff\ in kK, \logg\
  and \loggc\ in \cmsecsec, \rstar\ in \rsun, \lstar\ in \lsun,
  \vturb\ and \vsini\ in \kmsec, \mdot\ in \msunyr, \Ms\ and \Mev\ in
  \msun\ and \Qnull\ in number of photons per second. Note that
  ``NGC'' is omitted from the identifications of the FLAMES targets.}
  \label{tab:errors}
  \begin{center}
  \begin{tabular}{lcccccllclllc}
  \hline\\[-9pt] \hline \\[-7pt]
  ID & $\Delta$\teff & $\Delta$\loggc & $\Delta$\rstar & $\Delta \log
  \lstar$ & $\Delta$\yhe & \multicolumn{1}{c}{$\Delta$\vturb} &
  \multicolumn{1}{c}{$\Delta$\vsini} &
  $\log \Delta$\mdot & \multicolumn{1}{c}{$\Delta$$\beta$} &
  \multicolumn{1}{c}{$\Delta$\Ms} & \multicolumn{1}{c}{$\Delta$\Mev} & $\Delta \log \Qnull$ \\[1pt]
 \hline \\[-9pt]
346-001 & $^{-0.6}_{+0.6}$ & $^{-0.12}_{+0.17}$ & $\pm$2.0 & $\pm$0.06 & $^{-0.03}_{+0.06}$ & \hspace{6pt}$^{-2.8}_{\rm+ND}$  & \hspace{10pt}$^{-9}_{+15}$  & $^{-0.04}_{+0.05}$  & $^{-0.09}_{+0.06}$ & \hspace{4pt}$^{-19}_{+35}$ & \hspace{4pt}$^{-6}_{+6}$    & $^{-0.17}_{+0.16}$ \\[3.5pt]
346-007 & $^{-0.7}_{+1.5}$ & $^{-0.04}_{+0.08}$ & $\pm$0.7 & $\pm$0.08 & $^{-0.01}_{+0.01}$ & \hspace{6pt}$^{-8.7}_{+4.7}$    & \hspace{10pt}$^{-13}_{+10}$  & $^{-1.30}_{+0.23}$  & $-$ & \hspace{4pt}$^{-4}_{+8}$ & \hspace{4pt}$^{-3}_{+5}$                    & $^{-0.12}_{+0.20}$ \\[3.5pt]
346-010 & $^{-1.0}_{+1.3}$ & $^{-0.08}_{+0.13}$ & $\pm$0.7 & $\pm$0.09 & $^{-0.03}_{+0.04}$ & \hspace{6pt}$^{-7.6}_{\rm+ND}$  & \hspace{10pt}$^{-23}_{+27}$  & $^{-0.14}_{+0.09}$  & $-$ & \hspace{4pt}$^{-3}_{+6}$ & \hspace{4pt}$^{-3}_{+3}$                    & $^{-0.23}_{+0.30}$ \\[3.5pt]
346-012 & $^{-0.5}_{+0.8}$ & $^{-0.05}_{+0.10}$ & $\pm$0.8 & $\pm$0.08 & $^{-0.01}_{+0.01}$ & \hspace{6pt}$^{-1.3}_{+1.9}$    & \hspace{10pt}$^{-4}_{+4}$  & $^{-0.20}_{+0.52}$  & $^{-0.03}_{+0.41}$ & \hspace{4pt}$^{-2}_{+4}$ & \hspace{4pt}$^{-1}_{+1}$       & $^{-0.15}_{+0.20}$ \\[3.5pt]
346-018 & $^{-1.3}_{+1.1}$ & $^{-0.14}_{+0.15}$ & $\pm$0.8 & $\pm$0.09 & $^{-0.03}_{+0.05}$ & \hspace{6pt}$^{-0.0}_{+10.5}$   & \hspace{10pt}$^{-30}_{+38}$  & $^{-1.60}_{+0.56}$  & $-$ & \hspace{4pt}$^{-3}_{+4}$ & \hspace{4pt}$^{-2}_{+3}$                    & $^{-0.29}_{+0.25}$ \\[3.5pt]
346-022 & $^{-0.8}_{+0.9}$ & $^{-0.16}_{+0.20}$ & $\pm$0.5 & $\pm$0.07 & $^{-0.01}_{+0.04}$ & \hspace{6pt}$^{-8.4}_{+5.2}$    & \hspace{10pt}$^{-9}_{+11}$  & $^{-1.27}_{+0.50}$  & $-$ & \hspace{4pt}$^{-10}_{+19}$ & \hspace{4pt}$^{-1}_{+2}$                   & $^{-0.13}_{+0.15}$ \\[3.5pt]
346-025 & $^{-0.8}_{+1.2}$ & $^{-0.08}_{+0.24}$ & $\pm$0.5 & $\pm$0.08 & $^{-0.02}_{+0.03}$ & \hspace{6pt}$^{-6.1}_{+8.3}$    & \hspace{10pt}$^{-14}_{+17}$  & $^{-1.39}_{+0.41}$  & $-$ & \hspace{4pt}$^{-5}_{+17}$ & \hspace{4pt}$^{-2}_{+2}$                   & $^{-0.13}_{+0.17}$ \\[3.5pt]
346-026 & $^{-1.2}_{+0.4}$ & $^{-0.17}_{+0.05}$ & $\pm$0.6 & $\pm$0.09 & $^{-0.01}_{+0.03}$ & \hspace{6pt}$^{-3.1}_{+4.4}$    & \hspace{10pt}$^{-5}_{+9}$  & $^{-1.35}_{+0.27}$  & $-$ & \hspace{4pt}$^{-6}_{+3}$ & \hspace{4pt}$^{-2}_{+2}$                      & $^{-0.28}_{+0.13}$ \\[3.5pt]
346-028 & $^{-1.4}_{+1.1}$ & $^{-0.12}_{+0.17}$ & $\pm$0.4 & $\pm$0.08 & $^{-0.05}_{+0.04}$ & \hspace{6pt}$^{-8.7}_{+6.1}$    & \hspace{10pt}$^{-13}_{+12}$  & $^{-1.40}_{+0.44}$  & $-$ & \hspace{4pt}$^{-4}_{+7}$ & \hspace{4pt}$^{-3}_{+3}$                    & $^{-0.19}_{+0.16}$ \\[3.5pt]
346-031 & $^{-1.4}_{+1.2}$ & $^{-0.18}_{+0.24}$ & $\pm$0.5 & $\pm$0.08 & $^{-0.04}_{+0.06}$ & \hspace{6pt}$^{-3.7}_{+6.3}$    & \hspace{10pt}$^{-9}_{+10}$  & $^{-1.70}_{+0.63}$  & $-$ & \hspace{4pt}$^{-6}_{+12}$ & \hspace{4pt}$^{-3}_{+3}$                    & $^{-0.18}_{+0.17}$ \\[3.5pt]
346-033 & $^{-0.8}_{+1.6}$ & $^{-0.16}_{+0.16}$ & $\pm$0.5 & $\pm$0.09 & $^{-0.01}_{+0.02}$ & \hspace{6pt}$^{-9.8}_{\rm+ND}$  & \hspace{10pt}$^{-26}_{+35}$  & $^{-0.23}_{+0.15}$  & $-$ & \hspace{4pt}$^{-14}_{+21}$ & \hspace{4pt}$^{-3}_{+3}$                  & $^{-0.13}_{+0.20}$ \\[3.5pt]
346-046 & $^{-1.8}_{+1.7}$ & $^{-0.29}_{+0.23}$ & $\pm$0.4 & $\pm$0.10 & $^{-0.03}_{+0.06}$ & \hspace{6pt}$^{-13.0}_{\rm+ND}$ & \hspace{10pt}$^{-27}_{+45}$  & $^{-1.00}_{+0.42}$  & $-$ & \hspace{4pt}$^{-8}_{+11}$ & \hspace{4pt}$^{-2}_{+3}$                   & $^{-0.22}_{+0.21}$ \\[3.5pt]
346-050 & $^{-1.0}_{+1.3}$ & $^{-0.21}_{+0.20}$ & $\pm$0.4 & $\pm$0.08 & $^{-0.03}_{+0.05}$ & \hspace{6pt}$^{-10.0}_{\rm+ND}$ & \hspace{10pt}$^{-31}_{+33}$  & $^{-1.85}_{+0.46}$  & $-$ & \hspace{4pt}$^{-6}_{+9}$ & \hspace{4pt}$^{-1}_{+2}$                    & $^{-0.15}_{+0.18}$ \\[3.5pt]
346-051 & $^{-1.6}_{+0.9}$ & $^{-0.24}_{+0.17}$ & $\pm$0.4 & $\pm$0.09 & $^{-0.02}_{+0.04}$ & \hspace{6pt}$^{-6.6}_{+6.4}$    & \hspace{10pt}$^{-9}_{+14}$  & $^{-1.48}_{+0.31}$  & $-$ & \hspace{4pt}$^{-9}_{+11}$ & \hspace{4pt}$^{-3}_{+3}$                    & $^{-0.20}_{+0.15}$ \\[3.5pt]
346-066 & $^{-1.4}_{+1.8}$ & $^{-0.16}_{+0.25}$ & $\pm$0.4 & $\pm$0.10 & $^{-0.02}_{+0.03}$ & \hspace{6pt}$^{-10.8}_{\rm+ND}$ & \hspace{10pt}$^{-22}_{+22}$  & $^{-0.99}_{+0.40}$  & $-$ & \hspace{4pt}$^{-6}_{+14}$ & \hspace{4pt}$^{-2}_{+2}$                   & $^{-0.32}_{+0.38}$ \\[3.5pt]
346-077 & $^{-1.0}_{+1.3}$ & $^{-0.08}_{+0.24}$ & $\pm$0.4 & $\pm$0.09 & $^{-0.01}_{+0.03}$ & \hspace{6pt}$^{-10.0}_{\rm+ND}$ & \hspace{10pt}$^{-19}_{+14}$  & $^{-1.58}_{+0.40}$  & $-$ & \hspace{4pt}$^{-2}_{+7}$ & \hspace{4pt}$^{-2}_{+2}$                    & $^{-0.15}_{+0.18}$ \\[3.5pt]
346-090 & $^{-1.4}_{+0.9}$ & $^{-0.20}_{+0.13}$ & $\pm$0.4 & $\pm$0.09 & $^{-0.01}_{+0.03}$ & \hspace{6pt}$^{-11.1}_{+7.3}$   & \hspace{10pt}$^{-18}_{+25}$  & $^{-1.98}_{+0.18}$  & $-$ & \hspace{4pt}$^{-7}_{+7}$ & \hspace{4pt}$^{-1}_{+2}$                    & $^{-0.32}_{+0.22}$ \\[3.5pt]
346-093 & $^{-2.2}_{+1.0}$ & $^{-0.29}_{+0.18}$ & $\pm$0.4 & $\pm$0.13 & $^{-0.03}_{+0.04}$ & \hspace{6pt}$^{-11.0}_{\rm+ND}$ & \hspace{10pt}$^{-32}_{+42}$  & $^{-1.48}_{+0.26}$  & $-$ & \hspace{4pt}$^{-13}_{+14}$ & \hspace{4pt}$^{-2}_{+2}$                  & $^{-0.46}_{+0.24}$ \\[3.5pt]
346-097 & $^{-1.0}_{+1.2}$ & $^{-0.18}_{+0.21}$ & $\pm$0.4 & $\pm$0.08 & $^{-0.01}_{+0.03}$ & \hspace{6pt}$^{-8.3}_{+5.1}$    & \hspace{10pt}$^{-13}_{+15}$  & $^{-0.99}_{+0.34}$  & $-$ & \hspace{4pt}$^{-12}_{+24}$ & \hspace{4pt}$^{-2}_{+2}$                  & $^{-0.15}_{+0.17}$ \\[3.5pt]
346-107 & $^{-1.4}_{+1.1}$ & $^{-0.27}_{+0.22}$ & $\pm$0.3 & $\pm$0.09 & $^{-0.01}_{+0.05}$ & \hspace{6pt}$^{-4.8}_{+6.6}$    & \hspace{10pt}$^{-13}_{+18}$  & $^{-1.89}_{+0.38}$  & $-$ & \hspace{4pt}$^{-5}_{+7}$ & \hspace{4pt}$^{-1}_{+1}$                    & $^{-0.31}_{+0.26}$ \\[3.5pt]
346-112 & $^{-1.3}_{+1.9}$ & $^{-0.21}_{+0.29}$ & $\pm$0.3 & $\pm$0.11 & $^{-0.02}_{+0.06}$ & \hspace{6pt}$^{-15.4}_{\rm+ND}$ & \hspace{10pt}$^{-30}_{+25}$  & $^{-1.28}_{+0.78}$  & $-$ & \hspace{4pt}$^{-4}_{+9}$ & \hspace{4pt}$^{-2}_{+2}$                    & $^{-0.30}_{+0.40}$ \\[1pt]  
  \hline\\[-9pt]
330-013 & $^{-0.9}_{+0.8}$ & $^{-0.15}_{+0.14}$ & $\pm$1.0 & $\pm$0.07 & $^{-0.03}_{+0.05}$ & \hspace{6pt}$^{-3.2}_{\rm+ND}$  & \hspace{10pt}$^{-11}_{+9}$  & $^{-0.12}_{+0.20}$  & $^{-0.30}_{+0.24}$ & \hspace{4pt}$^{-6}_{+8}$ & \hspace{4pt}$^{-3}_{+3}$      & $^{-0.23}_{+0.20}$ \\[3.5pt]
330-052 & $^{-1.6}_{+2.4}$ & $^{-0.20}_{+0.54}$ & $\pm$0.4 & $\pm$0.13 & $^{-0.07}_{+0.06}$ & \hspace{6pt}$^{-11.1}_{\rm+ND}$ & \hspace{10pt}$^{-43}_{+37}$  & $^{-1.91}_{+0.61}$  & $-$ & \hspace{4pt}$^{-3}_{+17}$ & \hspace{4pt}$^{-2}_{+3}$                   & $^{-0.35}_{+0.50}$ \\[1pt]  
  \hline\\[-9pt]																									  
\azv14 & $^{-1.0}_{+1.7}$ & $^{-0.10}_{+0.12}$ & $\pm$1.0 & $\pm$0.09 & $^{-0.01}_{+0.03}$  & \hspace{6pt}$^{-12.2}_{\rm+ND}$ & \hspace{10pt}$^{-22}_{+26}$  & $^{-1.33}_{+0.28}$  & $-$ & \hspace{4pt}$^{-20}_{+32}$ & \hspace{4pt}$^{-7}_{+10}$                 & $^{-0.15}_{+0.21}$ \\[3.5pt]
\azv15 & $^{-1.5}_{+1.9}$ & $^{-0.14}_{+0.20}$ & $\pm$1.3 & $\pm$0.10 & $^{-0.02}_{+0.03}$  & \hspace{6pt}$^{-2.7}_{+12.3}$   & \hspace{10pt}$^{-24}_{+20}$  & $^{-0.32}_{+0.19}$  & $^{-0.23}_{+0.53}$ & \hspace{4pt}$^{-17}_{+37}$ & \hspace{4pt}$^{-7}_{+8}$   & $^{-0.20}_{+0.23}$ \\[3.5pt]
\azv26 & $^{-1.0}_{+1.8}$ & $^{-0.08}_{+0.15}$ & $\pm$1.8 & $\pm$0.10 & $^{-0.02}_{+0.02}$  & \hspace{6pt}$^{-0.7}_{+12.5}$   & \hspace{10pt}$^{-26}_{+20}$  & $^{-0.20}_{+0.14}$  & $^{-0.17}_{+0.30}$ & \hspace{4pt}$^{-26}_{+59}$ & \hspace{4pt}$^{-13}_{+15}$ & $^{-0.15}_{+0.22}$ \\[3.5pt]
\azv95 & $^{-1.2}_{+0.6}$ & $^{-0.12}_{+0.12}$ & $\pm$0.9 & $\pm$0.08 & $^{-0.02}_{+0.04}$  & \hspace{6pt}$^{-7.2}_{\rm+ND}$  & \hspace{10pt}$^{-9}_{+9}$  & $^{-0.11}_{+0.17}$  & $^{-0.26}_{+0.24}$ & \hspace{4pt}$^{-8}_{+11}$ & \hspace{4pt}$^{-3}_{+5}$      & $^{-0.17}_{+0.12}$ \\[3.5pt]
\azv243 & $^{-0.6}_{+0.8}$ & $^{-0.07}_{+0.09}$ & $\pm$0.9 & $\pm$0.07 & $^{-0.02}_{+0.02}$ & \hspace{6pt}$^{-0.0}_{+9.2}$    & \hspace{10pt}$^{-6}_{+8}$  & $^{-0.22}_{+0.11}$  & $^{-0.16}_{+0.43}$ & \hspace{4pt}$^{-10}_{+14}$ & \hspace{4pt}$^{-4}_{+4}$     & $^{-0.12}_{+0.13}$ \\[3.5pt]
\azv372 & $^{-1.2}_{+0.7}$ & $^{-0.17}_{+0.16}$ & $\pm$2.0 & $\pm$0.09 & $^{-0.03}_{+0.04}$ & \hspace{6pt}$^{-4.6}_{\rm+ND}$  & \hspace{10pt}$^{-16}_{+20}$  & $^{-0.09}_{+0.09}$  & $^{-0.15}_{+0.19}$ & \hspace{4pt}$^{-16}_{+22}$ & \hspace{4pt}$^{-6}_{+7}$   & $^{-0.27}_{+0.18}$ \\[3.5pt]
\azv388 & $^{-0.7}_{+0.9}$ & $^{-0.03}_{+0.11}$ & $\pm$0.7 & $\pm$0.07 & $^{-0.01}_{+0.01}$ & \hspace{6pt}$^{-9.0}_{+1.6}$    & \hspace{10pt}$^{-12}_{+7}$  & $^{-0.31}_{+0.10}$  & $-$ & \hspace{4pt}$^{-5}_{+11}$ & \hspace{4pt}$^{-4}_{+4}$                    & $^{-0.12}_{+0.14}$ \\[3.5pt]
\azv469 & $^{-0.5}_{+0.4}$ & $^{-0.06}_{+0.09}$ & $\pm$1.4 & $\pm$0.06 & $^{-0.03}_{+0.03}$ & \hspace{6pt}$^{-4.6}_{\rm+ND}$  & \hspace{10pt}$^{-8}_{+10}$  & $^{-0.12}_{+0.09}$  & $^{-0.12}_{+0.21}$ & \hspace{4pt}$^{-7}_{+11}$ & \hspace{4pt}$^{-3}_{+4}$     & $^{-0.15}_{+0.14}$ \\[1pt]  
  \hline
  \end{tabular}
  \end{center}
\end{table*}

\subsection{Error estimates}
We define error estimates for the fit parameters by estimating the
width of the optimum in parameter space associated with the global
optimum. This width defines, as was argued in \citetalias{mokiem05},
the region in parameters space which contains models with comparable
fit quality. Consequently, by determining the maximum variation of the
parameters within this group of models, the error is estimated
\citepalias[see][]{mokiem05}.

Table~\ref{tab:errors} contains the optimum width based error
estimates of the fit parameters for each analysed object. Based on
these estimates the errors on the derived parameters (\rstar, \lstar,
\Ms\ and \Mev) in this table were calculated using the same approach
as was used in \citetalias{mokiem05}. The single difference is the
adopted uncertainty in the absolute visual magnitude. Here we adopt an
uncertainty of 0.14\magn. This value is equal to the sum of the
statistical and systematic error in the determination of the SMC
distance modules by \cite{harries03}. The method used to determine the
uncertainty in \Qnull\ is explained in Sect.~\ref{sec:ion_flux}.

\section{Fundamental parameters}
\label{sec:fun_par}
In this section we will discuss the stellar properties of the
investigated sample. Table~\ref{tab:results} lists the values
determined for the seven free parameters as well as quantities derived
from these. Error estimates on the parameters are given in
Tab.~\ref{tab:errors}. The fits of the spectra together with comments
on the individual objects are presented in the appendix.

\subsection{Effective temperatures}
\label{sec:teff}

\begin{figure}[t]
  \centering
  \resizebox{8.8cm}{!}{ \includegraphics{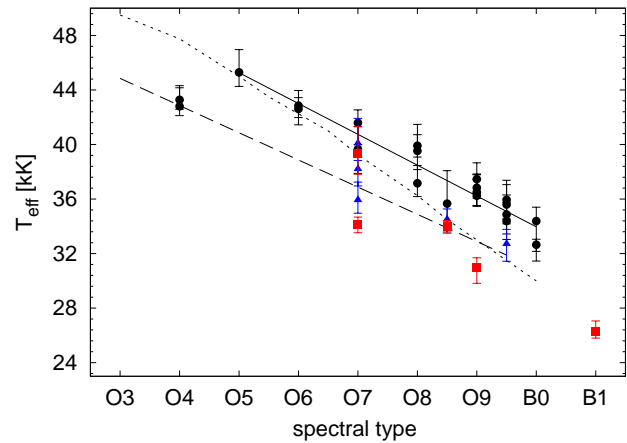}}
  \caption{Effective temperatures as a function of spectral type for
  SMC objects studied in this paper. Dwarfs, giants and supergiants
  are denoted by, respectively, solid circles, triangles and
  squares. Shown with a solid line is the average \teff\ relation for
  the dwarfs with spectral type later than O4. SMC dwarfs are found to
  be typically 3 to 4~kK hotter than the \teff\ calibration for
  Galactic O-type dwarfs according to \citet[][dashed
  line]{martins05a}. The dotted line corresponds to the SMC \teff\
  scale derived by \cite{massey05} for luminosity class V and III
  objects.}
  \label{fig:teff}
\end{figure}

The cumulative opacity of all spectral lines, referred to as line
blanketing, has a strong effect on both the structure and the emergent
spectrum of hot star atmospheres. It has been shown by several authors
that line blanketing changes the relation between spectral type (set
by the ionisation balance of mainly helium and/or silicon) and
effective temperature (related to the gas temperature of the line
forming layers). As line blanketing enhances the diffuse radiation
field, because lines ``trap'' the photons and introduce additional
back scattering \cite[see e.g.][]{schaerer97, repolust04}, models that
account for it can suffice with a lower temperature to match a given
spectral type \citep{dekoter98, martins02, crowther02, herrero02}. As
iron group lines dominate the line blanketing, the spectral type
vs. \teff\ relation is expected to depend on metallicity, i.e.\ SMC
stars of given spectral type will have a higher effective temperature
compared to their galactic counterparts.

Figure~\ref{fig:teff} shows the distribution of the effective
temperature as a function of O and early-B spectral sub-type for the
investigated sample. The different luminosity classes are denoted
using circles for the dwarfs and subgiants, triangles for the giants,
and squares for the bright giants and supergiants. We now concentrate
on dwarf stars only, which account for the vast majority in our
sample. For these objects the figure shows a well defined relation,
the mean of which is represented by the solid line. At the earliest
spectral types (O4--O6) the relation seems to flatten out. However, the
small number of objects analysed in this range make this part of the
diagram uncertain. Moreover, two of the objects at spectral type O4
and O5 suffer from strong nebular contamination, complicating the
determination of their effective temperature (see the appendix).

Shown with a dashed line in Fig.~\ref{fig:teff} is the \teff\
``observational'' calibration for Galactic O-type dwarf stars as
derived by \cite{martins05a}. A clear offset is apparent between this
relation and the average calibration for the SMC O dwarfs. This offset
is approximately 3.3\,kK for the latest types up to approximately
4.4\,kK for spectral type O5. These differences in \teff\ correspond
to a shift in spectral sub-type of about 2 for the late type objects
and $\sim$1.5 sub-types for the earliest types. \cite{mokiem04}, using
\cmfgen\ \citep{hillier98} and covering the metallicity range from 2
times Solar to 1/10 Solar, find typical shifts of one sub-type, though
we should add that this comparison is not as extensive as the one
presented here.

Figure~\ref{fig:teff} also shows a clear separation between objects of
different luminosity class. Compared to dwarfs, the giants, bright
giants, and supergiants systematically have lower effective
temperatures. The reason for this separation is twofold. First, the
latter group of objects represent more evolved evolutionary
phases. Their lower gravities result in an increased helium ionisation
\citep[e.g.][]{kudritzki83}, reducing the effective temperature
associated with a given spectral type. Second, these objects are
expected to have stronger stellar winds. This induces an increased
line blanketing effect, further reducing the \teff\ for a given
spectral type.

\cite{massey05} also report a Sp.Type(\teff) calibration for SMC
stars. As can be seen in Fig.~\ref{fig:teff} their relation
essentially agrees with ours at spectral types earlier than
O8. However, at later types their results suggest a rather sharp turn
towards the \citeauthor{martins05a} relation for Galactic stars which
is not observed in our sample. The reason for this apparent
discrepancy is that our calibration employs dwarfs, whilst
\citeauthor{massey05} had to rely upon giant stars at the latest O
subtypes.

\subsection{Ionising fluxes}
\label{sec:ion_flux}

The ionising fluxes of massive stars are important quantities that are
used in the study of, for instance, \ion{H}{ii} regions and
starburst galaxies \citep[e.g.][]{vacca94}. The parameter that is used
to characterise the ionising output is the number of photons present
in the Lyman continuum, \Qnull. It is defined as
\begin{equation}
  \Qnull = 4\pi^2 R_{\star}^{2} \int_0^{\lambda_0}
          \frac{\pi \lambda F_\lambda}{hc}d\lambda~,
\end{equation}
where $F_\lambda$ and $\lambda_0$ are, respectively, the stellar flux
in \mbox{erg\,s$^{-1}$\,cm$^{-2}$\,\AA$^{-1}$} and the limiting
wavelength for photons able to ionise hydrogen, i.e.\ 912~\AA.

Tables \ref{tab:results} and \ref{tab:errors} list the number of
ionising photons for the individual objects and the associated error
estimates. These errors are dominated by the uncertainty in stellar
radius and effective temperature. To calculate the error estimates we
adopted an uncertainty of 0.03~dex in \rstar, which is dominated by
the estimate for $\Delta$\Mv. The error introduced by $\Delta$\teff\
was estimated from Fig.~16 from \cite{martins05a}, which shows the
number of ionising Lyman continuum photons as a function of effective
temperature for the line blanketed stellar atmosphere codes \cmfgen,
{\sc wm-basic} and {\sc tlusty}. The difference between the
predictions of these three codes is relatively small and,
consequently, we estimated the uncertainty introduced in \Qnull\ to be
0.09~dex/kK for $\teff > 36$~kK and 0.18~dex/kK for $\teff < 36$~kK.

\begin{figure}[t]
  \centering
  \resizebox{8.8cm}{!}{ \includegraphics{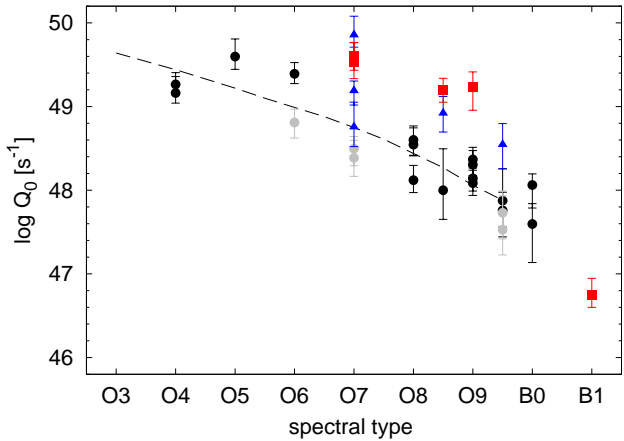}}
  \caption{Number of Lyman continuum ionising photons \Qnull\ as a
  function of spectral type. Different luminosity classes are shown
  using circles, triangles and squares for, respectively, IV-V, III
  and I-II class objects. Grey circles correspond to stars located on
  or left of the ZAMS. Indicated using a dashed line is the Galactic
  calibration for dwarfs from \cite{martins05a}.}
  \label{fig:Q_0}
\end{figure}

In Fig.~\ref{fig:Q_0} the distribution of \Qnull\ as a function of
spectral type for our programme stars is presented. Different
luminosity classes are indicated using circles, triangles and squares
for respectively, IV-V, III and I-II class objects. On average the
more evolved stars are found to produce more ionising photons for a
given spectral type because of their larger radii, hence higher
luminosities.  Also shown in this figure as a dashed line is the
``observational'' \Qnull\ calibration for Galactic dwarfs from
\cite{martins05a}. In general we find that the ionising fluxes of the
SMC dwarfs are in good agreement with this calibration \citep[also
see][]{mokiem04}. The most pronounced differences are for the O4 and
O7 stars. With respect to the earliest spectral type this can be
explained by nebular contamination hampering the \teff\ determination
(see previous section). The low ionising fluxes of the two O7 dwarfs,
however, cannot be explained in this way, as they are found to follow
the \teff\ trend in Fig.~\ref{fig:teff}. Instead, we believe that the
discrepancy is related to their location in the HR-diagram, i.e.\
evolutionary phase. The two dwarfs are found to locate a position on
or to the left of the ZAMS (see Sect.~\ref{sec:zams_stars}). As a
result of this they are less luminous compared to objects of the same
spectral type that are located to the right of the ZAMS. Consequently,
the total number of ionising photons produced by these objects is
smaller than the average associated with their spectral type.

Apart from the two dwarfs at spectral type O7 three additional stars
were found to lie on or to the left of the ZAMS (see
Sect.~\ref{sec:zams_stars}). As can be seen in Fig.~\ref{fig:Q_0},
where we have highlighted all ZAMS stars using grey circles, these
stars produce on average less ionising photons. Note that two dwarfs
at O8 and O8.5 also seem to lie below the average. The O8.5 star has a
below average \teff\ for its spectral type (see
Fig.~\ref{fig:teff}). This explains its somewhat peculiar \Qnull\
behaviour. The O8 star below the average is \ngc346-050. It also has a
temperature that is lower than the average for its spectral type,
though not to the same extent as \ngc330-052. Interestingly, this
object lies closest to the ZAMS (see Fig.~\ref{fig:hrd_age}) of all
non-ZAMS stars, and seems to behave in terms of \Qnull\ in a similar
manner as the ZAMS objects. We conclude that the ZAMS stars for given
spectral type have $\sim$0.4~dex lower Lyman continuum photons.

\subsection{Gravities}
\label{sec:gravities}

\begin{figure}[!t]
  \centering
  \resizebox{8.8cm}{!}{ \includegraphics{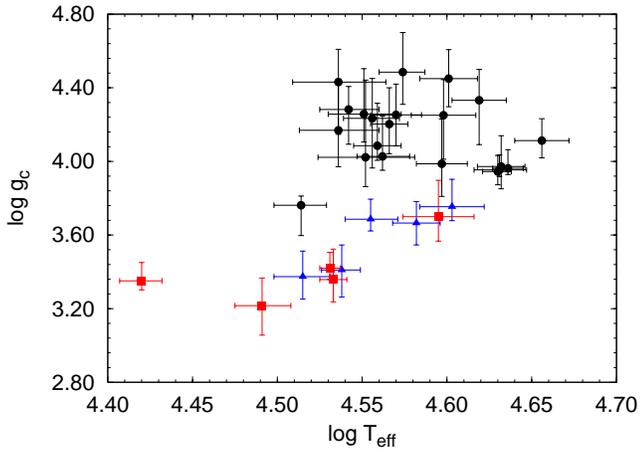}}
  \caption{Distribution of the analysed SMC objects in the $\log
  \teff$ -- \loggc\ plane. Indicated with different symbols are
  luminosity class IV-V (circles), III (triangles) and I-II
  (squares). With exception of the subgiant \ngc346-026, the first
  group of objects are clearly separated from the more evolved
  objects. These more evolved objects seem to follow a trend of
  increasing surface gravity with increasing effective temperature.}
  \label{fig:logg}
\end{figure}

In Fig.~\ref{fig:logg} we present the distribution of SMC objects in
the log \teff\ -- \loggc\ plane. The surface gravity corrected for
centrifugal acceleration (\loggc) was calculated according to the
method discussed by \cite{herrero92} and \cite{repolust04}. Different
luminosity classes are indicated using circles, triangles, and squares
for type IV-V, III, and I-II objects, respectively. With the exception
of one object, the dwarf type objects are all above $\loggc \simeq
3.9$. The object with a lower gravity, \ngc346-026 is the only
subgiant. The luminosity class I-III objects occupy a strip below the
dwarfs, reflecting the evolutionary path of hot massive stars in this
diagram. Note that no separation between luminosity class III and I-II
objects is visible in this secular behaviour.

\begin{figure}[t]
  \centering
  \resizebox{8.8cm}{!}{ \includegraphics{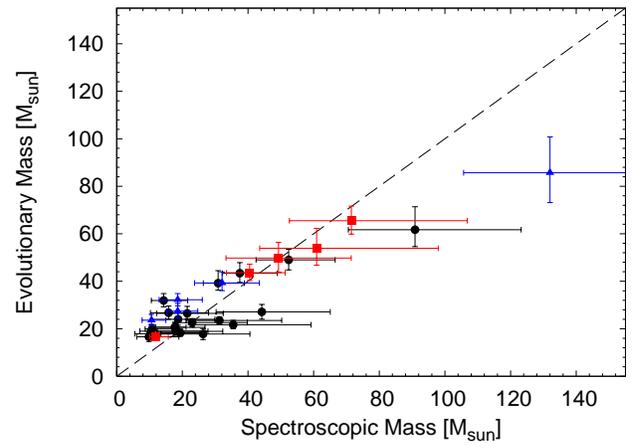}}
  \caption{Comparison of spectroscopically determined masses with
    masses derived from the evolutionary tracks of \cite{charbonnel93}
    for $Z=0.2\,\zsun$. Denoted by circles, triangles and squares are,
    respectively, luminosity classes IV-V, III and I-II. The dashed
    line corresponds to the one-to-one relation between the two mass
    scales. With exception of the two most massive objects, no
    systematic discrepancy between the spectroscopic and evolutionary
    masses is observed.}
  \label{fig:mass}
\end{figure}

A comparison of the masses based on the spectroscopically determined
surface gravities and those derived from predictions of massive star
evolution is presented in Fig.~\ref{fig:mass}. The different
luminosity classes are distinguished using the same symbols as in
Fig.~\ref{fig:logg}. To determine the evolutionary masses evolutionary
tracks from \cite{charbonnel93}, for $Z = 0.2\,\zsun$ were used. The
errors in evolutionary mass reflect the mass interval allowed within
the error box spanned by the stellar luminosity and effective
temperature. As the tracks of \citeauthor{charbonnel93} do {\em not}
account for the effects of rotation, this source of error is not
included. Predictions accounting for \vrot\ show complicated tracks
including loops during the secular redward evolution. Therefore, one
can no longer assign an unambiguous $M(L,\teff)$. Still, assessing the
impact of rotation using the \cite{maeder01} and \cite{meynet05}
computations that adopt an initial rotational velocity $\vr =
300\,\kmsec$ shows that the error in the evolutionary mass will not
increase by more than $\sim$13 percent. The errors in the
spectroscopic mass are much larger than those in \Mev, and primarily
reflect the error in gravity.

Inspection of Fig.~\ref{fig:mass} reveals no convincing systematic
discrepancy between the spectroscopic and evolutionary mass, even
though some objects do not agree within their standard deviation with
the one-to-one relation. At the low mass end there appears to be some
tendency for the spectroscopic masses to be less than those from
evolutionary tracks. This behaviour is similar to the ``mass
discrepancy problem'' reported and discussed by e.g.\ \cite{herrero02}
and \cite{repolust04}. Note that most of this classical problem has
been resolved, i.e.\ it has been attributed to limitations of the
stellar atmosphere models \citep{herrero93} and biases in the fitting
process (see \citetalias{mokiem05}). With respect to the stars at the
low mass end, we note that star \ngc346-107 occupies a location in the
HR-diagram left of the ZAMS (see Fig.~\ref{fig:hrd}). As the
evolutionary status of this object is formally not defined, the mass
that is given is based on an extrapolation of the tracks. This could
lead to an erroneous value of \Mev. Interestingly, for the other stars
at the low mass end the helium abundances listed in
Tab.~\ref{tab:results} seem to correlate with the mass discrepancy. We
will investigate this in detail in the next section. Also note that
\cite{massey05} identify a mass discrepancy in a sample of Magellanic
Cloud stars for objects with $\teff \gtrsim 45$~kK, which they
attribute to a possible underestimation of \logg\ by the model
atmospheres. Unfortunately, our sample does not contain enough
early-type O stars to corroborate the findings of these authors.

For the two brightest objects, \azv14 and \azv26, the spectroscopic
masses are much larger than the implied evolutionary masses. The
profile fits, in particular those of the gravity sensitive hydrogen
Balmer lines, (see the appendix) are good. So, it is not likely that
the reason for the discrepancy is an overestimation of the
spectroscopically determined gravity. One could speculate about a
possible binary nature of both stars, as the spectroscopic mass is
more sensitive to changes in luminosity. An indication of binarity may
be the conspicuously high luminosities. However, using spectroscopic
and spectral morphological arguments, \cite{massey04} tentatively rule
out a composite explanation for both objects. In contrast to this we
note that the luminosity and derived mass-loss rate for these two
stars (see Sect.~\ref{sec:wind_param}) imply a position in the
modified wind momentum vs. luminosity diagram which is well below what
is expected from theory, i.e.\ the mass loss for these two stars would
be in better agreement with predictions if their luminosity would be
lower.

\subsection{Helium abundances}
\label{sec:yhe}

\begin{figure}[t]
  \centering
  \resizebox{8.8cm}{!}{ \includegraphics{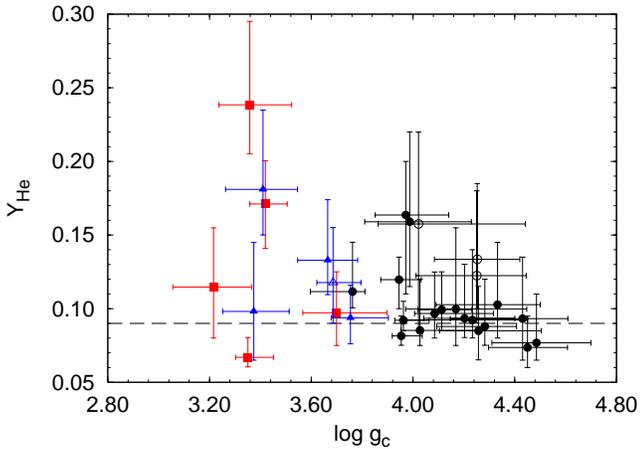}}
  \caption{Helium abundances as a function of surface
  gravity. Luminosity classes IV-V, III and I-II are denoted using
  circles, triangles and squares, respectively. Open symbols
  correspond to fast rotators ($\vsini > 250\,\kmsec$). The average
  helium abundance of the investigated sample is 0.13. Shown with a
  dashed line is a measure for the initial \yhe, which is taken to be
  the average of the helium abundances of the dwarfs with a \yhe\
  smaller than the sample average. Compared to this ``initial''
  abundance of 0.09 an increase in \yhe\ is visible for decreasing
  surface gravity.}
  \label{fig:yhe}
\end{figure}

The automated method also treats the helium abundance as a {\em
continuous} free parameter. Usually, spectroscopic analyses assume an
initial Solar value for \yhe, which is only modified when no
satisfying fit can be obtained \citep[e.g.][]{herrero02, repolust04,
massey04}. Because of the automated treatment and the extent of our
sample we can, for the first time rigorously, investigate possible
correlations between the surface helium abundance and other
fundamental parameters.

In Fig.~\ref{fig:yhe} we show \yhe\ as a function of surface gravity
\loggc. The horizontal dashed line represents the initial helium
abundance of the SMC stars investigated. It corresponds to the average
of the helium abundances of the dwarf objects with \yhe\ smaller than
the total sample average. Using this ``initial'' abundance of
$0.09^{+0.009}_{-0.004}$ as a reference, the overall trend is that the
average \yhe\ increases for decreasing surface gravity. This is
consistent with the standard picture that more evolved objects may
have their atmospheres enriched with primary helium. Interestingly,
however, one may immediately spot two deviating objects from the
overall trend. First, the supergiant \ngc346-012 has a helium
abundance lower than the ``initial'' value. This object has $\loggc =
3.35$ and $\yhe = 0.07$. The reason for the low surface helium
abundance is unclear and we will exclude this star from the remainder
of this discussion.

Second, some of the unevolved objects have enhanced helium abundances.
It can therefore be suspected that more parameters are involved in
controlling the enrichment displayed in Fig.~\ref{fig:yhe}.
One such parameter could be stellar rotation. \cite{meynet00}, for
instance, predicted that extensive mixing in fast rotators could
result in significant surface helium enhancement relatively early in
the evolution. To probe this possibility we have highlighted the fast
rotators, defined as having $\vsini > 250\,\kmsec$, in
Fig.~\ref{fig:yhe} using open symbols. In case of all four fast
rotators we see that their helium abundances are enhanced with respect
to the ``initial'' value, suggesting that the helium enhancement may
be (partly) related to fast rotation. This may imply that two other
dwarf objects with a clear helium enhancement ($\yhe \approx 0.16$)
but with low projected rotational velocity are in fact fast rotators
seen pole on.

\begin{figure}[t]
  \centering
  \resizebox{8.74cm}{!}{\includegraphics{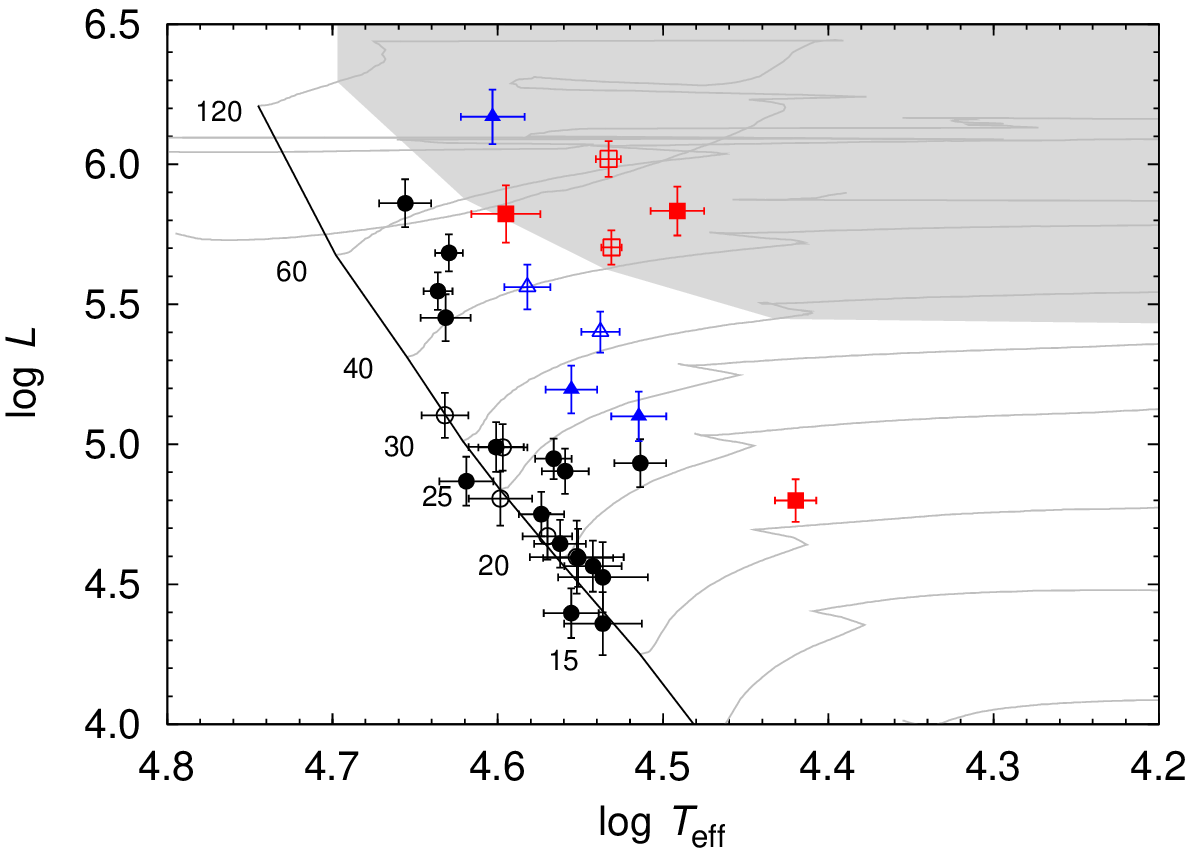}}
  \caption{HR-diagram of the analysed SMC sample. Different luminosity
  classes are distinguished using identical symbols as used in
  Fig.~\ref{fig:yhe}. Open symbols correspond to objects with a helium
  abundance of at least 0.12. Over-plotted in grey are the
  evolutionary tracks of \cite{maeder01} and \cite{meynet05} with
  $Z=0.2\,\zsun$ and $\vr = 300\,\kmsec$. The ZAMS corresponding to
  these tracks is shown as a black solid line. The grey area in the
  top part of the diagram corresponds to the region in which the
  evolutionary models exhibit an enhancement of the surface
  helium abundance by ten percent or more.}
  \label{fig:hrd}
\end{figure}

How well do our spectroscopically derived helium abundances compare to
those predicted by evolutionary models? This is a fundamental but
complicated question. Important to realise is that one of the effects
of rotation is that it introduces a wide bifurcation in the
evolutionary tracks. Stars rotating above a threshold \vr\ of about
30--50 percent of break-up (depending on initial mass;
\citealt{yoon05}) follow tracks which are essentially those of
chemically homogeneous evolution. Such tracks do not evolve towards
the red in the HRD, but evolve bluewards (from the zero age main
sequence) and upwards until the star enters the Wolf-Rayet phase
\citep{maeder87}. Stars rotating below this critical value evolve
along tracks which are about similar to non-rotating ones, though at
an earlier age the surface abundances of helium (and carbon, nitrogen
and oxygen) will be affected by rotation induced mixing.

Let us first compare our results with tracks for rotational velocities
below this critical value. In the HR-diagram shown in
Fig.~\ref{fig:hrd} the grey lines correspond to evolutionary
predictions by \cite{maeder01} and \cite{meynet05}, which were
calculated for $Z = 0.2\,\zsun$ and $\vrot = 300\,\kmsec$. Objects for
which a spectroscopic helium abundance of at least $\yhe = 0.12$ was
found are denoted using open symbols. Note the location of four helium
rich dwarf stars close to or even on the ZAMS. These will be discussed
in more detail in Sect.~\ref{sec:age}.  The grey area in the figure
corresponds to the region in which the evolutionary models predict a
surface helium enhancement of at least ten percent.

The regime in which the evolved stars showing significant helium
surface enrichment reside roughly coincides with the location of the
grey area. Not all evolved objects show evidence of \yhe\
enrichment. Many exhibit an abundance about equal to the initial
value. None of these stars are fast rotators (they all have $\vsini
\lesssim 140 \kmsec$). This can easily be explained using tracks for
non-rotating stars \citep{meynet00}. The fair consistency between
observed and predicted helium abundance {\em in evolved objects} is
reassuring, however, a more detailed comparison requires the
availability of tracks for several more values of \vrot.

\begin{figure}[t]
  \centering \resizebox{8.8cm}{!}{\includegraphics{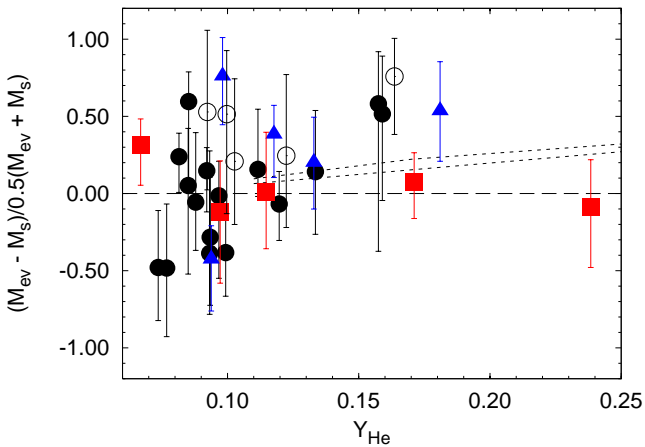}}
  \caption{The mass discrepancy (see also Fig.\,\ref{fig:mass}) as
  a function of surface helium abundance. To compute the discrepancy
  non-rotating tracks of \cite{charbonnel93} are used. Circles,
  triangles, and squares denote dwarfs, giants, and supergiants,
  respectively. The open circles indicate stars on or left of the
  ZAMS. The bulk of the stars has the initial helium abundance $\yhe =
  0.09$ where the discrepancy seems random and of the order of the
  error. For helium enriched stars the discrepancy always favours a
  higher \Mev, qualitatively consistent with models accounting for
  mixing increasing the $M/L$-ratio and bringing helium to the
  surface. A quantitative prediction (short dashed lines) based on
  ZAMS stars of 20 (upper curve) and 30~\msun\ (lower curve) varying
  helium abundance shows, however, a more modest effect than implied
  by our findings.}
  \label{fig:massdisc-yhe}
\end{figure}

We mentioned above that for modest rotation the tracks do not differ
greatly from non-rotating ones. This is not completely correct as
rotation tends to make the star somewhat more luminous.
\cite{langer92} proposed that this might explain the mass discrepancy
problem identified by \cite{herrero92}.  Although recent analyses no
longer suggest a convincing systematic discrepancy (see
e.g. Sect.\,\ref{sec:gravities}), it is still interesting to look at
this idea in some more detail. Langer connects the apparent mass
problem to the helium abundance by showing that the $M/L$-ratio is a
monotonically decreasing function for increasing helium
enrichment. Consequently, if mixing brings primary helium to the
surface one expects the star to be overluminous, leading to an
overestimate of the evolutionary mass if non-rotating tracks are
adopted. If indeed this is the case, then the scatter around the
one-to-one relation in Fig.\,\ref{fig:mass} might reveal a trend {\em
when plotted against \yhe.}

The result of this exercise is shown in
Fig.\,\ref{fig:massdisc-yhe}. On the vertical axis a measure of the
mass discrepancy is given. Note that the mass difference is plotted
relative to the mean of the evolutionary and spectroscopic mass to
ensure that positive and negative discrepancies are shown on the same
linear scale. For the evolutionary masses, tracks that do not account
for rotation are used. The circles, triangles, and squares denote
dwarfs, giants, and supergiants, respectively. The open circles
indicate stars on or left of the ZAMS. At the ``initial'' helium
abundance of $\yhe = 0.09 \pm 0.01$ -- where most of the stars reside
and dwarfs dominate -- the scatter around the $\Mev = \Ms$ relation
appears random. This essentially reflects that there is no systematic
mass discrepancy. At $0.13 \lesssim \yhe \lesssim 0.19$ the scatter is
not random, as all objects show a positive mass discrepancy. In
principle this is qualitatively consistent with the above described
idea. However, is it also quantitatively consistent? To assess this we
have computed the mass discrepancy for ZAMS stars with a variable
helium abundance. This should reflect the maximum effect of rotation,
i.e. such effective mixing that it leads to chemically homogeneous
evolution. The results for stellar masses of 20 and 30~\msun\ --
typical for the bulk of our sample -- are shown (short dashed
lines). These predictions clearly show a more modest mass discrepancy,
though the error bars on the mass discrepancy for the programme stars
do reach these predictions.  {\em We conclude that stars with an
enriched helium surface abundance tend to show a systematic mass
discrepancy that is qualitatively consistent with predictions of
chemically homogeneous evolution.}  We finally note that the
supergiants in our sample, which can be explained using evolutionary
models including rotation, show the best agreement between \Mev\ and
\Ms\ (see also Fig.\,\ref{fig:mass}).

\subsection{Microturbulence}
\label{sec:vturb}

Similar to the helium abundance the fact that the microturbulent
velocity is treated as a free parameter allows us for the first time
to investigate possible correlations for this parameter. However, in
contrast to the former parameter, this investigation did not yield any
clear relation between \vturb\ and any other parameter. For each
parameter given in Tab.~\ref{tab:results} a comparison with the
microturbulent velocity basically results in a scatter diagram. This
null result is similar to the findings in \citetalias{mokiem05} and
reflects the uncertainty with which \vturb\ can be determined from the
hydrogen and helium spectrum. Apparently the line profiles are not
very sensitive to this parameter.

\begin{figure*}[t]
  \centering
  \resizebox{14cm}{!}{\includegraphics{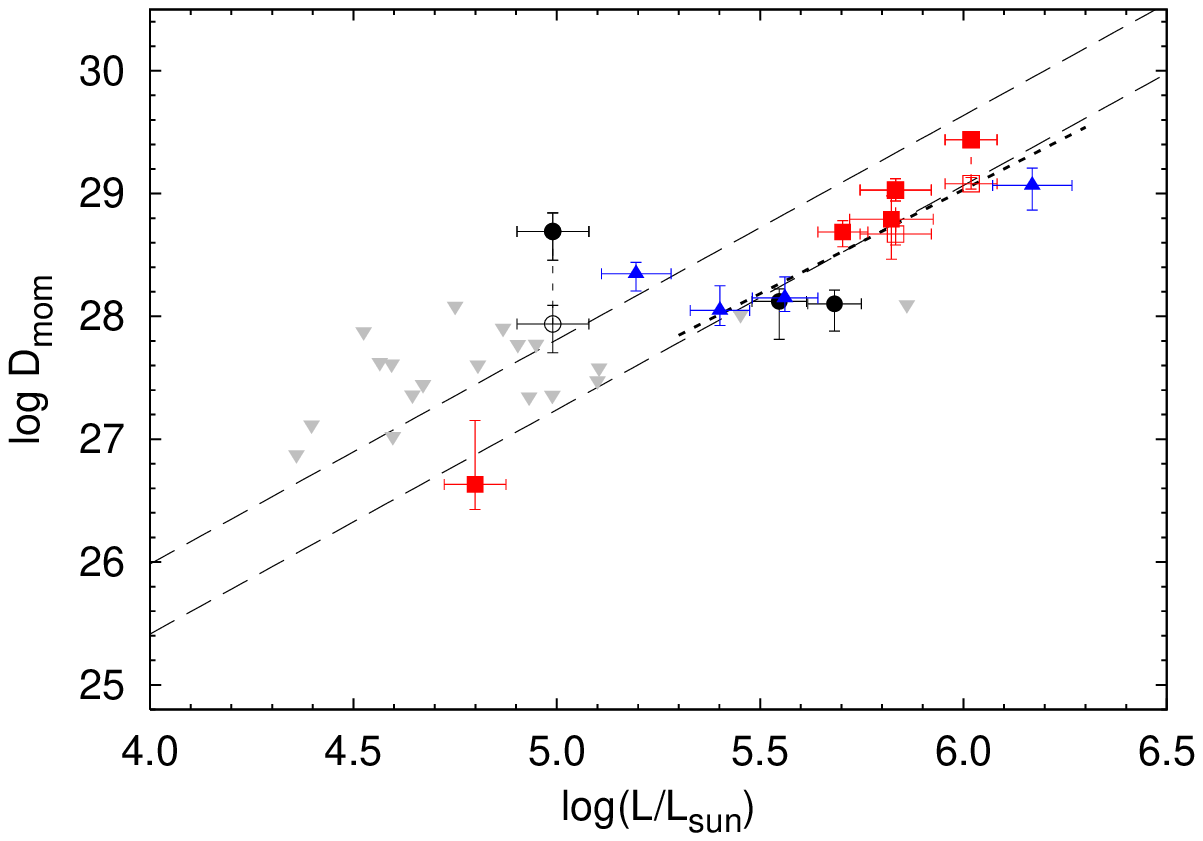}}
  \caption{Modified wind momentum (\Dmom) in units of ${\rm
  g\,cm\,s^{-2}}\,\rsun$ as a function of luminosity for the analysed
  SMC sample. Luminosity classes are distinguished using identical
  symbols as in Fig.~\ref{fig:yhe}. Indicated using grey inverted
  triangles are upper limits. At $\log \lstar/\lsun \approx 5.0$ the
  open symbol indicates the location \ngc346-033 would have if its
  true \vinf\ would be lower by a factor two compared to the adopted
  value. Using open squares modified wind momenta corrected for
  clumping are shown for the two supergiants with \ha\ emission
  profiles. The upper and lower dashed lines correspond to the
  theoretical wind-momentum luminosity relations (WLR) as derived by
  \cite{vink00, vink01} for, respectively, Galactic and SMC
  metallicity. Shown with a dotted line is the empirical WLR
  constructed for objects with $\log \lstar/\lsun \gtrsim 5.4$.}
  \label{fig:wlr}
\end{figure*}

We should also consider the fact that the error estimates determined
for \vturb, as given in Tab.~\ref{tab:errors} are on average
considerably large. As a result of this, one could argue that for many
objects it simply was not possible to accurately determine
\vturb. Consequently, it is not possible to find any correlation when
the total sample is considered. To avoid this potential problem we
also investigated possible correlations using a subset of 14 objects
for which the microturbulent velocity was determined relatively
well. The selection criterion for this subset was that the \vturb\
error bars should be well confined within the search domain, spanning
the range of 0 up to 20~\kmsec. The comparison within the subset again
did not result in a correlation for \vturb\ with any of the other
parameters.

\subsection{Wind parameters}
\label{sec:wind_param}

Our sample is dominated by late O-type dwarf stars, which are expected
to have relatively weak winds.  These are so weak that they challenge
the sensitivity of \ha\ as a mass-loss diagnostic. With nebular
contamination as an additional complicating factor, we could not
derive reliable \mdot\ for all objects. For nineteen stars (see
Tab.~\ref{tab:results}) we can only derive upper limits, i.e.\ the
downward error bars on our fit extend to the lower limit of the \mdot\
regime in which the automated method was allowed to search for a
solution. It happens to be so that these can be identified by an error
bar $-\log \Delta \mdot \gtrsim 1.0$ dex (see
Tab.~\ref{tab:results}). For many stars we could also not determine
the acceleration behaviour of the wind, expressed by the exponent
$\beta$ of the velocity law. For these objects we adopted $\beta =
0.8$ consistent with theoretical expectations \citep{pauldrach86}.

Despite the large number of upper limits, it is still possible to
quantitatively investigate the SMC stellar winds. We do this by
studying the distribution of the analysed objects in the so-called
modified stellar wind momentum vs.\ luminosity diagram. The modified
wind-momentum, which is defined as $\Dmom \equiv \mdot \vinf
R^{1/2}_{\star}$, is predicted to behave as a power-law as a function
of stellar luminosity \citep{kudritzki95, puls96}
\begin{equation}
  \log \Dmom = x \log \left(\lstar/\lsun\right) + \log D_{\circ}~.
\end{equation}
In the above equation $x$ corresponds to the inverse of the slope of
the line-strength distribution function corrected for ionisation
effects \citep{puls00} and $D_{\circ}$ is a measure for the effective
number of lines contributing to the acceleration of the wind.

In Fig.~\ref{fig:wlr} we present the distribution of the modified wind
momenta of our sample. The upper dashed curve is the theoretical
prediction for a Galactic metal content; the lower dashed curve is
that for an SMC metallicity (\citealt{vink01}; see also below), which
is predicted to be shifted downward by 0.57~dex with respect to the
Galactic relation. Before confronting theory with observations, we
first discuss a few individual objects. The \Dmom\ upper limit at
$\log \lstar/\lsun \approx 5.9$ corresponds to the object \azv14. As
was argued in Sect.~\ref{sec:gravities}, this object might be
over-luminous because of a binary nature.

The O8\,V star \ngc346-033 (at $\log \lstar/\lsun \approx 5.0$) is
positioned far above the SMC prediction. We suspect this is connected
to the anomalously high terminal velocity of 4100~\kmsec, which
results from a scaling with \vesc\ as no direct UV measurement is
available. If this \vinf\ would be overestimated by a factor of two
(for an O8\,V star one would expect $\vinf \sim 1900\,\kmsec$, cf.\
\citealt{kudritzki00}) the \Dmom\ would be reduced by approximately a
factor of six. This is so because \Dmom\ scales directly with \vinf\
and indirectly with $\vinf^{3/2}$ through the invariant wind-strength
parameter ($Q \propto \mdot / \vinf^{3/2}$, e.g.\ see
\citealt{puls05}). The open circle shows the effect of such a decrease
in terminal velocity. We decided to exclude this object from the
remainder of this discussion.

The two O-type supergiants in our sample, \azv372\ and \ngc346-001,
are the only two exhibiting an \ha\ emission profile. \cite{markova04}
and \cite{repolust04} have argued that for such stars the mass loss
may be overestimated relative to dwarf stars due to wind clumping
effects. For dwarf stars the \ha\ absorption line is formed relatively
close to the stellar surface, where clumping may be negligible. The
\ha\ emission line is typical for supergiants and in contrast reflects
that the line is formed over a larger volume, where -- they propose --
clumping has set in. Indeed, for \ngc346-001 \cite{crowther02} present
evidence for wind clumping based on the analysis of the UV phosphorus
lines. \cite{repolust04} derive a correction for this clumping by
multiplying the mass loss by a factor 0.44, which we have applied for
the two objects. The corrected modified wind-momenta are shown in
Fig.~\ref{fig:wlr} using open squares.

For SMC objects, Fig.~\ref{fig:wlr} represents the best populated
modified wind-momentum diagram presented so far (12 mass loss
determinations and 19 upper limits). In particular at $\log
\lstar/\lsun \gtrsim 5.4$ we find excellent agreement with the
\citeauthor{vink01} predictions, {\em establishing that at these
luminosities the winds of SMC stars are weaker, in accord to
theoretical expectations}. At $\log \lstar/\lsun \lesssim 5.2$ the
situation is less clear as for many stars we could only set upper
limits and for those for which \mdot\ could be derived the error bars
are large. For the weak wind regime we therefore cannot draw firm
conclusions.

We have constructed an empirical WLR by fitting a linear function to
the objects in Fig.~\ref{fig:wlr} with $\log \lstar/\lsun \gtrsim
5.4$, while accounting for the symmetric errors in luminosity and the
asymmetric errors in \Dmom. Using the clumping corrected \Dmom\ values
for \azv~372 and \ngc346-001 this results in the following relation
\begin{equation}
  \log \Dmom = (1.69 \pm 0.34) \log \left(\lstar/\lsun\right)
               + (18.87 \pm 2.00)~,
\end{equation}
which is shown in Fig.~\ref{fig:wlr} as a dotted line. Within the
error bars the fit parameters agree with the theoretical calculations
of \cite{vink01}, who predict $x = 1.83$ and $\log D_{\circ} = 18.11$
for SMC metallicity. To illustrate the agreement we compare the
predicted and fitted wind momenta at the boundaries of the fitting
range. For $\log \lstar/\lsun = 5.4$ and $\log \lstar/\lsun = 6.2$ the
differences are $-0.02$~dex and $0.08$~dex, respectively. Given the
fact that the typical uncertainty in \Dmom\ is of the order of
0.2~dex, we find the agreement excellent. We note that the weak wind
regime is discussed in more detail in Sect.~\ref{sec:weak_winds}.

\subsection{Projected rotational velocities}
\label{sec:vsini}

\begin{figure*}
  \centering
  \resizebox{15cm}{!}{ \includegraphics{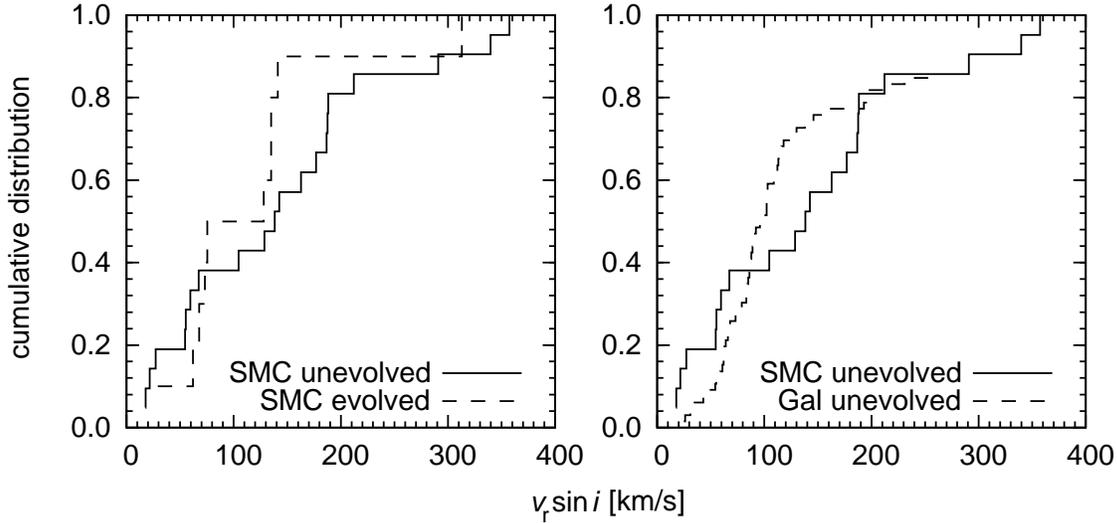}}
  \caption{Comparison of \vsini\ distributions. {\it Left panel:}
  cumulative distribution functions (cdf) of \vsini\ for unevolved
  objects, i.e.\ luminosity class IV and V, and evolved objects, i.e.\
  luminosity class I, II, III, in the SMC. The group of unevolved SMC
  objects are found to contain relatively both more fast rotators
  ($\vsini \gtrsim 100\,\kmsec$) as well as more slow rotators
  ($\vsini \lesssim 50\,\kmsec$) with respect to the group of evolved
  objects. {\it Right panel:} cdf of unevolved SMC objects and
  unevolved Galactic objects \citep{penny96}. Compared to the Galactic
  sample the SMC sample contains a larger fraction of intermediately
  fast rotating stars ($120 \gtrsim \vsini \lesssim 190\,\kmsec$).}
  \label{fig:obs_vsini}
\end{figure*}

The projected rotational velocities as determined with the automated
fitting method are listed in Tab.~\ref{tab:results}. The associated
uncertainties range from approximately 10~\kmsec\ for the slow
rotators up to about 40~\kmsec\ for the fast rotators, indicating that
we were able to accurately determine \vsini\ with our self consistent
method. Note that the low \vsini\ of \ngc346-031 and \ngc346-051 are
close to the effective resolution of the FLAMES observation
($\sim$15~\kmsec), and should be interpreted as upper limits.

If a star rotates at more than approximately 80\% of critical rotation
the assumption of spherical symmetry may break down
\citep[e.g.][]{maeder00}. Checking the ratio $\vsini/v_{\rm crit}$ for
all our target stars, we find a maximum of about $\sim$0.65. We
therefore do not expect any significant deviations from sphericity,
nor uncertainties in the derived parameters due to rotational effects.

\subsubsection{Rotation vs.\ macroturbulence}
In our calculations we consider the possibility of small scale
velocity fields. These thermal and microturbulent motions have a
coherence length that is small compared to the line forming region. We
do not include the possibility of motions that have a coherence length
that is comparable to or is larger than the line forming zone. Such
motions are denoted with the term macroturbulence (\vmac). If a large
macroturbulent velocity field is present, it may be expected that the
projected rotational velocity that we determine from the hydrogen and
helium profiles is overestimated.

To assess the importance of macroturbulence we have also determined
\vsini\ by means of a Fourier technique \citep{simon06} using weak
metal lines as an independent diagnostics. The Fourier technique
allows to discriminate between macroturbulence and rotation, as it is
sensitive to differences in the rotational and macroturbulent profiles
\citep[see e.g.][]{gray78}. For details on this method we refer to
\citeauthor{simon06} In all but one case we get \vsini\ values that
are consistent to within the error estimates. As in first order the
\vsini\ and \vmac\ broadening should be added quadratically, this does
not necessarily imply that macroturbulent motions are absent. For
\vsini\ of order 100 to 150 \kmsec, macroturbulent fields with
characteristic velocities of up to several tens of kilometres per
second, i.e.\ comparable to the small scale velocity component, remain
a possibility. Obviously, rapid rotators may show larger \vmac\
components.

For the O8.5 bright giant \azv469 the Fourier method recovers a
\vsini\ that is about 30 percent lower than the 81~\kmsec\ derived
from the hydrogen and helium profiles. This could indicate the
presence of a macroturbulent velocity field with $\vmac \sim
50\,\kmsec$. Significant macroturbulent fields have been reported for
B supergiants \citep[e.g.][]{ryans02}, but not for early O-type
stars. The fact that this star is of relatively late O sub-type seems
consistent with these findings. Consequently, in the \vsini\
distribution analysis presented below we adopt the projected
rotational velocity for \azv469 as determined using the Fourier
method.

\subsubsection{Observed \vsini\ distributions}

For a meaningful comparison of the distribution of the projected
rotational velocities of the SMC objects with other observations and
with theory we use cumulative distribution functions (cdfs). The cdf
describes the distribution of \vsini\ by simply giving for every
observed \vsini\ the fraction of objects with lower or equal
velocities. In Fig.~\ref{fig:obs_vsini} the cdfs of the SMC sample are
presented and compared to cdfs of Galactic O-type stars.

The left panel in Fig.~\ref{fig:obs_vsini} compares the cdfs of
unevolved objects, i.e.\ luminosity class IV and V, and evolved
objects, i.e.\ luminosity class I, II and III, in the SMC. This
comparison shows that compared to the evolved objects the group of
unevolved objects contain relatively more fast rotators. For instance,
approximately ten percent of the evolved objects have a \vsini\ in
excess of 150~\kmsec, whereas approximately 40~percent of the
unevolved objects exhibit velocities larger than this \vsini. Note
that about 20 percent of the group of unevolved objects is rotating
slowly ($\vsini \lesssim 50\,\kmsec$), while this is only 10 percent
for the evolved stars.

Using the Kolmogorov-Smirnov (K-S) test we have determined that the
probability that the two samples are drawn from the same underlying
distribution is 23 percent. Therefore, the differences between the
cdfs of the unevolved and evolved SMC stars may be significant and
possibly not due to statistical fluctuations. The trend that is found
here is also seen in Galactic O-type stars. Using a similar approach,
\cite{howarth97} and \cite{penny04} also find relatively more slow and
fast rotators among the unevolved stars \citep[also
see][]{conti77b}. \citeauthor{howarth97} ascribe the reduced number of
fast rotators to spin down as a result of an increased radius for the
evolved stars, as well as to loss of angular momentum through the
stellar wind. This explanation may also apply to the SMC case. They
further suggest that the apparent lack of slow rotating evolved stars
is a spurious result, caused by erroneously assigning turbulent
broadening -- which is more pronounced for evolved stars
\citep{ryans02} -- to rotational broadening. As a result, the derived
\vsini\ are overestimates, therefore some \vsini\ larger than 50
\kmsec\ (causing the steep gradient of the dashed curve in the left
panel at about $70 - 80\,\kmsec$) in reality reflect projected
rotational velocities below 50 \kmsec. For this to be the correct
explanation, the required turbulent velocities should be large (order
$40 - 70\,\kmsec$). Whether such an explanation is valid for our
sample is doubtful, as we have found no indication for the existence
of significant macroturbulent velocities in the objects with $\vsini
\lesssim 100\,\kmsec$.

In the right panel of Fig.~\ref{fig:obs_vsini} the 21 unevolved SMC
objects are compared to 66 unevolved Galactic stars as measured by
\cite{penny96}. Again using the Kolmogorov-Smirnov test we determined
that the probability that the two samples have the same underlying
distribution is 13\%. Therefore, we tentatively assume that the SMC
distribution of unevolved stars is significantly different from the
Galactic distribution. However, this should be verified using a larger
sample of SMC objects.

\begin{figure*}[t]
  \centering \resizebox{15cm}{!}{ \includegraphics{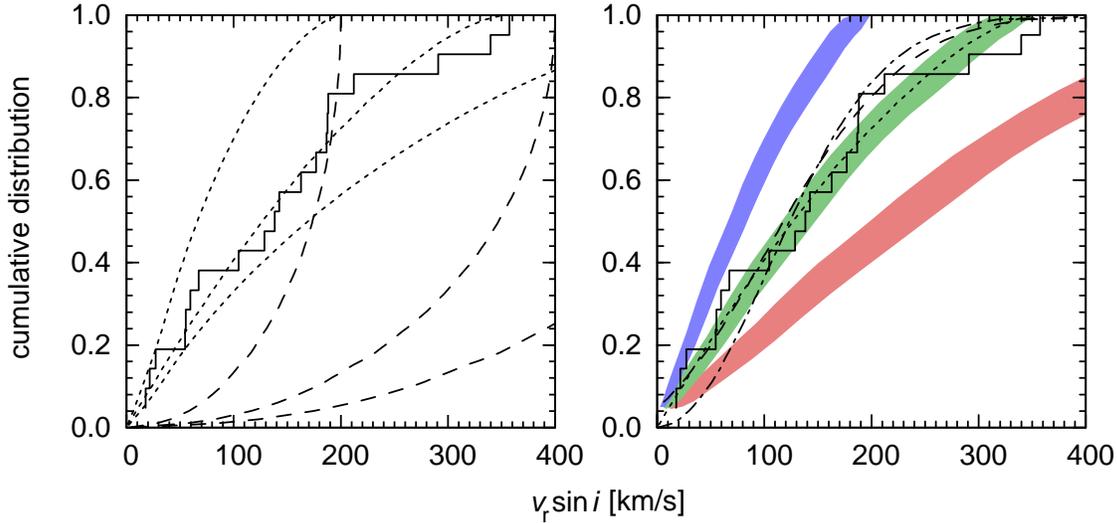}}
  \caption{{\it Left panel:} cdfs of observed (solid line) and
  theoretical initial \vsini\ distributions. Dotted and dashed lines,
  respectively, correspond to a constant \vr\ distribution, i.e.\
  block function, and a single value \vr\ distribution. {\it Right
  panel:} filled areas indicate the one~$\sigma$ probability
  distribution for theoretical \vsini\ cdfs for the limited number of
  21 objects calculated from a constant \vr\ distribution with $\vmin
  = 0\,\kmsec$ and \vmax\ of 200, 352 or 600~\kmsec. Shown with a
  dotted, dashed and dashed-dotted curve are, respectively, the best
  fitting block function, Gaussian and Maxwellian distribution
  cdf. See text and Fig.~\ref{fig:theo_dist} for further comments and
  explanation.}
  \label{fig:theo_vsini}
\end{figure*}

A marked difference between the two curves is the behaviour between
the intersection points at $\sim$90 and $\sim$190~\kmsec. The fact
that these curves intersect at these two points reflects that for both
the SMC and Galaxy the same fraction of stars show projected
rotational velocities in this range. However, the fact that in between
90 and 190~\kmsec\ the Galactic curve lies above the SMC one implies
that the Galactic stars show preferably lower \vsini\ (i.e.\ closer to
90~\kmsec) than in the SMC. This behaviour is consistent with the SMC
stars suffering less from spin down due to mass loss, as the SMC stars
are expected to have weaker winds. The behaviour outside of the above
velocity range cannot be understood within the context of spin down
through winds. At $\vsini \lesssim 90\,\kmsec$ the SMC shows preferably
larger projected rotational velocities; at $\vsini \gtrsim
190\,\kmsec$ both galaxies show the same distribution. If the Galactic objects
indeed suffer from a stronger spin down, the latter behaviour could in
principle imply that star formation in a relatively metal rich
environment results in larger initial rotational velocities. However,
note that the SMC cdf for $\vsini > 200\,\kmsec$ only contains four
objects, which makes this last statement very uncertain.

The scenario in which SMC stars suffer less from spin down is
consistent with the recent analysis performed by \cite{dufton06} for
the young Galactic cluster \ngc6611.  They find that the \vr\
distribution of the O-type stars can be characterised by a Gaussian
with a mean of $\sim$125~\kmsec. In the next section we will show that
the underlying \vr\ distribution of our SMC stars can be fitted by a
Gaussian with a mean of $\sim$160~\kmsec. Consequently, as the age of
\ngc6611 ($\sim$2~Myr) is comparable to the age of \ngc346, it appears
that the weaker winds of the SMC objects result in less spin down in
the first few million years of evolution.

It would be interesting to compare the evolved SMC and Galactic cdfs
as the effect metallicity has on stellar rotation would likely be much
more pronounced. However, for our sample this is not possible. The
reason is that in the Galactic sample of \cite{penny96} the ratio of
luminosity class I-II to class III objects is a factor two larger
compared to our SMC sample. Consequently, when making this comparison
we could be confusing evolutionary effects with metallicity effects.

\section{Parent rotational velocity distribution}
\label{sec:par_vr_dist}

The effect of rotation on the evolution of massive stars has been
studied by e.g.\ \cite{heger00}, \cite{meynet00} and
\cite{maeder01}. This has shown that rotation may cause extensive
mixing, changing the size of the reservoir of nuclear fuel available
for evolution, and thus the lifetimes and tracks
\citep[e.g.][]{langer95}. The effect of metal content on the time
evolution of the surface equatorial velocity as a function of metal
content is nicely illustrated in Figure~10 of \cite{meynet00} and
Figure~3 in \cite{maeder01}. The first figure shows this evolution for
different stellar masses in a Galactic environment, using an initial
surface equatorial velocity of 300~\kmsec. The second figure is
identical, but now for a SMC environment. Ignoring the first
$\sim$$10^{5}$~yrs -- which show a rapid decrease of \veq\ from
$\sim$300 to $\sim$250~\kmsec, reflecting the initial convergence of
the rotation law --
the main sequence phase shows a monotonic decline of the equatorial
rotational velocity as a result of an increasing radius and mass
loss. Using the 20~\msun\ track as an example, \veq\ has reduced from
$\sim$250 to $\sim$120~\kmsec\ at the end of the main sequence phase in
the case of Galactic stars, but to only 200~\kmsec\ for SMC stars. The
main fraction of the SMC decline occurs near the end of the main
sequence phase; halfway the main sequence, $\veq \sim
230\,\kmsec$. The reason for the modest decline of the SMC star is the
fact that its stellar wind is weaker than that of its Galactic
counterparts. Note that for an initially 60~\msun\ SMC star, which has
a stronger wind than the 20~\msun\ star, wind effects do play an
important role. Most stars in our unevolved SMC sample, however, have
masses of about $\sim$15--30~\msun, therefore are representative for
the discussed case. Given that the age of the \ngc346 cluster is about
1--3~Myr (see Sect.~\ref{sec:age}), we may conclude that -- based on
the evolution models -- the observed cdf of the unevolved SMC objects
should lie close to the initial cdf. This allows to address the
interesting question: what is the initial \vsini\ distribution?

\subsection{initial \vr\ distributions}

We compare the cdf of the unevolved SMC objects with the theoretical
predictions. To calculate these theoretical cdfs we constructed simple
models of the underlying unprojected rotational velocity distribution.
For each distribution function we synthesised a theoretical
distribution using a large number of objects ($N = 10^{4}$), while
assuming randomly oriented rotation axes. The four adopted models are:
\begin{enumerate}

\item a delta function, i.e.\ one single possible value of \vr\ for all
      objects;

\item a block function, i.e.\ a constant distribution of objects
      between a minimum velocity \vmin\ and a maximum velocity \vmax;

\item a Gaussian distribution, with a mean velocity \vrgauss\ and a
      standard deviation $\sigma$;

\item a Maxwellian distribution, specified by the most likely velocity
      \vrmaxwell;

\end{enumerate} 

The normalised distributions of (best fits of the) latter three models
are shown in the right panel of Fig.~\ref{fig:theo_vsini}. Results for
the delta and block function are presented in the left panel of
Fig.~\ref{fig:theo_vsini}. The top, middle, and bottom dashed curve
correspond to a population with a single rotational velocity \vr\ =
200, 400, and 600~\kmsec, respectively. From a comparison with the SMC
distribution, we find that models assuming a single valued underlying
\vr\ distribution fail to reproduce the observed cdf. Consequently, it
is very unlikely that the velocity distribution of massive stars in
this cluster can be characterised by a single rotational
velocity. This immediately implies that {\em the initial \vr\
distribution of this young cluster is not such that, for instance,
massive stars are all born rotating at critical velocity}. Such a
supposition would not be completely unreasonable as the initial
angular momentum of Jeans-unstable molecular cloud fragments is so
large that -- if one assumes it to remain conserved during collapse
leading to the formation of a single object -- all stars would have to
rotate super-critical. Such rapid rotation must lead to the ejection
of material and angular momentum, at least until the star is rotating
at break-up velocity. As pointed out, the observed \vsini\
distribution does not support such a scenario \citep[also
see][]{herrero05}.

The dotted curves in the left panel of Fig.~\ref{fig:theo_vsini}
correspond to the theoretical cdfs calculated for the second model,
i.e.\ the block function. The best fit to the observations requires
\vmin\ = 0~\kmsec\ and \vmax\ = 352~\kmsec. For comparison, we also
plot the result for \vmax\ = 200 (top curve) and 600~\kmsec (bottom
curve). We see that the overall shape of this model is in better
agreement with the observed cdf and that the best fit model gives a
good fit to the SMC stars.

Given the fact that the SMC cdf is constructed from a limited number
of objects, we should also try to account for the effect that a small
sample size has on the distribution in the theoretical cdfs. To do
this we use the following approach. Instead of using a large number of
simulated objects to calculate the theoretical cdf, we use a number
equal to the amount of observed objects. Using different sets of
random inclination angles we then calculate an ensemble of theoretical
cdfs. The resulting distribution of these cdfs in the ensemble then
describes the effect of statistical fluctuations due to a limited
sample size. In the right panel of Fig.~\ref{fig:theo_vsini} the
results of this approach are shown for the theoretical cdfs with
underlying constant \vr\ distributions. The filled areas in this panel
correspond to the ranges in the diagram containing one~$\sigma$, i.e.\
68 percent, of the theoretical cdfs. In other words the surfaces
correspond to the area in which one can expect a theoretical cdf to be
located within a one~$\sigma$ probability. The top, middle and bottom
areas, again, correspond to distributions with \vmax\ equal to,
respectively, 200, 352, and 600~\kmsec.

\begin{figure}[t]
  \centering
  \resizebox{8.8cm}{!}{\includegraphics{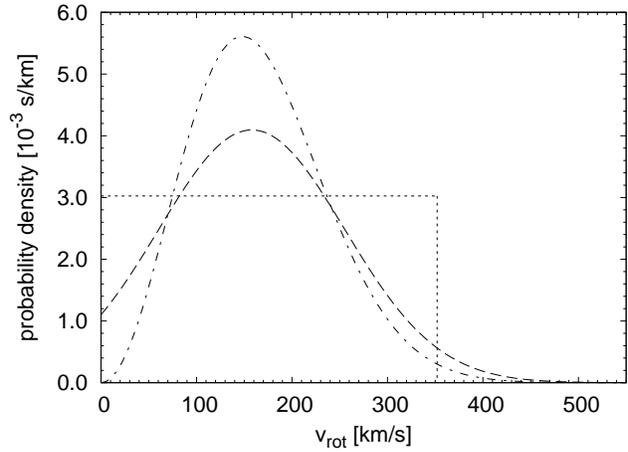}}
  \caption{Best fitting model \vr\ distributions for the \vsini\
  distribution of the unevolved SMC objects. Shown with a dotted,
  dashed and dashed-dotted line are, respectively, the constant, the
  Gaussian and the Maxwellian distribution.}
  \label{fig:theo_dist}
\end{figure}

So, may we conclude that the underlying rotational velocity
distribution is indeed a block function? This would be premature. To
illustrate this we show two additional models in the right panel of
Fig.~\ref{fig:theo_vsini}. The dashed curve is the best fit for a
Gaussian \vr\ distribution, with a mean velocity $\vrgauss =
158$~\kmsec\ and $\sigma = 97$~\kmsec. The dashed-dotted curve is the
best fit for a Maxwellian distribution, characterised by a most likely
velocity $\vrmaxwell = 148$~\kmsec. For reference, the dotted line
gives the best fit for the constant distribution. The Gaussian and
Maxwellian models have an uncertainty which is similar to that of the
block function (the dotted line, with the middle grey area
representing the uncertainty). Given the modest difference between the
block and the Gaussian curves -- and in view of their error bars -- it
is clear that we cannot distinguish between these two models. The
Maxwellian function is also very similar, however, here the agreement
with the observations seems a bit off, in particular at low
velocities.

Why are the block and Gaussian distribution so indistinguishable?
The reason must be that the underlying intrinsic distributions are
essentially similar. This is shown in Fig.~\ref{fig:theo_dist}. Though
the models obviously show differences, they are both characterised by
a mean velocity of about $150-180$~\kmsec\ and a comparable effective
half width of roughly $100-150$~\kmsec. Therefore, this is currently
the most meaningful and robust specification we can assign to the
underlying rotational velocity distribution.  The similarity between
the models in the cdf plot shows that even with a significantly larger
set of observed rotational velocities it will be difficult to better
define the exact shape of the parent population of \vr. The Maxwellian
distribution is a bit worse, essentially because it is lacking in
stars with low ($\lesssim 50$~\kmsec\ say) rotational velocities. We
conclude that at birth the massive star population of \ngc346 must
have included $\sim$5--15 percent slow rotators.

\section{The weak wind problem}
\label{sec:weak_winds}

The majority of our programme stars have luminosities less than
$\lstar \sim 10^{5.25}\,\lsun$. The winds of these SMC O- and B-type
stars are currently at the focus of attention as they do not appear to
agree with theoretical predictions \citep{bouret03, martins04}.
Starting from this luminosity the observed rates rapidly fall below
the predicted ones, leading to a discrepancy of about two orders of
magnitude at $\lstar \sim 10^{5}\,\lsun$. The reason for this
discrepancy is not understood so far. The cause may be ill treated or
missing physics in the \mdot\ predictions, e.g.\ because of a break
down of the adopted Sobolev approximation for low density winds
\citep{owocki99} or possibly the neglect of ion-decoupling
\citep{krticka03}. Whether the last hypothesis is indeed a viable
option seems doubtful. Computations of \citeauthor{krticka03} appear
to indicate that this effect only starts to play a role at a
metallicity $Z \lesssim 1/30\,\zsun$. Moreover, \cite{martins05b} also
report a weak wind problem for Galactic stars making ion-decoupling
even more unlikely. The \citeauthor{martins05b} result also suggests
that metallicity effects can be excluded. An alternative explanation
may be that the dwarfs showing the weak wind problem represent an
earlier evolutionary state, i.e.\ a state in which the wind is not yet
``fully developed''. The problem with this hypothesis is that the
characteristic timescale for the wind to develop -- of order of the
dynamical timescale of the wind -- is very short compared to
evolutionary timescales, and that the wind properties are determined
by global and atmospheric properties only.

The cause of the discrepancy may also be connected to a defect in the
spectroscopic derivation of the mass loss rate. The winds of these SMC
dwarf O-type stars are quite weak ($\sim$$10^{-8}- 10^{-6}\,\msunyr$),
and reach the limits of sensitivity of \ha\ as a mass loss
diagnostic. Therefore, the \citeauthor{bouret03} and
\citeauthor{martins05b} study use unsaturated ultraviolet resonance
lines of carbon, nitrogen, and oxygen, which are much more sensitive
mass loss indicators. However, these lines typically belong to trace
ions, for which the ionisation is extremely sensitive to the local
(shock generated) X-ray radiation field. Though significant progress
in our understanding of the processes leading to non-thermal X-rays in
stellar winds has been gained in the last years
\citep[e.g.][]{feldmeier97, pauldrach01, kramer03}, we cannot yet
exclude problems in the UV based \mdot\ determination as the cause for
the weak wind discrepancy.

\begin{figure}[t]
  \centering
  \resizebox{8.8cm}{!}{\includegraphics{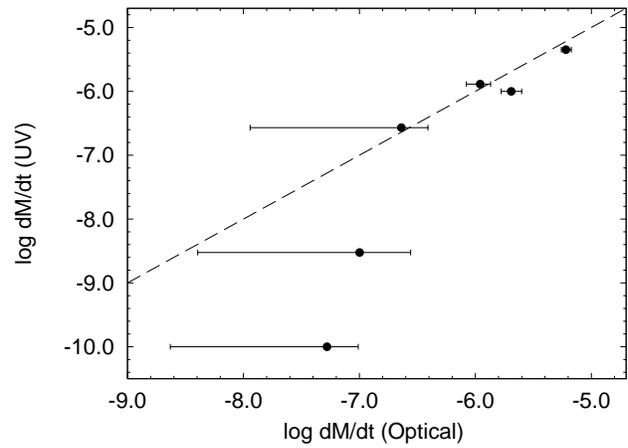}}
  \caption{Comparison of mass loss rates determined in this study
  using the optical spectrum with values determined from the UV
  spectrum. For the three objects with the lowest mass loss rates the
  values determined from the optical spectra correspond to upper
  limits.}
  \label{fig:mdot_comp}
\end{figure}

Our genetic algorithm based fitting method allows to determine the
mass loss at the ``upper end'' of the weak wind regime.
Figure~\ref{fig:mdot_comp} shows a comparison of six O stars for which
the UV and, save for \ngc346-026 and \ngc346-028, optical spectra have
been used to determine \mdot\ \citep{crowther02, bouret03} and which
we have analysed here using optical spectra only. The four stars with
the highest mass loss all have a luminosity $L > 10^{5.3} \lsun$,
i.e.\ they are in a regime in which the theory agrees with
observations. In these cases the UV+optical studies all include
\ha. The \mdot\ results reported in these studies compare very well
with our findings.

The two remaining stars, \ngc346-026 and \ngc346-028 have been
analysed by \cite{bouret03}. Their luminosities are $10^{4.93}$ and
$10^{5.10}$ \lsun, respectively. This is in the regime where observed
and predicted mass loss disagree. Though the \citeauthor{bouret03}
analysis includes part of the optical spectrum (from 3910 to
5170~\AA), it does not include \ha. For \ngc346-026 they derive $\mdot
= 3.2 \times 10^{-9} \msunyr$; we find a $\sim$30 times higher
rate. For \ngc346-028 they obtain $\mdot = 10^{-10} \msunyr$, where we
find a $\sim$500 times larger value. Note, however, that for both
stars the downward error bars on our fit values extend to the lower
limit of the mass loss regime in which the automated method was
allowed to search for a solution. This implies that our results are
{\em de facto} upper limits. Therefore, we cannot (yet) conclude
whether or not the UV and \ha\ diagnostic agree or disagree in weak
wind situations.

\begin{figure*}[t]
   \centering
   \resizebox{16cm}{!}{\includegraphics{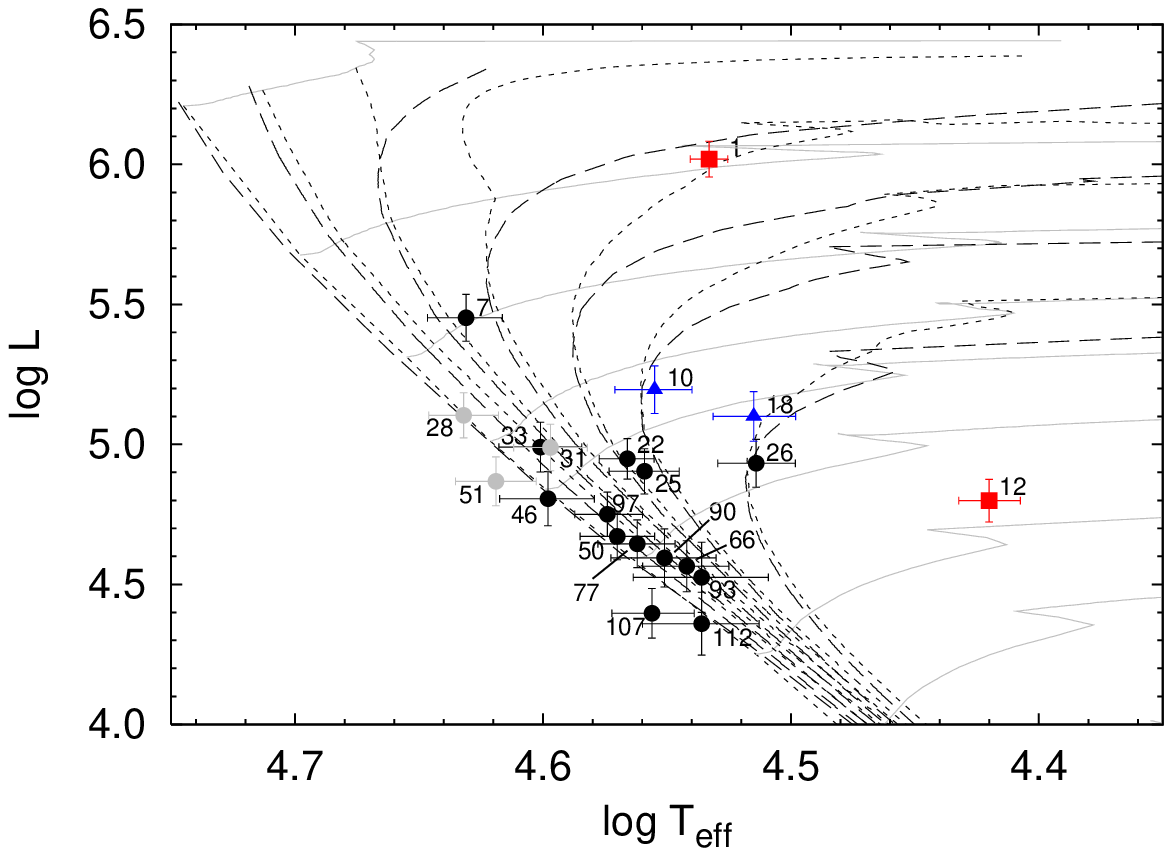}}
   \caption{Comparison of the programme stars located in \ngc346 with
   isochrones for zero age, 1, 2, 3, 4, 5 and 7~Myr derived from the
   non rotating evolutionary models of \cite{charbonnel93} (dashed
   lines) and models including rotation from \cite{maeder01} and
   \cite{meynet05} (dotted lines). For reference the tracks of the
   rotating models are also shown (grey lines, see Fig.~\ref{fig:hrd}
   for initial masses of these tracks), where for clarity purpose only
   the blueward part of the evolution of the most massive tracks is
   plotted. Shown with identical symbols as in Fig.~\ref{fig:yhe} are
   our programme stars, which are located in \ngc346. Luminosity class
   Vz objects are denoted using grey circles. See text for a
   discussion.}
   \label{fig:hrd_age}
\end{figure*}

Note that though we interpret these \mdot\ as upper limits, the
automated fitting method did return a best fit value.
Figure~\ref{fig:wlr} shows the strength of the stellar winds of our
entire sample in terms of the modified wind momentum (see
Sect.~\ref{sec:wind_param} for a discussion). The inverted triangles
represent upper limits. All these upper limits are well above (up to
two orders of magnitude) the mass loss rates derived on the basis of
UV profile fitting; not a single one ``by chance'' gives a value
comparable to what is found with the UV method. Why is this so? We can
identify two potential reasons. First, the automated fitting procedure
scans the mass loss dimension in a linear way. For weak wind cases
typical lower and upper bounds of this dimension are $10^{-9}$ and
$10^{-7}$~\msunyr. In most cases it is reasonable to assume that only
above $10^{-7}$~\msunyr\ the profile of \ha\ is sensitive to mass
loss. In that case it is, e.g.\ nine times as likely that the
automated method more-or-less by chance settles for a mass loss above
$10^{-8}$~\msunyr, and not for the much lower values indicated by UV
analysis. Second, small errors in the broadening function of \ha\ may
perhaps be ``corrected'' in the automated fitting process by settling
for a mass loss which is ``as high as possible''. Consequently, we
conclude that we cannot assign any importance to the best fit value
returned by the automated method when the downward errors extend to
the lower bound of \mdot\ allowed in the fitting process.

\section{On the evolutionary status of \ngc346}
\label{sec:age}

Undoubtedly the large number of early-type stars in \ngc346
\citep{massey89} is indicative for a young age. Further evidence for
the youthful nature of this cluster may be given by the presence of
several so-called Vz stars, which are hypothesised to represent the
earliest stages of main sequence evolution of massive stars
\citep[e.g.][]{walborn92, parker92}. In this section we will determine
an age estimate for \ngc346 and try to quantify the uncertainty
rotation introduces in this estimate. We will also discuss how well
the spectroscopic Vz designation correlates with ZAMS evolution.

\subsection{The age of \ngc346}

Rotation affects both the lifetimes and the tracks of stellar
evolution models. Consequently, before we can estimate the age of
\ngc346 we need to assess the systematic uncertainty rotation will
introduce in the age estimate. We do this by considering the
differences between isochrones derived from the non rotating models of
\cite{charbonnel93} and those based on the rotating models of
\cite{maeder01} and \cite{meynet05} adopting an initial \vr\ of 300
\kmsec. In Fig.~\ref{fig:hrd_age} the two sets are shown as,
respectively, dashed and dotted curves for zero age, 1, 2, 3, 4, 5 and
7~Myr. For reference the evolutionary tracks of the models accounting
for rotation are also shown in this figure using grey lines.

\begin{figure}[t]
   \centering
   \resizebox{8.8cm}{!}{\includegraphics{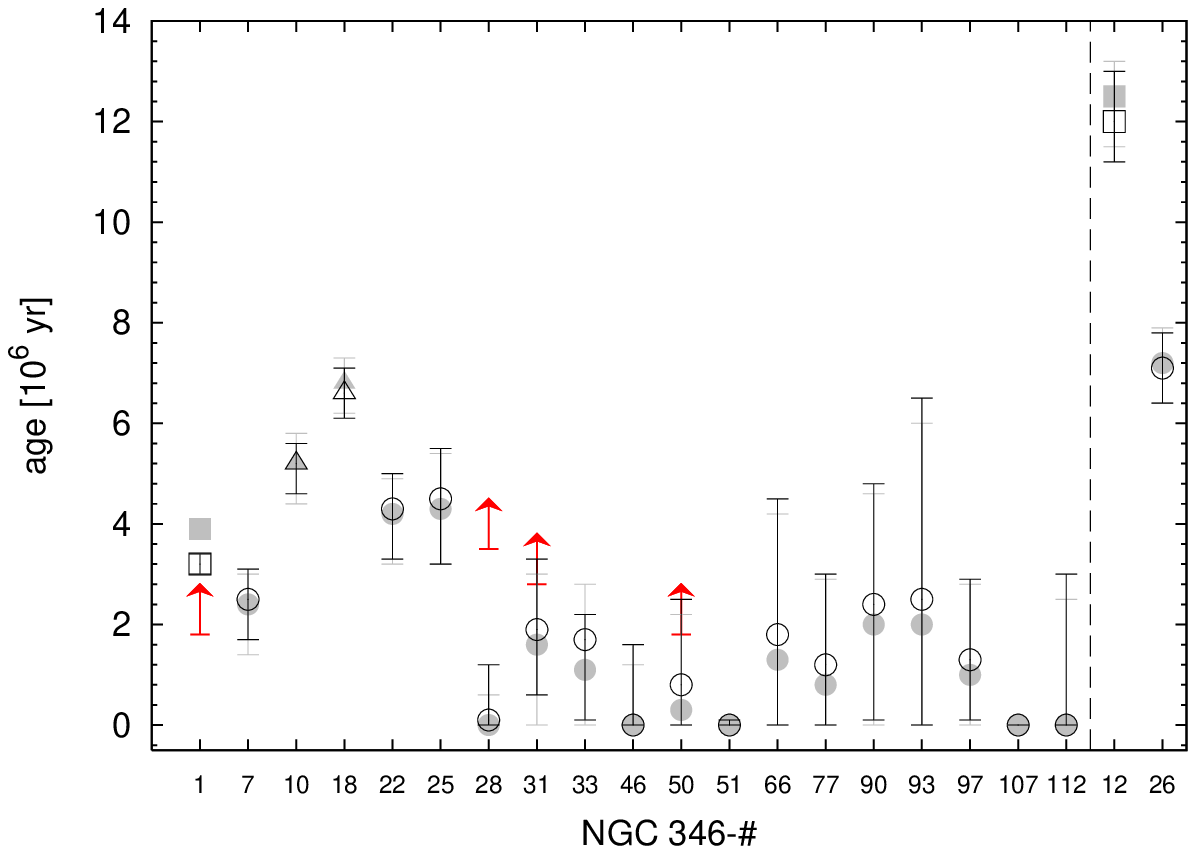}}
   \caption{Age determination of the programme stars in \ngc346\ using
   the non-rotating SMC models of \cite{charbonnel93} (open symbols)
   and those including rotation \citep{maeder01, meynet05} (closed
   symbols). Circles, triangles, and squares denote dwarfs, giants,
   and supergiants, respectively.  The horizontal axis gives the ID
   number of the star (see e.g. Tab.~\ref{tab:results}); the vertical
   axis the age in Myr. The lower limits (upwards pointing arrows)
   provide age estimates from tracks of chemically homogeneous
   evolution (Yoon et al.\ in preparation); see text for a
   discussion.}
   \label{fig:age_ngc346}
\end{figure}

Age determinations of individual stars, based on the evolutionary
tracks, are given in Fig.\,\ref{fig:age_ngc346}. The error bars on
these life times account for the uncertainties in temperature and
luminosity. Open symbols (see the figure caption for details) refer to
non-rotating tracks; grey symbols for those accounting for
rotation. Note that they differ only very modestly for relatively
unevolved objects.  For instance, stars near the main sequence are
judged to be $\sim$$10^{5}$ years younger if rotation is taken into
account. We estimate that for main sequence stars having luminosities
larger than $\log \lstar/\lsun \approx 4.5$ this systematic difference
may increase up to approximately 0.5~Myr.  For the more evolved
phases, i.e.\ approximately giant phase and later, this systematic
discrepancy switches sign and rapidly becomes larger. Considering the
supergiant \ngc346-001 at $\log \lstar/\lsun = 6.0$ we estimate its
age to be $\sim$1~Myr {\em more} if it is initially rotating with $\vr
\sim 300 \kmsec$ compared to the non-rotating case.

Ignoring the supergiant \ngc346-012 at $\log \teff \approx 4.4$, for
which cluster membership is very uncertain \citep{evans06}, we find
that the objects populate the region between the ZAMS and the 7~Myr
isochrone, with typical error bars (in both directions) of 2--5~Myr.

So, what could explain the apparent age scatter? 
First, some objects in our sample might actually not belong to the
cluster. In addition to \ngc346-012 this is, for instance, also
suggested by \cite{walborn00} for the B0 dwarf \ngc346-026. Based on
its spatial location, discrepant radial velocity and stellar
parameters these authors conclude that this object is not a coeval
member of the cluster. Its large age of about 7~Myr seems to support
this conclusion. Because of cluster membership issues we have placed
both \ngc346-012 and \ngc346-026 to the right in
Fig.\,\ref{fig:age_ngc346} and have separated them from the other
stars using a vertical dashed line.  For the remainder of the
discussion we treat them as not belonging to the cluster.
Second, the Oe star \ngc346-018 is poorly fitted because of
contamination by circumstellar material. This implies that the
parameters derived for this star should be taken with considerable
care, and therefore also the apparently large age of about 7~Myr that
we derive for this object. Ignoring the above three discussed objects
the oldest investigated star in \ngc346 is $\sim$ 5~Myr (\ngc346-010).

What remains is an age spread of stars from zero to the above
motivated 5~Myr. Can such an age spread be explained? We first checked
whether there is a relation between the age of the stars and their
distance to the cluster centre. Though we do see that the oldest stars
(\ngc346-001 through \ngc346-025) are all at some distance from the
centre this also holds for some that are assigned very young
ages. Also two relatively old stars (\ngc346-007 and \ngc346-022) are
located relatively close to the cluster core. Therefore, we can not
claim such a relation.

Next we consider the possibility that the adopted tracks are not
appropriate for some stars. As we have explained, when it comes to age
determination there is no fundamental difference between tracks
without rotation and those that do account for it starting the
evolution from an initial \vr\ = 300 \kmsec. However, as already
discussed in Sect.\,\ref{sec:yhe}, if \vr\ exceeds some critical value
the evolution will show a bifurcation behaviour, leading to tracks
that evolve bluewards (from the ZAMS) and upwards. This is the result
of such efficient mixing that the HRD behaviour proceeds as for
chemically homogeneous evolution. Adopting such tracks from Yoon et
al.\ (in prep.) \citep[also see][]{yoon05} it turned out to be
difficult to derive ages as for near critical rotation the effective
temperature changes so rapidly (as a function of \vr) that one can in
practise only rely on luminosity as an age discriminator. Given the
error bars on $L$ and \Ms\ this did not lead to meaningful constraints
for most of the objects. Moreover, in a few cases the mass discrepancy
turned out to be so large that the luminosity range implied by \Ms\
(plus errors) did not coincide with the observed \lstar. This also
prevents one from deriving an age estimate.

A different approach we tried is based on the fact that for chemically
homogeneous evolution the surface helium abundance is essentially a
measure of stellar age, for given initial mass. Taking into account
the errors on \yhe\ and \Ms, this may yield lower limits to the age of
the stars. One derives lower limits because if evolution is not fully
chemically homogeneous the measured helium abundance is a lower limit
for the mean stellar \yhe. The method will only provide useful results
for those stars that show enhanced helium {\em even when correcting
for the negative error in \yhe}. This is the case for four objects
(\ngc346-001, \ngc346-028, \ngc346-31, and \ngc346-050). The derived
lower age limits are shown in Fig.~\ref{fig:age_ngc346} (arrow
symbols). The lower limit of \ngc346-001 (the most massive star) is
consistent with the age as derived using the standard
method. Interestingly, the lower limits of the ages of the three
remaining stars are all significantly above what we derived using
modestly rotating or non-rotating models. We have no indication that
our \yhe\ determinations are flawed or that these stars had an initial
helium fraction significantly higher than 0.09. Consequently, it is
likely that these stars evolve along chemically homogeneous tracks and
have ages comparable to the oldest stars in the cluster.

In view of the above discussion we have to conclude that we can not
derive a unique age for \ngc346. The distribution of ages seems to
indicate two preferred values: 1--3~Myr and 3--5~Myr. We add that some
stars appear to reside on or even to the left of the ZAMS. Though, it
could be that these latter stars evolve along tracks for rapidly
rotating stars (following tracks for chemically homogeneously evolving
stars). If so, we could not constrain their ages from such models.
The ages estimated from modestly rotating or non-rotating evolutionary
models of three helium enriched dwarf objects are so low (less than
2~Myr) that it is not possible to bring them into agreement with the
derived \yhe\ values. Chemically homogeneous evolutionary models
instead predict an age comparible with the oldest cluster population.
Given/assuming that all stars belong to the same compact cluster it
seems very unlikely that the range of ages implies that \ngc346
experienced a prolonged burst of star formation, or a series of short
bursts resulting in multiple coeval populations.

%
%
%

The age of \ngc346 was also estimated by \cite{kudritzki83},
\cite{massey89}, \cite{walborn00} and \cite{bouret03} \citep[also
see][]{massey05}. In the first, second and fourth studies ages of
2.6~Myr, 2--4~Myr and 3~Myr were found, which in principle are
consistent with our findings.  \citeauthor{walborn00} estimate an age
of only 1~Myr. This agrees with our lower age limit. The reason for
their young age estimate is that their four star \ngc346 sample
predominantly contained the stars in the cluster closest to the
ZAMS. Combined with the \teff\ calibration from \cite{vacca96} adopted
by these authors, which resulted in relatively high estimates for this
parameter, for instance, compared to the analysis of \cite{bouret03},
explains their relatively young age estimate.

\subsection{ZAMS stars}
\label{sec:zams_stars}

The defining characteristic of the luminosity class designation Vz is
a very intense \heii~$\lambda$4686 absorption line that is
substantially stronger than any of the other helium absorption
lines. The distinct behaviour of this line is believed to be linked to
a luminosity effect connected to an early evolutionary phase. As Vz
stars are {\em hypothesised} to lie relatively close to the ZAMS they
are {\em anticipated} to be less luminous compared to normal dwarf
stars. As a result of this, the relatively strong $\lambda$4686 line
can be explained as a lack of partial filling in of the line profile
due to (wind) emission which normal dwarf stars do experience
\citep{parker92, walborn97}.

In the HR-diagram shown in Fig.~\ref{fig:hrd_age} we have highlighted
the location of the three \ngc346 Vz stars in our sample using grey
circles. From these three objects, \ngc346-028, \ngc346-031 and
\ngc346-051, we find that the first and last are indeed found to lie
on or left of the ZAMS. \ngc346-031 in contrast occupies a location
between the one and two million year isochrones. We suspect that that
the strong \heii~$\lambda$4686 line in the spectrum of this object is
in fact related to its peculiar chemical composition (see
Sect.~\ref{sec:yhe}). Moreover, the spatial separation of \ngc346-031
of $\sim$5 arcminutes from the cluster centre, makes it unlikely that
this object is formed recently as part of the cluster. Consequently,
care should be taken with the correct interpretation of Vz stars as
near-ZAMS objects, based on optical classification criteria only.

In addition to the two Vz stars Fig.~\ref{fig:hrd_age} reveals the
presence of three more stars located on or left of the ZAMS. Even
though the spectra of these objects, \ngc346-046, \ngc346-107 and
\ngc346-112, do not exhibit the Vz characteristics, their location in
the HR-diagram clearly suggest an early evolutionary phase. Why is
this so? We suspect that the reason for the lack of a strong
\heii~$\lambda$4686 line in the spectra of these objects is related to
their relatively low effective temperatures. Compared to the neutral
helium lines the relative strength of the \heii\ lines decreases for
decreasing \teff. Consequently, even though its line profile is not
filled in it becomes more unlikely for the \heii~$\lambda$4686 to be
stronger than all the \hei\ lines for relatively low effective
temperatures. In Fig.~\ref{fig:hrd_age} we see that these normal dwarf
stars are indeed cooler than the Vz stars at the ZAMS. Interestingly,
it turns out that in the spectrum of the hottest of these three
objects, \ngc346-046, the \heii~$\lambda$4686 is the strongest \heii\
line, which has a strength that is approximately equal to the strength
of the strongest \hei\ line. We therefore argue that the effective
temperature of this object, i.e. $\sim$40~kK, corresponds to the lower
limit of \teff\ for SMC metallicity for which a Vz luminosity class
designation is at all possible.

\section{Summary and conclusions}
\label{sec:sum}

Using the automated fitting method developed by \cite{mokiem05}, which
combines the stellar atmosphere code \fastwind\ with the genetic
algorithm based optimisation routine \pikaia, we have performed a
quantitative analysis of 31 O- and early B-type stars located in the
SMC. This sample was mostly drawn from the targets observed in the SMC
clusters \ngc346 and \ngc330 as part of the VLT-FLAMES survey of
massive stars \citep[see][]{evans05}, and is the largest sample of
early-type SMC stars analysed so far. Even though many of the observed
spectra show nebular contamination and have signal-to-noise ratios as
low as 50, the fitting method did not encounter convergence
problems. Instead we find that the quality of the data is naturally
reflected in the errors that are estimated for the fit parameters
using the width of the global optimum.

Concerning the stellar properties of the objects in our sample we can
draw the following conclusions:

\begin{enumerate}
   
   \item[\it i)] The average effective temperature for a given
   spectral type for SMC dwarf stars is hotter by typically 3 to 4~kK
   than for their Galactic counterparts \citep{martins05a}. We attribute
   this hotter temperature scale to a reduced line blanketing in the
   SMC. In contrast to \cite{massey05} we find that this blanketing
   effect on \teff\ extends to spectral types as late as B0.

   \item[\it ii)] The helium enrichment found in our stars does not
   appear to be consistent with predictions of the evolution of
   massive stars ignoring rotation. Comparison with evolutionary
   tracks that do account for rotation shows a qualitative agreement.
   In the HR-diagram the region in which a helium enrichment is
   predicted roughly coincides with the location of evolved objects
   showing evidence of \yhe\ enrichment. The actual degree of
   enrichment is in many cases still strongly under predicted by the
   most recent evolutionary models.

   \item[\it iii)] From our three programme stars with a Vz luminosity
   class designation, we find that only two are true ZAMS
   objects. Moreover, three additional objects, which are not
   classified as Vz stars, are found to lie on or left of the ZAMS. We
   argue that this is related to a temperature effect inhibiting
   relatively cool stars located on the ZAMS to exhibit the spectral
   features characteristic for the Vz luminosity class.

   \item[\it iv)] We find good agreement between our spectroscopically
   derived masses and those derived from evolutionary calculations. A
   mild mass discrepancy is still found for stars with $\Ms \lesssim
   20\,\msun$. This discrepancy correlates with the surface helium
   abundance and is qualitatively consistent with the predictions of
   chemically homogeneous evolution.

\end{enumerate}

\noindent
The stars in the young cluster \ngc346 allow to study the role of mass
loss and angular momentum loss during the early evolution of massive
stars in a low metallicity environment. Regarding these issues we can
conclude that:

\begin{enumerate}
   \item[\it v)] The observed distribution of \vsini\ for evolved
   stars (luminosity classes I-II) contains relatively less fast
   rotators and slow rotators compared to the distribution for
   unevolved stars (luminosity classes IV-V). These findings are in
   agreement with analyses of Galactic samples and can be interpreted
   as a spin down due to an increased radius and angular momentum loss
   through the stellar wind, and excess turbulent broadening among the
   evolved stars \citep[cf.][]{howarth97}. We also find that compared
   to the velocity distribution of unevolved Galactic objects the
   distribution for unevolved SMC objects shows a relative excess of
   stars rotating with projected velocities $120 \gtrsim \vsini
   \lesssim 190\,\kmsec$. This excess can be interpreted as evidence
   of a reduction of dissipation of angular momentum in weaker stellar
   winds resulting from the low $Z$ environment of the SMC.

   \item[\it vi)] Using a simple modelling approach we have
   constrained the underlying \vr\ distribution of the unevolved SMC
   stars. The exact shape of this underlying distribution -- for which
   we have tried a block, Gaussian and Maxwellian shape -- is found to
   be degenerate, however, it can be characterised by a mean velocity
   of about $150-180$~\kmsec\ and an effective half width of roughly
   $100-150$~\kmsec. The Maxwellian distribution appears a somewhat
   poorer representation of the \vr\ distribution, indicating that
   5--15 percent of the stars must be slow rotators ($\vr \lesssim
   50$~\kmsec).

   \item[\it vii)] It is not possible to derive a unique age for
   \ngc346 using moderately rotating or non-rotating evolutionary
   models. Though the bulk of the stars indicate a lifetime of
   1--3~Myr, part of the investigated stars seem somewhat older (3--5
   Myr). Clearly the ages depend on the adopted evolutionary tracks.
   However, we point out that predictions for non-rotating stars and
   those accounting for the rotational evolution of stars with an
   initial $\vr = 300$~\kmsec\ essentially give the same result. Three
   main sequence stars, all with an age determination of less than 2
   Myr, show significant helium enhancement. Assuming that these stars
   have the same initial \yhe\ this implies a lower limit to their age
   of 2 to 3.5~Myr.

   \item[\it viii)] Based on the predictions of the time evolution of
   the equatorial velocity for low metallicity \citep{maeder01} the
   above age considerations imply that the underlying \vr\
   distribution derived for our sample (conclusion {\it vi}) is
   representative for the initial rotational velocity distribution.

   \item[\it ix)] Mass loss rates determined for the SMC objects are
   found to be systematically lower than for Galactic objects. {\em
   For $\log\lstar/\lsun \gtrsim 5.4$ we find that the wind strengths
   are in excellent agreement with the theoretical predictions of
   \cite*{vink01}}. The empirical modified wind momentum luminosity
   relation constructed for this luminosity regime agrees to within
   the error bars with the predicted WLR.

   \item[\it x)] Down to $\log \lstar/\lsun \approx 5.3$ the mass loss
   rates determined from the optical spectra are in agreement with UV
   rates. For luminosities lower than this value the winds become too
   weak for an accurate determination of \mdot\ from the optical
   spectrum only. Consequently, we cannot (yet) conclude whether or
   not the UV and \ha\ diagnostics agree or disagree in weak wind
   situations.

\end{enumerate}

\begin{acknowledgements}
  We would like to thank Mike Irwin and Rens Waters for constructive
  discussions, George Meynet for providing the evolutionary models
  from the Geneva group and the anonymous referee for his/her
  constructive remarks that have helped to improve this paper. M.R.M.\
  acknowledges financial support from the NWO Council for Physical
  Sciences. A.H.\ and F.N.\ thank support to the Spanish MEC through
  project AYA2004-08271-C02. S.C.Y.\ is supported by the VENI grant
  (639.041.406) from the Netherlands Organization for Scientific
  Research (NWO). Spectral fits were calculated using the LISA compute
  cluster at SARA Computing \& Networking Services.
\end{acknowledgements}

\bibliographystyle{aa}

\begin{small}
\bibliography{flames_smc_an}

\begin{thebibliography}{94}
\expandafter\ifx\csname natexlab\endcsname\relax\def\natexlab#1{#1}\fi

\bibitem[{{Abbott}(1982)}]{abbott82}
{Abbott}, D.~C. 1982, \apj, 259, 282

\bibitem[{{Azzopardi} \& {Vigneau}(1975)}]{azzopardi75}
{Azzopardi}, M. \& {Vigneau}, J. 1975, \aaps, 19, 271

\bibitem[{{Azzopardi} \& {Vigneau}(1982)}]{azzopardi82}
{Azzopardi}, M. \& {Vigneau}, J. 1982, \aaps, 50, 291

\bibitem[{{Bouret} {et~al.}(2003){Bouret}, {Lanz}, {Hillier}, {Heap}, {Hubeny},
  {Lennon}, {Smith}, \& {Evans}}]{bouret03}
{Bouret}, J.-C., {Lanz}, T., {Hillier}, D.~J., {et~al.} 2003, \apj, 595, 1182

\bibitem[{{Charbonneau}(1995)}]{charbonneau95}
{Charbonneau}, P. 1995, \apjs, 101, 309

\bibitem[{{Charbonnel} {et~al.}(1993){Charbonnel}, {Meynet}, {Maeder},
  {Schaller}, \& {Schaerer}}]{charbonnel93}
{Charbonnel}, C., {Meynet}, G., {Maeder}, A., {Schaller}, G., \& {Schaerer}, D.
  1993, \aaps, 101, 415

\bibitem[{{Conti} \& {Ebbets}(1977)}]{conti77b}
{Conti}, P.~S. \& {Ebbets}, D. 1977, \apj, 213, 438

\bibitem[{{Crowther} {et~al.}(2002){Crowther}, {Hillier}, {Evans}, {Fullerton},
  {De Marco}, \& {Willis}}]{crowther02}
{Crowther}, P.~A., {Hillier}, D.~J., {Evans}, C.~J., {et~al.} 2002, \apj, 579,
  774

\bibitem[{{de Koter} {et~al.}(1998){de Koter}, {Heap}, \& {Hubeny}}]{dekoter98}
{de Koter}, A., {Heap}, S.~R., \& {Hubeny}, I. 1998, \apj, 509, 879

\bibitem[{{Dufton} {et~al.}(2006){Dufton}, {Smartt}, {Lee}, {Ryans}, {Hunter},
  {Evans}, {Herrero}, {Trundle}, {Lennon}, {Irwin}, \& {Kaufer}}]{dufton06}
{Dufton}, P.~L., {Smartt}, S.~J., {Lee}, J.~K., {et~al.} 2006, \aap, submitted

\bibitem[{{Evans} {et~al.}(2004{\natexlab{a}}){Evans}, {Crowther}, {Fullerton},
  \& {Hillier}}]{evans04b}
{Evans}, C.~J., {Crowther}, P.~A., {Fullerton}, A.~W., \& {Hillier}, D.~J.
  2004{\natexlab{a}}, \apj, 610, 1021

\bibitem[{{Evans} {et~al.}(2004{\natexlab{b}}){Evans}, {Howarth}, {Irwin},
  {Burnley}, \& {Harries}}]{evans04c}
{Evans}, C.~J., {Howarth}, I.~D., {Irwin}, M.~J., {Burnley}, A.~W., \&
  {Harries}, T.~J. 2004{\natexlab{b}}, \mnras, 353, 601

\bibitem[{{Evans} {et~al.}(2006){Evans}, {Lennon}, {Smartt}, \&
  {Trundle}}]{evans06}
{Evans}, C.~J., {Lennon}, D.~J., {Smartt}, S.~J., \& {Trundle}. 2006, \aap,
  accepted

\bibitem[{{Evans} {et~al.}(2004{\natexlab{c}}){Evans}, {Lennon}, {Trundle},
  {Heap}, \& {Lindler}}]{evans04a}
{Evans}, C.~J., {Lennon}, D.~J., {Trundle}, C., {Heap}, S.~R., \& {Lindler},
  D.~J. 2004{\natexlab{c}}, \apj, 607, 451

\bibitem[{{Evans} {et~al.}(2005){Evans}, {Smartt}, {Lee}, {Lennon}, {Kaufer},
  {Dufton}, {Trundle}, {Herrero}, {Sim{\' o}n-D{\'{\i}}az}, {de Koter},
  {Hamann}, {Hendry}, {Hunter}, {Irwin}, {Korn}, {Kudritzki}, {Langer},
  {Mokiem}, {Najarro}, {Pauldrach}, {Przybilla}, {Puls}, {Ryans}, {Urbaneja},
  {Venn}, \& {Villamariz}}]{evans05}
{Evans}, C.~J., {Smartt}, S.~J., {Lee}, J.-K., {et~al.} 2005, \aap, 437, 467

\bibitem[{{Feldmeier} {et~al.}(1997){Feldmeier}, {Kudritzki}, {Palsa},
  {Pauldrach}, \& {Puls}}]{feldmeier97}
{Feldmeier}, A., {Kudritzki}, R.-P., {Palsa}, R., {Pauldrach}, A.~W.~A., \&
  {Puls}, J. 1997, \aap, 320, 899

\bibitem[{{Garmany} {et~al.}(1987){Garmany}, {Conti}, \& {Massey}}]{garmany87}
{Garmany}, C.~D., {Conti}, P.~S., \& {Massey}, P. 1987, \aj, 93, 1070

\bibitem[{{Garmany} \& {Fitzpatrick}(1988)}]{garmany88}
{Garmany}, C.~D. \& {Fitzpatrick}, E.~L. 1988, \apj, 332, 711

\bibitem[{{Gray}(1978)}]{gray78}
{Gray}, D.~F. 1978, \solphys, 59, 193

\bibitem[{{Grevesse} \& {Sauval}(1998)}]{grevesse98}
{Grevesse}, N. \& {Sauval}, A.~J. 1998, Space Science Reviews, 85, 161

\bibitem[{{Harries} {et~al.}(2003){Harries}, {Hilditch}, \&
  {Howarth}}]{harries03}
{Harries}, T.~J., {Hilditch}, R.~W., \& {Howarth}, I.~D. 2003, \mnras, 339, 157

\bibitem[{{Haser} {et~al.}(1998){Haser}, {Pauldrach}, {Lennon}, {Kudritzki},
  {Lennon}, {Puls}, \& {Voels}}]{haser98}
{Haser}, S.~M., {Pauldrach}, A.~W.~A., {Lennon}, D.~J., {et~al.} 1998, \aap,
  330, 285

\bibitem[{{Heap} {et~al.}(2006){Heap}, {Lanz}, \& {Hubeny}}]{heap06}
{Heap}, S.~R., {Lanz}, T., \& {Hubeny}, I. 2006, \apj, 638, 409

\bibitem[{{Heger} \& {Langer}(2000)}]{heger00}
{Heger}, A. \& {Langer}, N. 2000, \apj, 544, 1016

\bibitem[{{Herrero}(1993)}]{herrero93}
{Herrero}, A. 1993, Space Science Reviews, 66, 137

\bibitem[{{Herrero} {et~al.}(1992){Herrero}, {Kudritzki}, {Vilchez}, {Kunze},
  {Butler}, \& {Haser}}]{herrero92}
{Herrero}, A., {Kudritzki}, R.~P., {Vilchez}, J.~M., {et~al.} 1992, \aap, 261,
  209

\bibitem[{{Herrero} {et~al.}(2002){Herrero}, {Puls}, \& {Najarro}}]{herrero02}
{Herrero}, A., {Puls}, J., \& {Najarro}, F. 2002, \aap, 396, 949

\bibitem[{{Herrero} \& {Najarro}(2005)}]{herrero05}
{Herrero}, J. \& {Najarro}, F. 2005, in Resolved Stellar Populations

\bibitem[{{Hilditch} {et~al.}(2005){Hilditch}, {Howarth}, \&
  {Harries}}]{hilditch05}
{Hilditch}, R.~W., {Howarth}, I.~D., \& {Harries}, T.~J. 2005, \mnras, 357, 304

\bibitem[{{Hillier} \& {Miller}(1998)}]{hillier98}
{Hillier}, D.~J. \& {Miller}, D.~L. 1998, \apj, 496, 407

\bibitem[{{Howarth} {et~al.}(1997){Howarth}, {Siebert}, {Hussain}, \&
  {Prinja}}]{howarth97}
{Howarth}, I.~D., {Siebert}, K.~W., {Hussain}, G.~A.~J., \& {Prinja}, R.~K.
  1997, \mnras, 284, 265

\bibitem[{{Johnson}(1966)}]{johnson66}
{Johnson}, H.~L. 1966, \araa, 4, 193

\bibitem[{{Keller}(2004)}]{keller04}
{Keller}, S.~C. 2004, Publications of the Astronomical Society of Australia,
  21, 310

\bibitem[{{Kramer} {et~al.}(2003){Kramer}, {Cohen}, \& {Owocki}}]{kramer03}
{Kramer}, R.~H., {Cohen}, D.~H., \& {Owocki}, S.~P. 2003, \apj, 592, 532

\bibitem[{{Krti{\v c}ka} {et~al.}(2003){Krti{\v c}ka}, {Owocki}, {Kub{\'a}t},
  {Galloway}, \& {Brown}}]{krticka03}
{Krti{\v c}ka}, J., {Owocki}, S.~P., {Kub{\'a}t}, J., {Galloway}, R.~K., \&
  {Brown}, J.~C. 2003, \aap, 402, 713

\bibitem[{{Kudritzki} \& {Puls}(2000)}]{kudritzki00}
{Kudritzki}, R. \& {Puls}, J. 2000, \araa, 38, 613

\bibitem[{{Kudritzki} {et~al.}(1995){Kudritzki}, {Lennon}, \&
  {Puls}}]{kudritzki95}
{Kudritzki}, R.-P., {Lennon}, D.~J., \& {Puls}, J. 1995, in Science with the
  VLT, 246--255

\bibitem[{{Kudritzki} {et~al.}(1987){Kudritzki}, {Pauldrach}, \&
  {Puls}}]{kudritzki87}
{Kudritzki}, R.~P., {Pauldrach}, A., \& {Puls}, J. 1987, \aap, 173, 293

\bibitem[{{Kudritzki} {et~al.}(1983){Kudritzki}, {Simon}, \&
  {Hamann}}]{kudritzki83}
{Kudritzki}, R.~P., {Simon}, K.~P., \& {Hamann}, W.-R. 1983, \aap, 118, 245

\bibitem[{{Lamers} \& {Pauldrach}(1991)}]{lamers91}
{Lamers}, H.~J.~G.~L.~M. \& {Pauldrach}, A.~W.~A. 1991, \aap, 244, L5

\bibitem[{{Lamers} {et~al.}(1995){Lamers}, {Snow}, \& {Lindholm}}]{lamers95}
{Lamers}, H.~J.~G.~L.~M., {Snow}, T.~P., \& {Lindholm}, D.~M. 1995, \apj, 455,
  269

\bibitem[{{Langer}(1992)}]{langer92}
{Langer}, N. 1992, \aap, 265, L17

\bibitem[{{Langer} \& {Maeder}(1995)}]{langer95}
{Langer}, N. \& {Maeder}, A. 1995, \aap, 295, 685

\bibitem[{{Maeder}(1987)}]{maeder87}
{Maeder}, A. 1987, \aap, 178, 159

\bibitem[{{Maeder} \& {Meynet}(2000)}]{maeder00}
{Maeder}, A. \& {Meynet}, G. 2000, \aap, 361, 159

\bibitem[{{Maeder} \& {Meynet}(2001)}]{maeder01}
{Maeder}, A. \& {Meynet}, G. 2001, \aap, 373, 555

\bibitem[{{Markova} {et~al.}(2004){Markova}, {Puls}, {Repolust}, \&
  {Markov}}]{markova04}
{Markova}, N., {Puls}, J., {Repolust}, T., \& {Markov}, H. 2004, \aap, 413, 693

\bibitem[{{Martins} {et~al.}(2002){Martins}, {Schaerer}, \&
  {Hiller}}]{martins02}
{Martins}, F., {Schaerer}, D., \& {Hiller}, D.~J. 2002, \aap, 382, 999

\bibitem[{{Martins} {et~al.}(2005{\natexlab{a}}){Martins}, {Schaerer}, \&
  {Hillier}}]{martins05a}
{Martins}, F., {Schaerer}, D., \& {Hillier}, D.~J. 2005{\natexlab{a}}, \aap,
  436, 1049

\bibitem[{{Martins} {et~al.}(2004){Martins}, {Schaerer}, {Hillier}, \&
  {Heydari-Malayeri}}]{martins04}
{Martins}, F., {Schaerer}, D., {Hillier}, D.~J., \& {Heydari-Malayeri}, M.
  2004, \aap, 420, 1087

\bibitem[{{Martins} {et~al.}(2005{\natexlab{b}}){Martins}, {Schaerer},
  {Hillier}, {Meynadier}, {Heydari-Malayeri}, \& {Walborn}}]{martins05b}
{Martins}, F., {Schaerer}, D., {Hillier}, D.~J., {et~al.} 2005{\natexlab{b}},
  \aap, 441, 735

\bibitem[{{Massey}(2002)}]{massey02}
{Massey}, P. 2002, \apjs, 141, 81

\bibitem[{{Massey} {et~al.}(2004){Massey}, {Bresolin}, {Kudritzki}, {Puls}, \&
  {Pauldrach}}]{massey04}
{Massey}, P., {Bresolin}, F., {Kudritzki}, R.~P., {Puls}, J., \& {Pauldrach},
  A.~W.~A. 2004, \apj, 608, 1001

\bibitem[{{Massey} {et~al.}(1989){Massey}, {Parker}, \& {Garmany}}]{massey89}
{Massey}, P., {Parker}, J.~W., \& {Garmany}, C.~D. 1989, \aj, 98, 1305

\bibitem[{{Massey} {et~al.}(2005){Massey}, {Puls}, {Pauldrach}, {Bresolin},
  {Kudritzki}, \& {Simon}}]{massey05}
{Massey}, P., {Puls}, J., {Pauldrach}, A.~W.~A., {et~al.} 2005, \apj, 627, 477

\bibitem[{{Meynet} \& {Maeder}(2000)}]{meynet00}
{Meynet}, G. \& {Maeder}, A. 2000, \aap, 361, 101

\bibitem[{{Meynet} \& {Maeder}(2005)}]{meynet05}
{Meynet}, G. \& {Maeder}, A. 2005, \aap, 429, 581

\bibitem[{{Mokiem} {et~al.}(2005){Mokiem}, {de Koter}, {Puls}, {Herrero},
  {Najarro}, \& {Villamariz}}]{mokiem05}
{Mokiem}, M.~R., {de Koter}, A., {Puls}, J., {et~al.} 2005, \aap, 441, 711

\bibitem[{{Mokiem} {et~al.}(2004){Mokiem}, {Mart{\'{\i}}n-Hern{\' a}ndez},
  {Lenorzer}, {de Koter}, \& {Tielens}}]{mokiem04}
{Mokiem}, M.~R., {Mart{\'{\i}}n-Hern{\' a}ndez}, N.~L., {Lenorzer}, A., {de
  Koter}, A., \& {Tielens}, A.~G.~G.~M. 2004, \aap, 419, 319

\bibitem[{{Momany} {et~al.}(2001){Momany}, {Vandame}, {Zaggia}, {Mignani}, {da
  Costa}, {Arnouts}, {Groenewegen}, {Hatziminaoglou}, {Madejsky}, {Rit{\'e}},
  {Schirmer}, \& {Slijkhuis}}]{momany01}
{Momany}, Y., {Vandame}, B., {Zaggia}, S., {et~al.} 2001, \aap, 379, 436

\bibitem[{{Niemela} {et~al.}(1986){Niemela}, {Marraco}, \&
  {Cabanne}}]{niemela86}
{Niemela}, V.~S., {Marraco}, H.~G., \& {Cabanne}, M.~L. 1986, \pasp, 98, 1133

\bibitem[{{Owocki} \& {Puls}(1999)}]{owocki99}
{Owocki}, S.~P. \& {Puls}, J. 1999, \apj, 510, 355

\bibitem[{{Parker} {et~al.}(1992){Parker}, {Garmany}, {Massey}, \&
  {Walborn}}]{parker92}
{Parker}, J.~W., {Garmany}, C.~D., {Massey}, P., \& {Walborn}, N.~R. 1992, \aj,
  103, 1205

\bibitem[{{Pauldrach} {et~al.}(1986){Pauldrach}, {Puls}, \&
  {Kudritzki}}]{pauldrach86}
{Pauldrach}, A., {Puls}, J., \& {Kudritzki}, R.~P. 1986, \aap, 164, 86

\bibitem[{{Pauldrach} {et~al.}(2001){Pauldrach}, {Hoffmann}, \&
  {Lennon}}]{pauldrach01}
{Pauldrach}, A.~W.~A., {Hoffmann}, T.~L., \& {Lennon}, M. 2001, \aap, 375, 161

\bibitem[{{Penny}(1996)}]{penny96}
{Penny}, L.~R. 1996, \apj, 463, 737

\bibitem[{{Penny} {et~al.}(2004){Penny}, {Sprague}, {Seago}, \&
  {Gies}}]{penny04}
{Penny}, L.~R., {Sprague}, A.~J., {Seago}, G., \& {Gies}, D.~R. 2004, \apj,
  617, 1316

\bibitem[{{Porter} \& {Rivinius}(2003)}]{porter03}
{Porter}, J.~M. \& {Rivinius}, T. 2003, \pasp, 115, 1153

\bibitem[{{Prinja} \& {Crowther}(1998)}]{prinja98}
{Prinja}, R.~K. \& {Crowther}, P.~A. 1998, \mnras, 300, 828

\bibitem[{{Puls} {et~al.}(1996){Puls}, {Kudritzki}, {Herrero}, {Pauldrach},
  {Haser}, {Lennon}, {Gabler}, {Voels}, {Vilchez}, {Wachter}, \&
  {Feldmeier}}]{puls96}
{Puls}, J., {Kudritzki}, R.-P., {Herrero}, A., {et~al.} 1996, \aap, 305, 171

\bibitem[{{Puls} {et~al.}(2000){Puls}, {Springmann}, \& {Lennon}}]{puls00}
{Puls}, J., {Springmann}, U., \& {Lennon}, M. 2000, \aaps, 141, 23

\bibitem[{{Puls} {et~al.}(2005){Puls}, {Urbaneja}, {Venero}, {Repolust},
  {Springmann}, {Jokuthy}, \& {Mokiem}}]{puls05}
{Puls}, J., {Urbaneja}, M.~A., {Venero}, R., {et~al.} 2005, \aap, 435, 669

\bibitem[{{Repolust} {et~al.}(2004){Repolust}, {Puls}, \&
  {Herrero}}]{repolust04}
{Repolust}, T., {Puls}, J., \& {Herrero}, A. 2004, \aap, 415, 349

\bibitem[{{Rolleston} {et~al.}(2003){Rolleston}, {Venn}, {Tolstoy}, \&
  {Dufton}}]{rolleston03}
{Rolleston}, W.~R.~J., {Venn}, K., {Tolstoy}, E., \& {Dufton}, P.~L. 2003,
  \aap, 400, 21

\bibitem[{{Ryans} {et~al.}(2002){Ryans}, {Dufton}, {Rolleston}, {Lennon},
  {Keenan}, {Smoker}, \& {Lambert}}]{ryans02}
{Ryans}, R.~S.~I., {Dufton}, P.~L., {Rolleston}, W.~R.~J., {et~al.} 2002,
  \mnras, 336, 577

\bibitem[{{Sanduleak}(1968)}]{sanduleak68}
{Sanduleak}, N. 1968, \aj, 73, 246

\bibitem[{{Schaerer} \& {de Koter}(1997)}]{schaerer97}
{Schaerer}, D. \& {de Koter}, A. 1997, \aap, 322, 598

\bibitem[{{Sim{\'o}n-D{\'{\i}}az} {et~al.}(2006){Sim{\'o}n-D{\'{\i}}az},
  {Herrero}, {Esteban}, \& {Najarro}}]{simon06}
{Sim{\'o}n-D{\'{\i}}az}, S., {Herrero}, A., {Esteban}, C., \& {Najarro}, F.
  2006, \aap, 448, 351

\bibitem[{{Smith} {et~al.}(2003){Smith}, {Davidson}, {Gull}, {Ishibashi}, \&
  {Hillier}}]{smith03}
{Smith}, N., {Davidson}, K., {Gull}, T.~R., {Ishibashi}, K., \& {Hillier},
  D.~J. 2003, \apj, 586, 432

\bibitem[{{Trundle} {et~al.}(2004){Trundle}, {Lennon}, {Puls}, \&
  {Dufton}}]{trundle04}
{Trundle}, C., {Lennon}, D.~J., {Puls}, J., \& {Dufton}, P.~L. 2004, \aap, 417,
  217

\bibitem[{{Vacca}(1994)}]{vacca94}
{Vacca}, W.~D. 1994, \apj, 421, 140

\bibitem[{{Vacca} {et~al.}(1996){Vacca}, {Garmany}, \& {Shull}}]{vacca96}
{Vacca}, W.~D., {Garmany}, C.~D., \& {Shull}, J.~M. 1996, \apj, 460, 914

\bibitem[{{van Boekel} {et~al.}(2003){van Boekel}, {Kervella}, {Sch{\"o}ller},
  {Herbst}, {Brandner}, {de Koter}, {Waters}, {Hillier}, {Paresce}, {Lenzen},
  \& {Lagrange}}]{vanboekel03}
{van Boekel}, R., {Kervella}, P., {Sch{\"o}ller}, M., {et~al.} 2003, \aap, 410,
  L37

\bibitem[{{Vink} {et~al.}(2000){Vink}, {de Koter}, \& {Lamers}}]{vink00}
{Vink}, J.~S., {de Koter}, A., \& {Lamers}, H.~J.~G.~L.~M. 2000, \aap, 362, 295

\bibitem[{{Vink} {et~al.}(2001){Vink}, {de Koter}, \& {Lamers}}]{vink01}
{Vink}, J.~S., {de Koter}, A., \& {Lamers}, H.~J.~G.~L.~M. 2001, \aap, 369, 574

\bibitem[{{Voels} {et~al.}(1989){Voels}, {Bohannan}, {Abbott}, \&
  {Hummer}}]{voels89}
{Voels}, S.~A., {Bohannan}, B., {Abbott}, D.~C., \& {Hummer}, D.~G. 1989, \apj,
  340, 1073

\bibitem[{{Walborn}(1977)}]{walborn77}
{Walborn}, N.~R. 1977, \apj, 215, 53

\bibitem[{{Walborn} \& {Blades}(1997)}]{walborn97}
{Walborn}, N.~R. \& {Blades}, J.~C. 1997, \apjs, 112, 457

\bibitem[{{Walborn} \& {Fitzpatrick}(1990)}]{walborn90}
{Walborn}, N.~R. \& {Fitzpatrick}, E.~L. 1990, \pasp, 102, 379

\bibitem[{{Walborn} {et~al.}(2002){Walborn}, {Fullerton}, {Crowther},
  {Bianchi}, {Hutchings}, {Pellerin}, {Sonneborn}, \& {Willis}}]{walborn02b}
{Walborn}, N.~R., {Fullerton}, A.~W., {Crowther}, P.~A., {et~al.} 2002, \apjs,
  141, 443

\bibitem[{{Walborn} {et~al.}(2000){Walborn}, {Lennon}, {Heap}, {Lindler},
  {Smith}, {Evans}, \& {Parker}}]{walborn00}
{Walborn}, N.~R., {Lennon}, D.~J., {Heap}, S.~R., {et~al.} 2000, \pasp, 112,
  1243

\bibitem[{{Walborn} \& {Parker}(1992)}]{walborn92}
{Walborn}, N.~R. \& {Parker}, J.~W. 1992, \apjl, 399, L87

\bibitem[{{Westerlund}(1997)}]{westerlund97}
{Westerlund}, B.~E. 1997, {The Magellanic Clouds} (Cambridge: Cambridge Univ.\
  Press)

\bibitem[{{Yoon} \& {Langer}(2005)}]{yoon05}
{Yoon}, S.-C. \& {Langer}, N. 2005, \aap, 443, 643

\end{thebibliography}
\end{small}

\clearpage

\appendix
\section{Fits and comments on individual objects}
\label{sec:fits}
The observed spectra shown in this section were corrected for radial
velocities. If not noted differently the lines that were fitted are
the hydrogen Balmer lines \ha, \hg\ and \hd; the \hei\ singlet line at
4387~\AA; the \hei\ triplet lines at 4026, 4471 and 4713~\AA; and the
\heii\ lines at 4200, 4541 and 4686~\AA. In general we find that all
lines could be fitted quite accurately including the
\hei~$\lambda$4471 for which we only encountered the ``generalised
dilution'' in one case. This is in accordance with the fact that the
majority of the sample consists of dwarf stars and the single star
where this problem is present, \azv372, is a late type supergiant.

\begin{figure*}[t]
  \centering \resizebox{17cm}{!}{
  \includegraphics{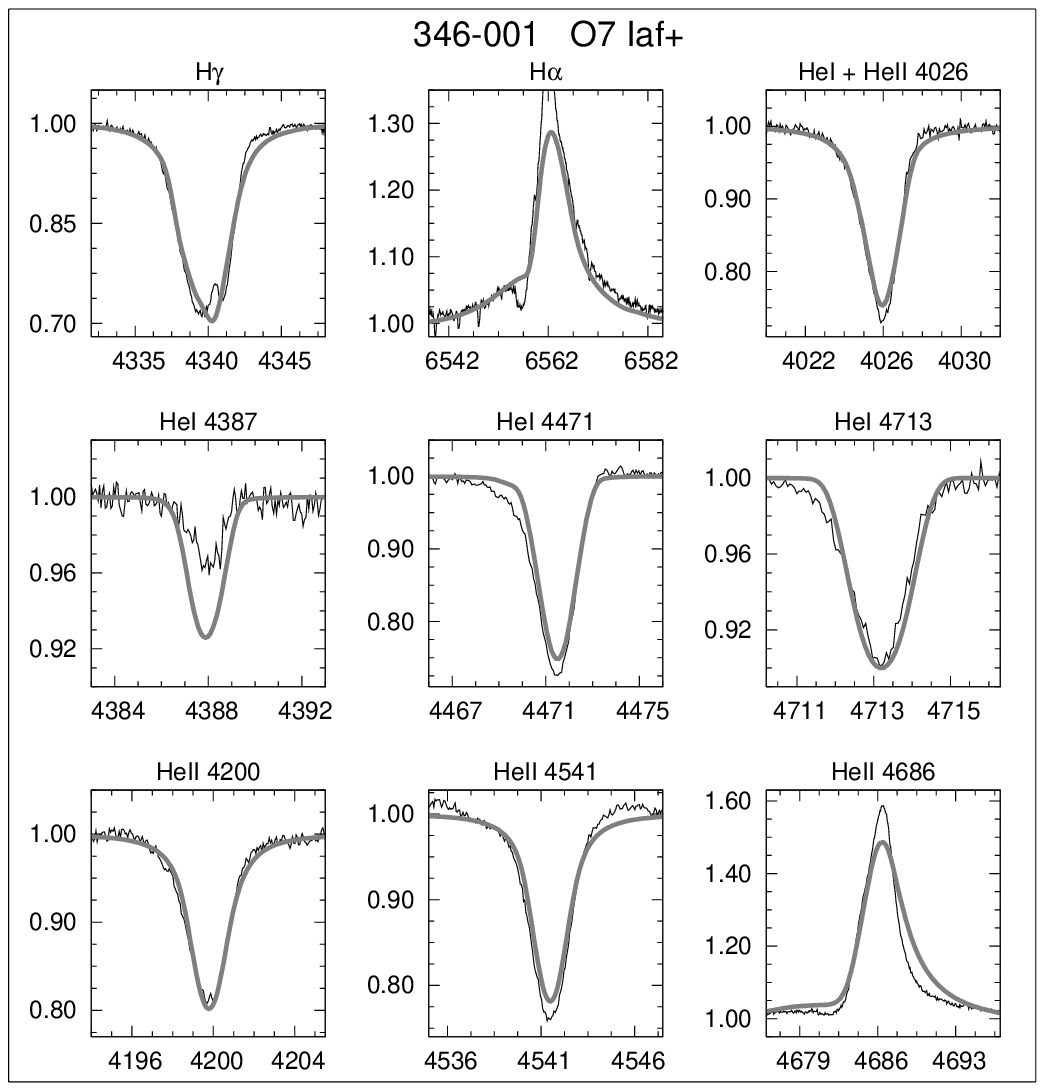}
  \includegraphics{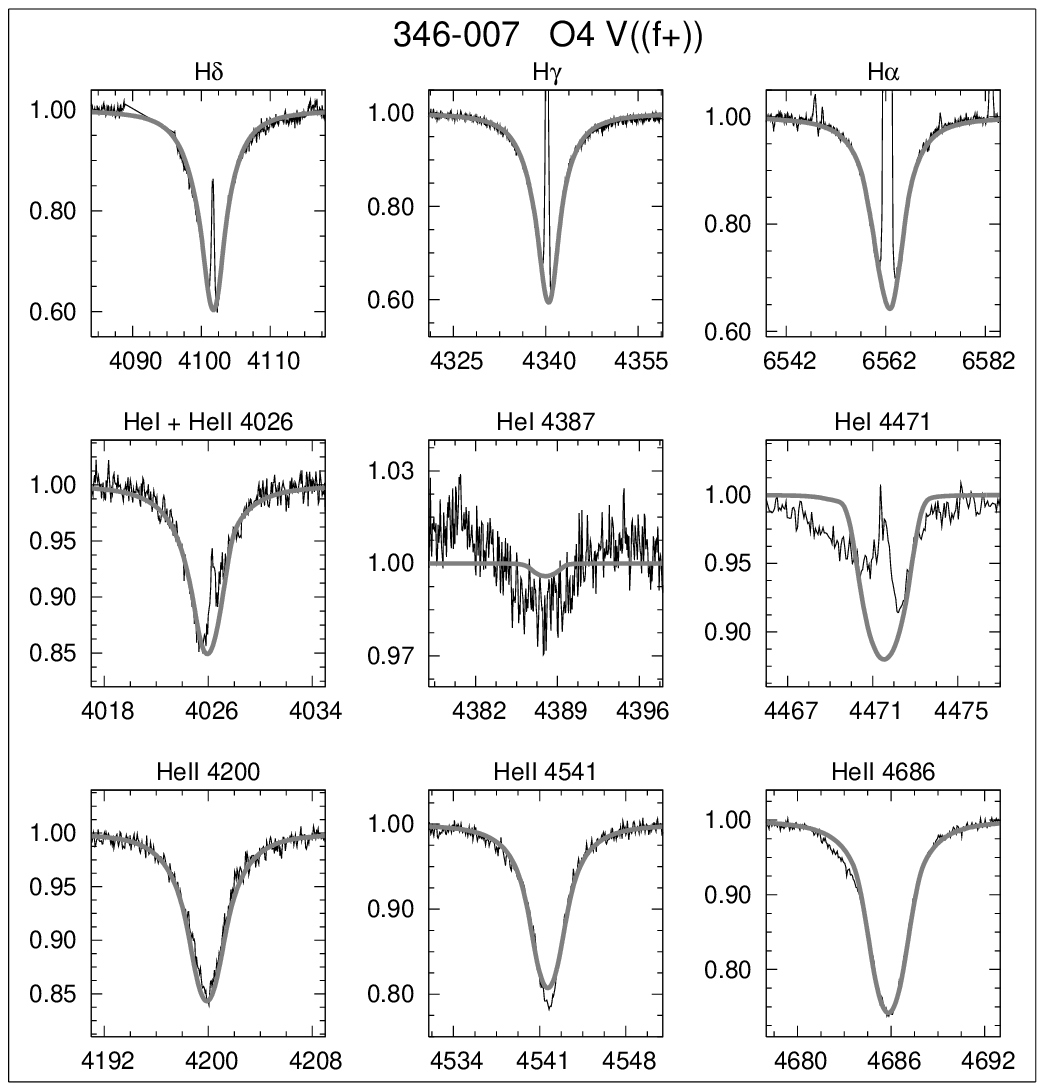} }

  \resizebox{1cm}{!}{ }

  \resizebox{17cm}{!}{
    \includegraphics{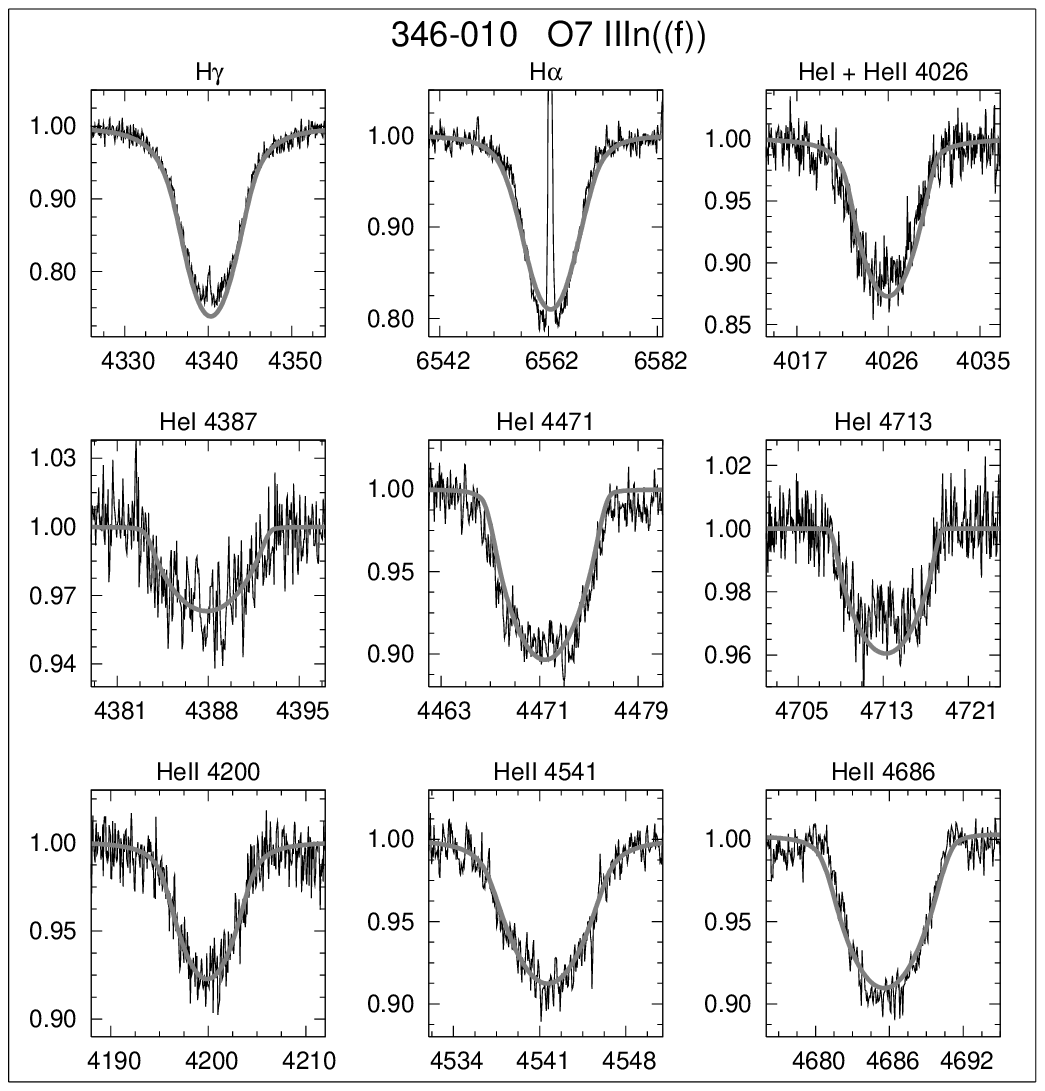}
    \includegraphics{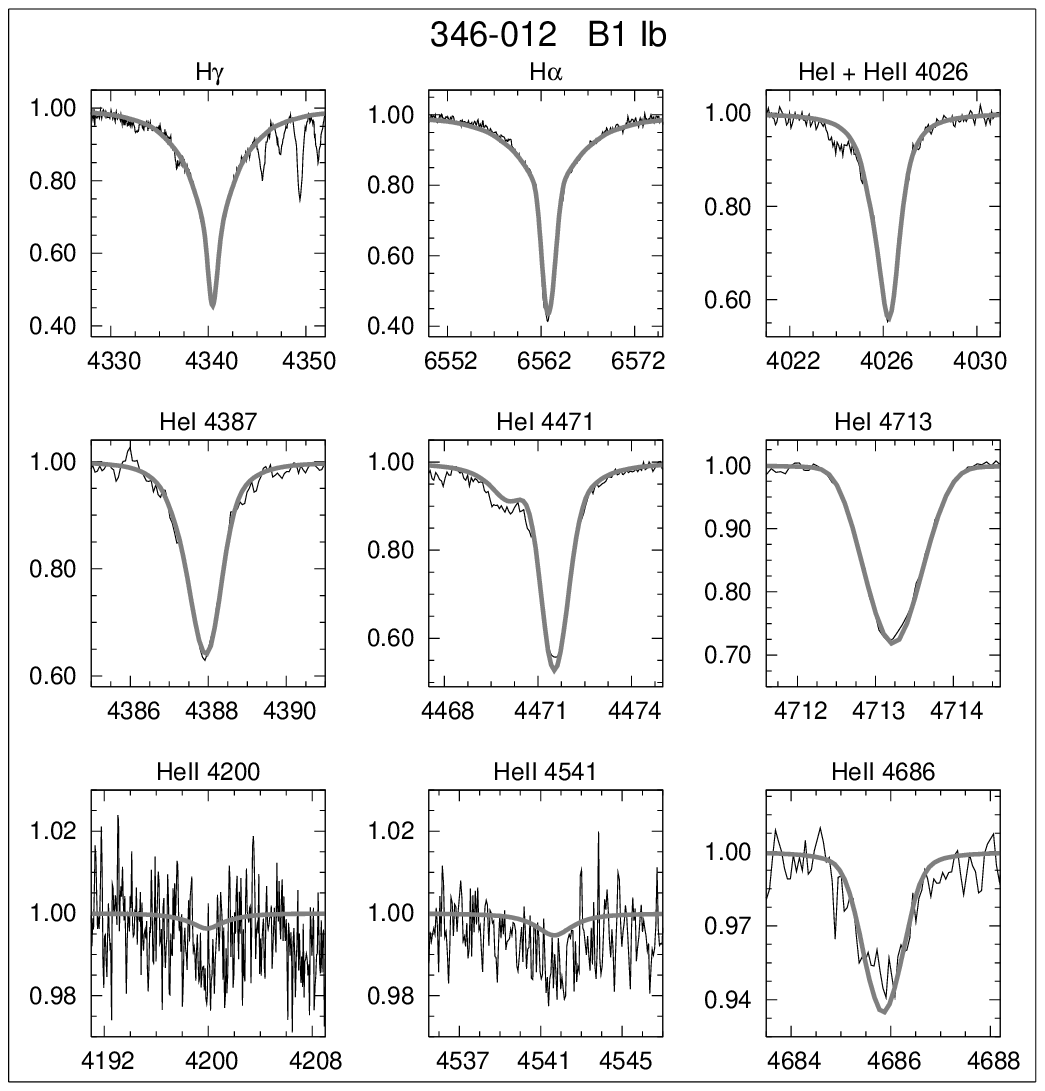}
  }
  \caption{Comparison of the observed line profiles of \ngc346-001,
    -007, -010 and -012 with best fitting synthetic line profiles
    obtained using the automated fitting method (grey
    lines). Wavelengths are given on the horizontal axis in \AA. The
    vertical axis gives the normalised flux. Note that this axis is
    scaled differently for each line.}
  \label{fig:fits_1}
\end{figure*}

\begin{figure*}[t]
  \centering
  \resizebox{17cm}{!}{
    \includegraphics{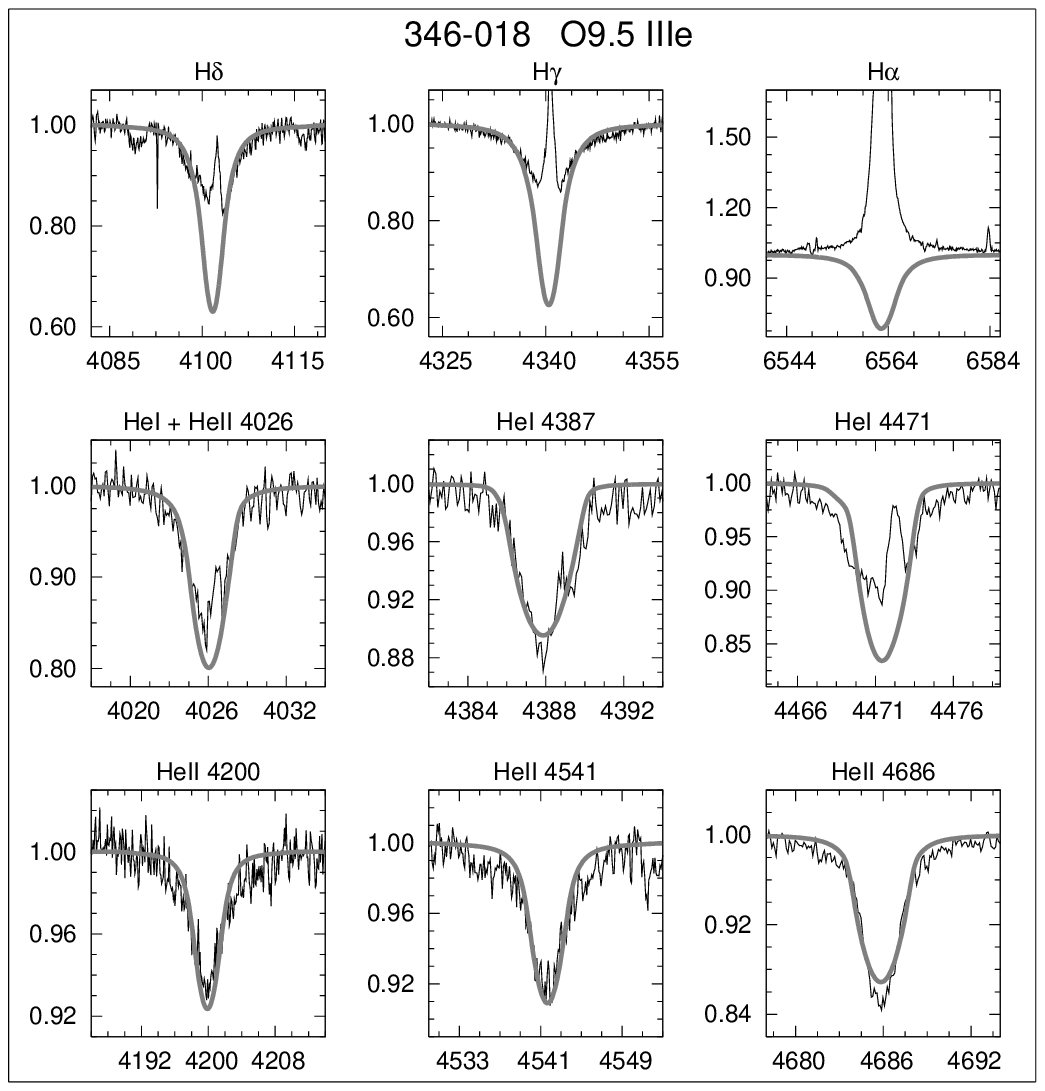}
    \includegraphics{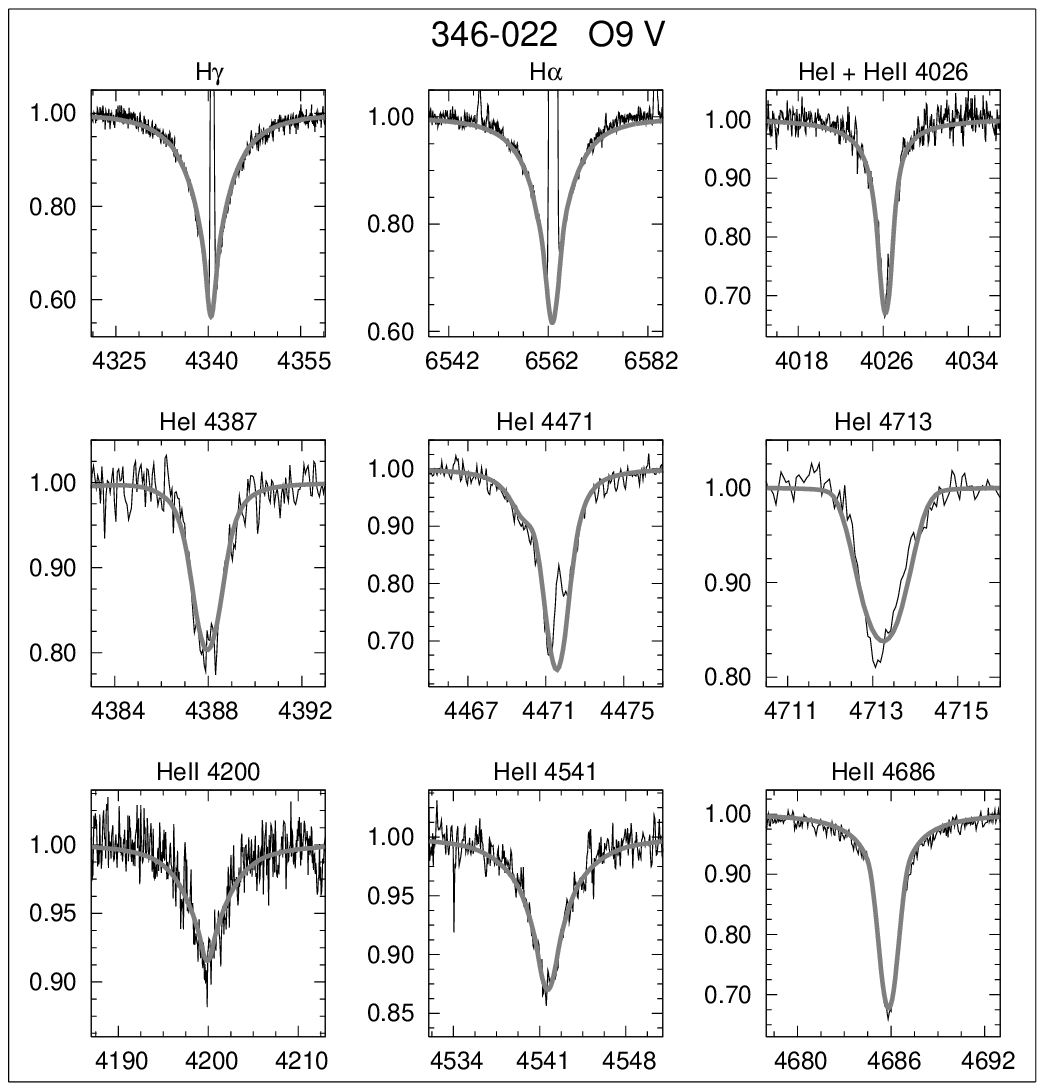}
  }

  \resizebox{1cm}{!}{ }

  \resizebox{17cm}{!}{
    \includegraphics{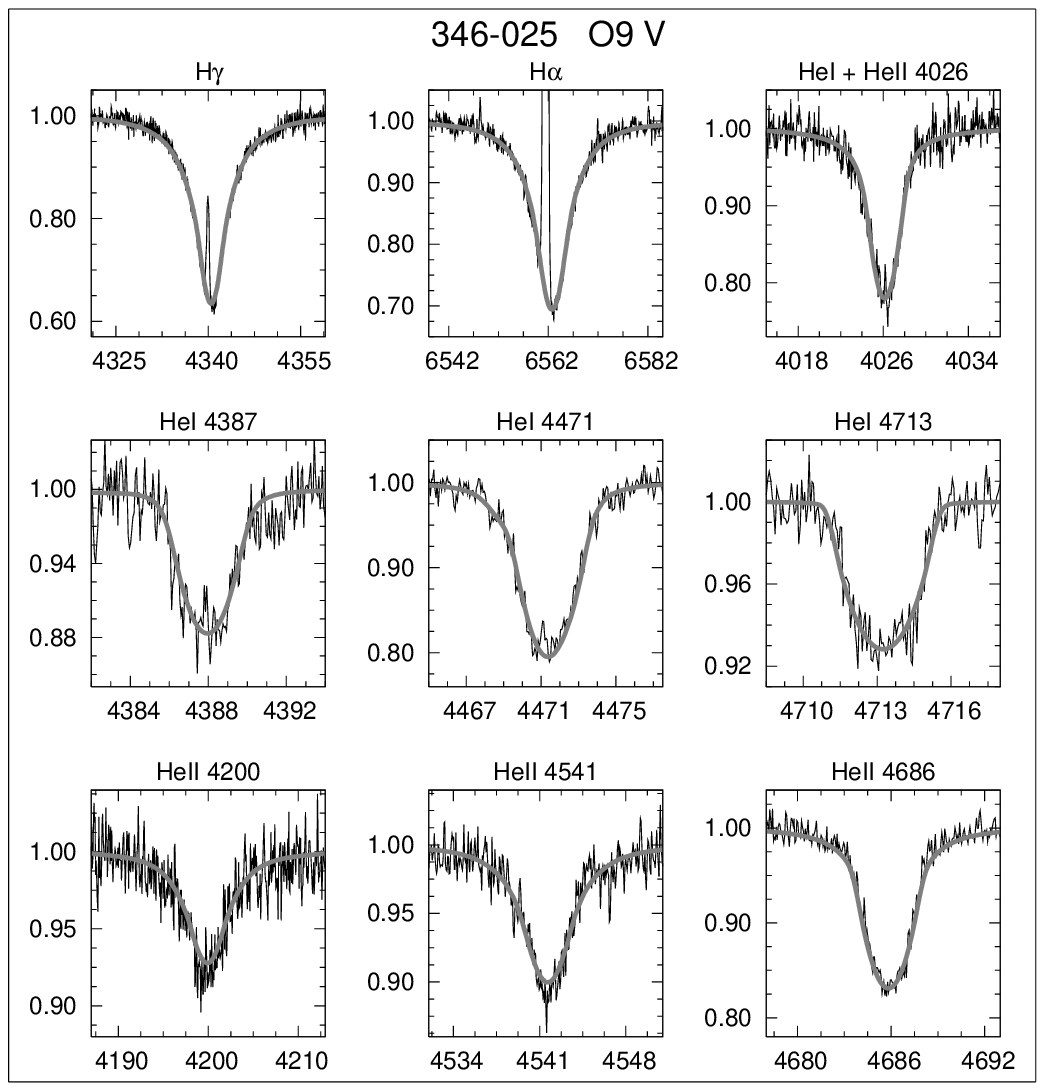}
    \includegraphics{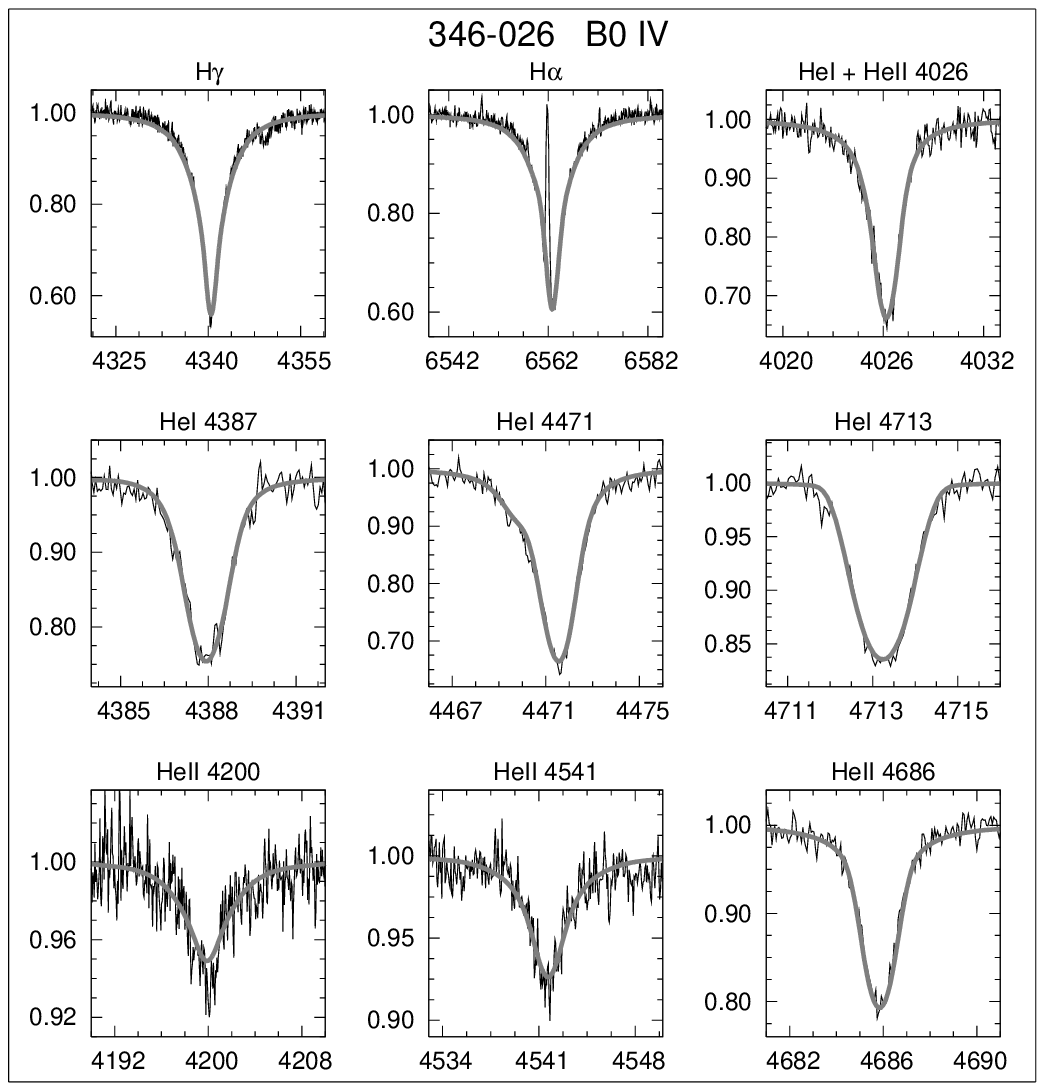}
  }
  \caption{Same as Fig.~\ref{fig:fits_1}, however, for \ngc346-018,
  -022, -025 and -026.}
  \label{fig:fits_2}
\end{figure*}

\begin{figure*}[t]
  \centering
  \resizebox{17cm}{!}{
    \includegraphics{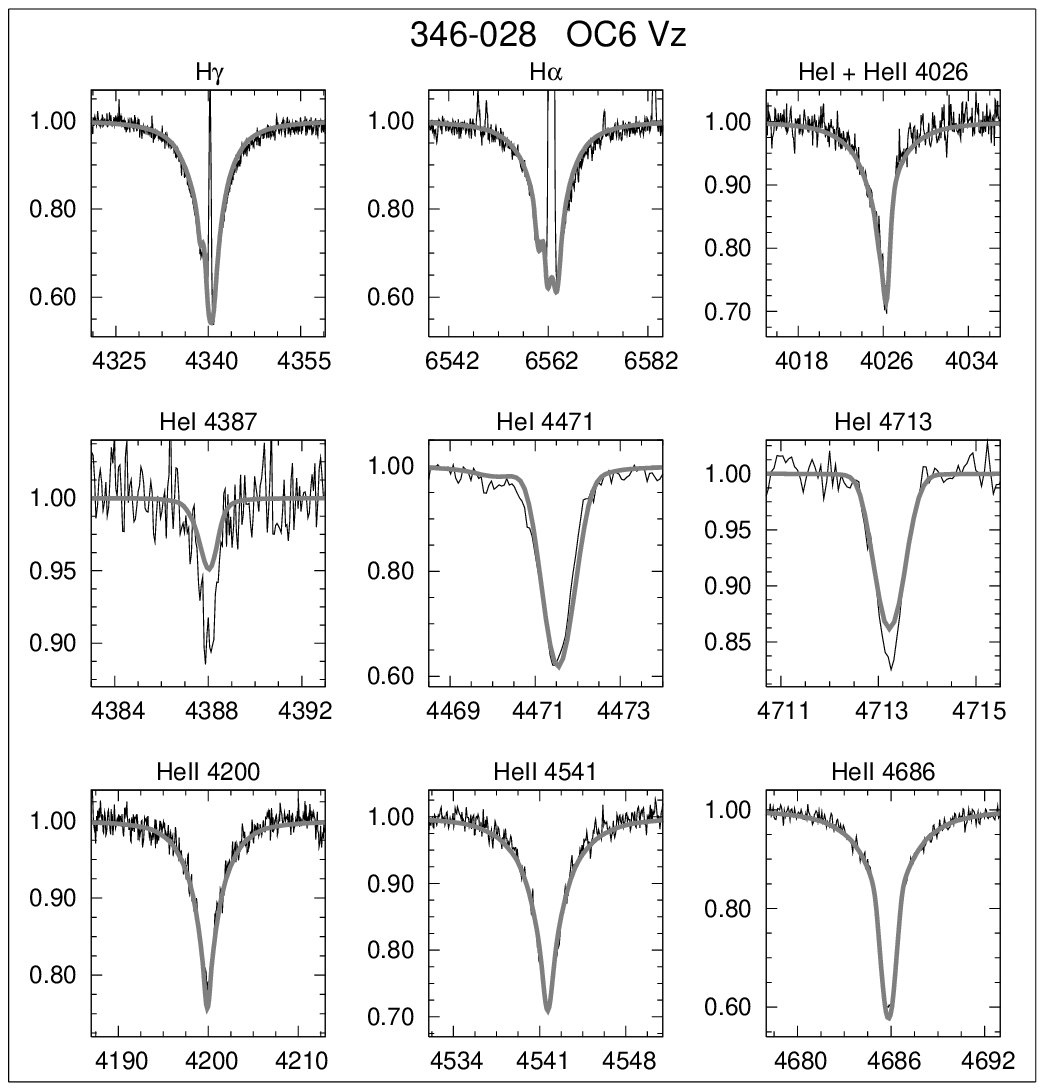}
    \includegraphics{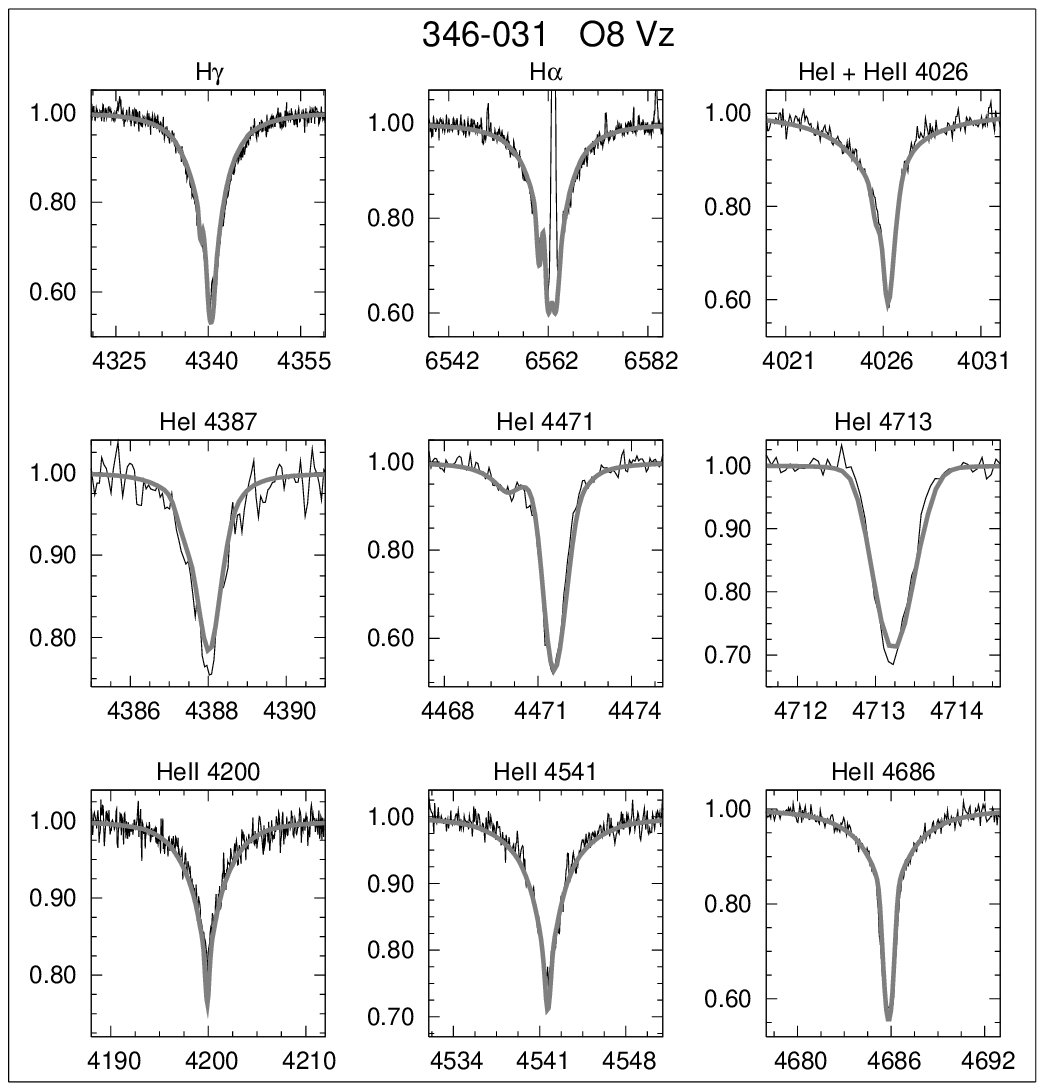}
  }

  \resizebox{1cm}{!}{ }

  \resizebox{17cm}{!}{
    \includegraphics{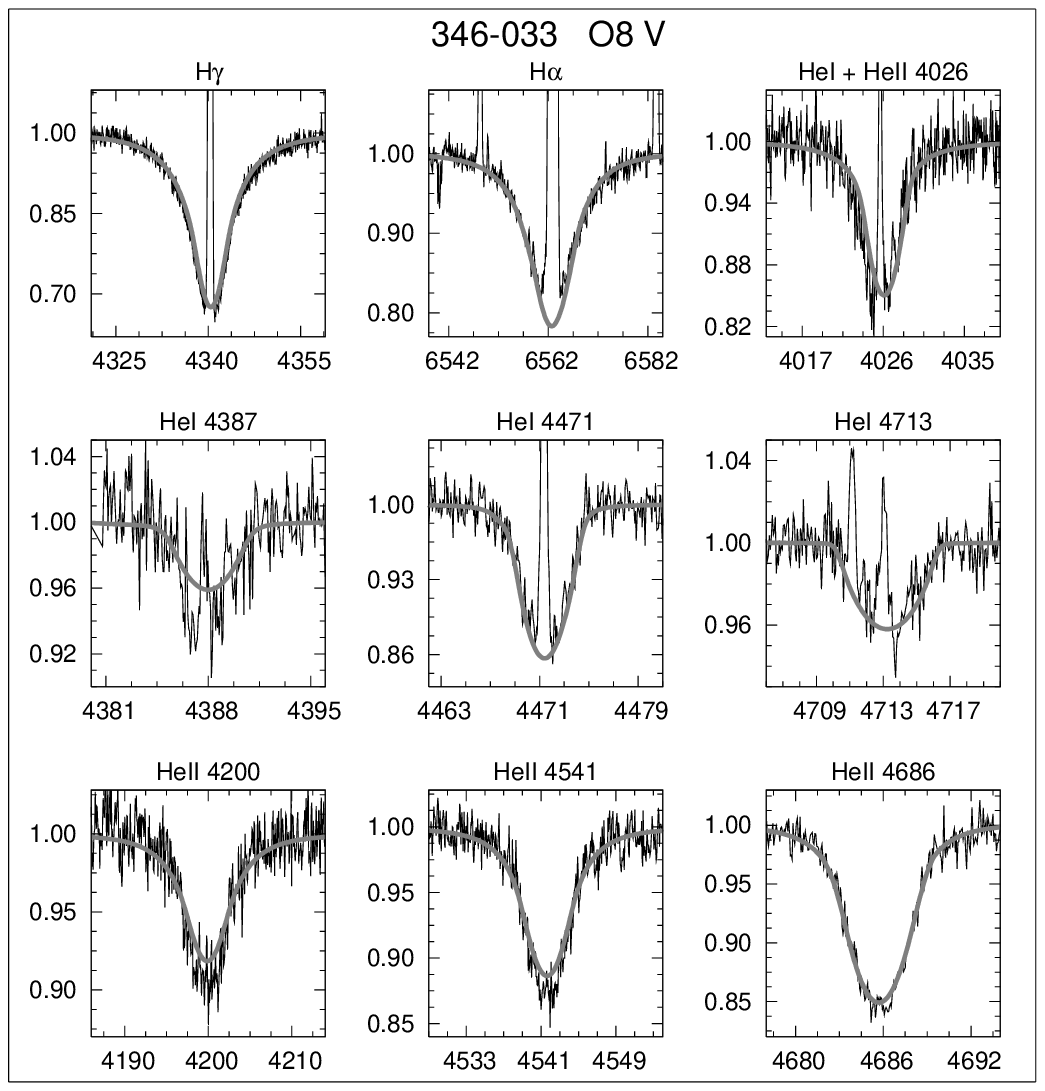}
    \includegraphics{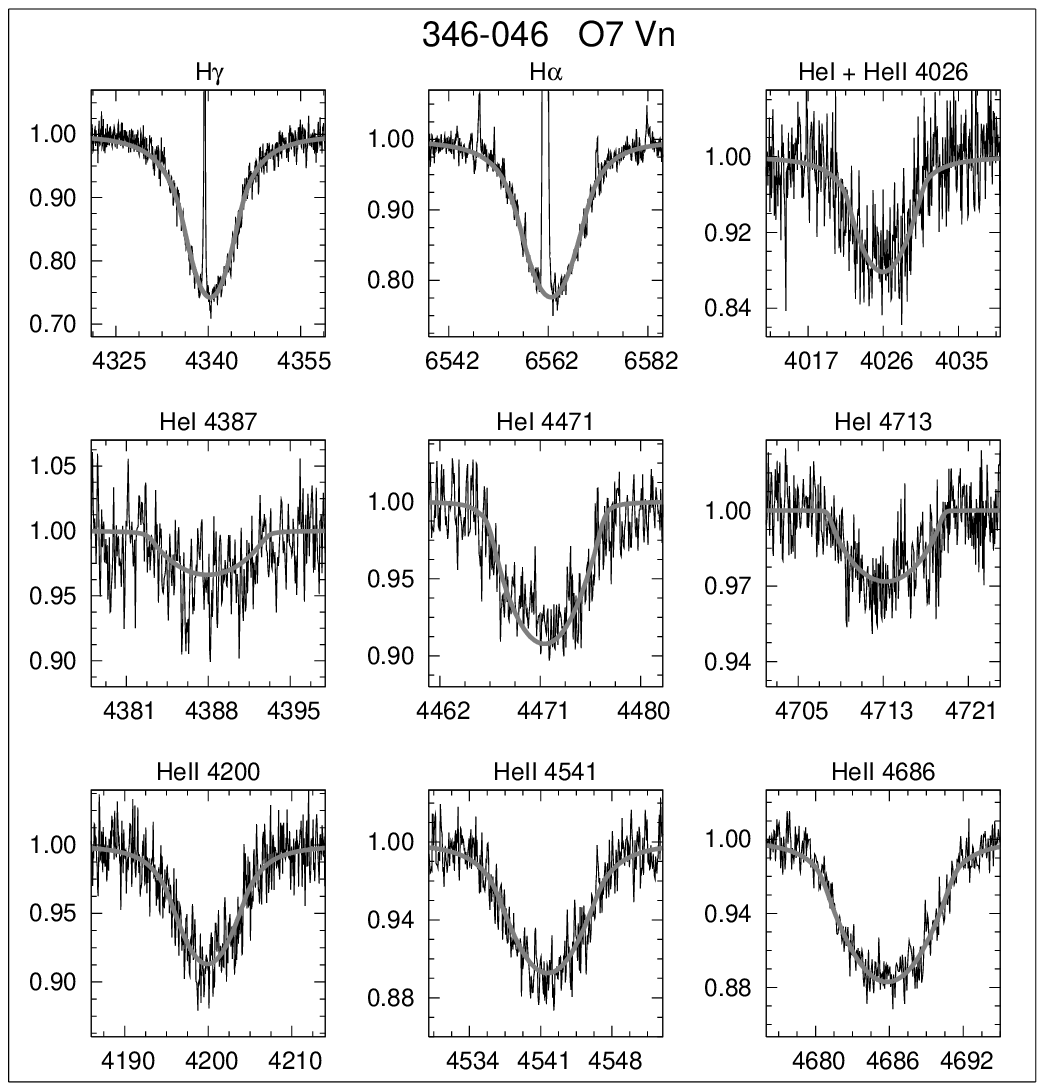}
  }
  \caption{Same as Fig.~\ref{fig:fits_1}, however, for \ngc346-028,
  -031, -033 and -046.}
  \label{fig:fits_3}
\end{figure*}

\begin{figure*}[t]
  \centering
  \resizebox{17cm}{!}{
    \includegraphics{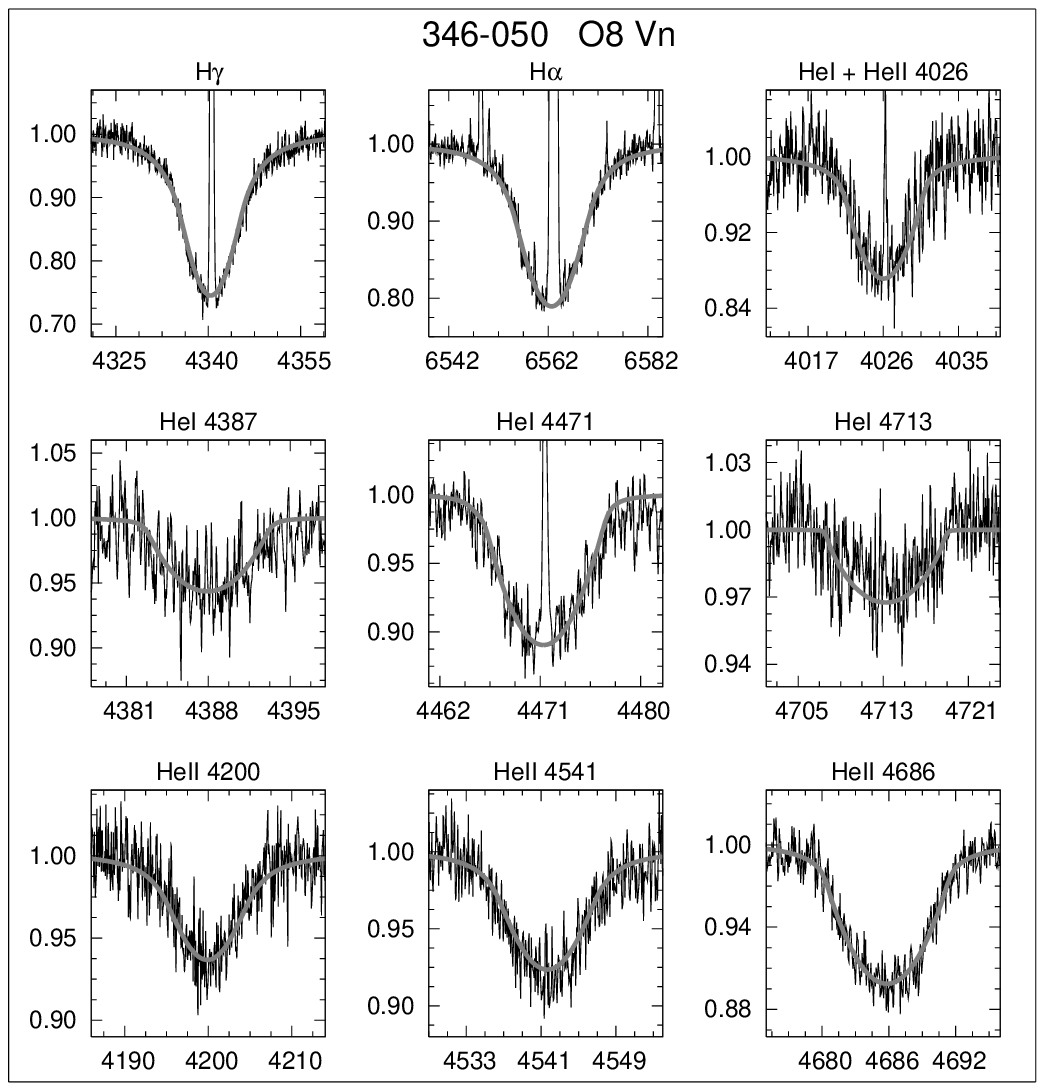}
    \includegraphics{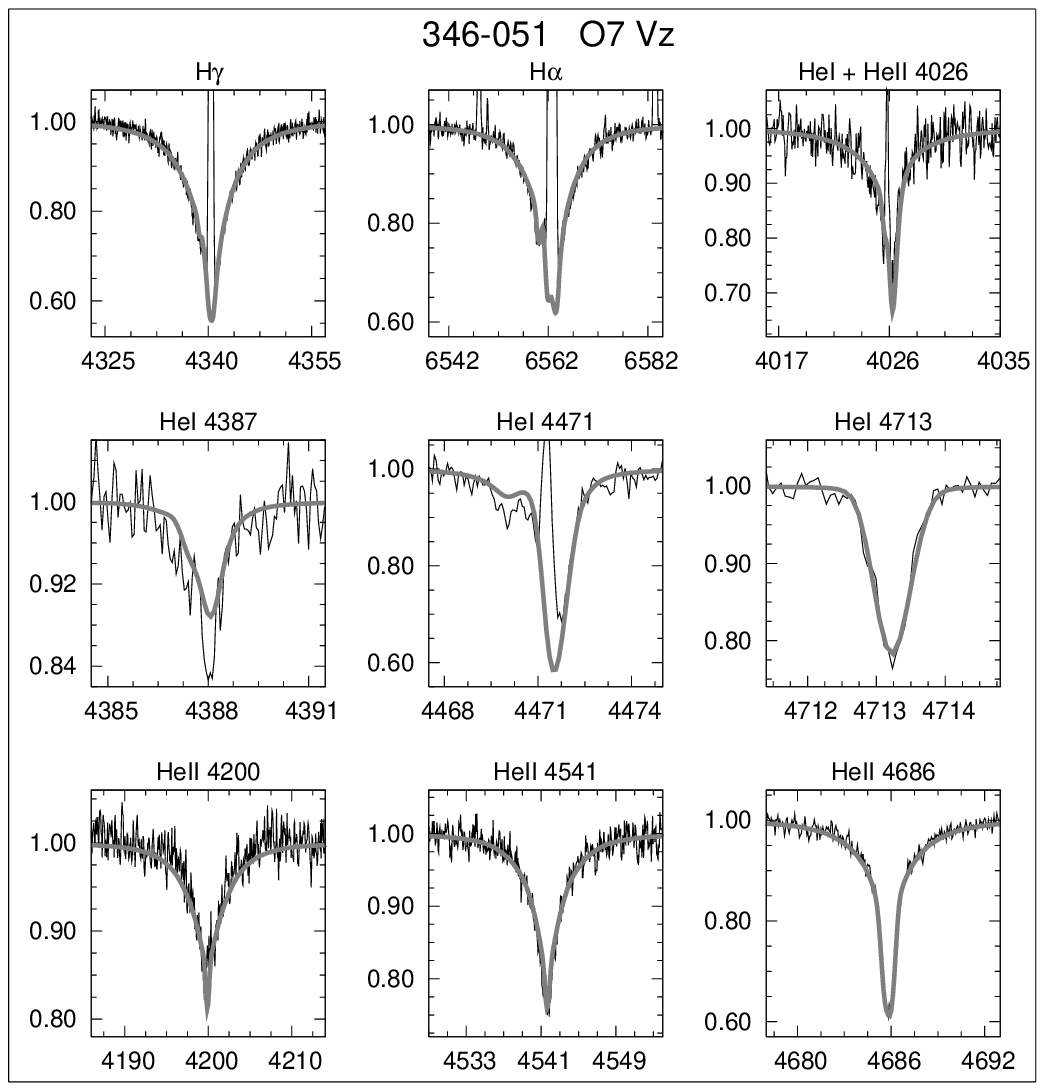}
  }
  \resizebox{1cm}{!}{ }

  \resizebox{17cm}{!}{
    \includegraphics{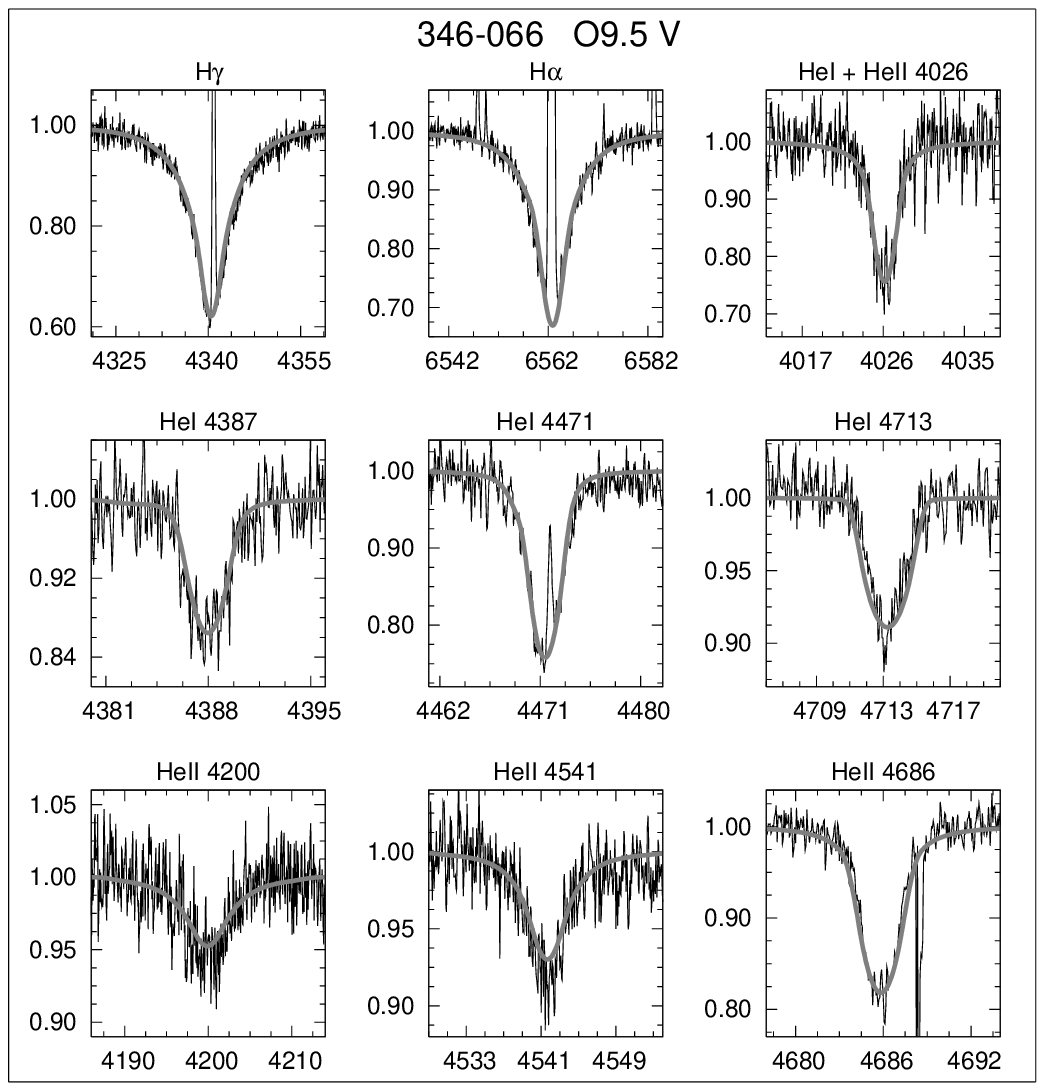}
    \includegraphics{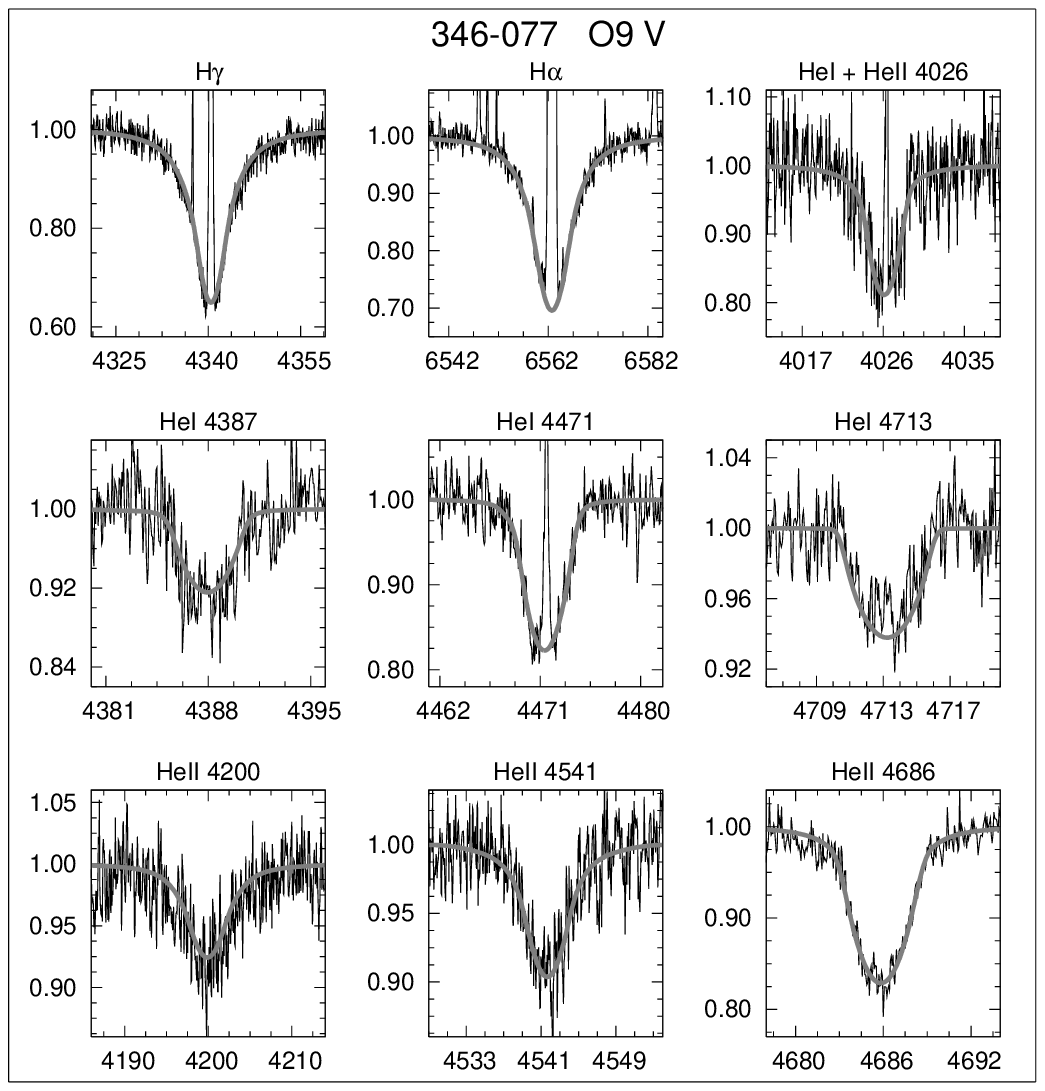}
  }
  \caption{Same as Fig.~\ref{fig:fits_1}, however, for \ngc346-050,
  -051, -066 and -077.}
  \label{fig:fits_4}
\end{figure*}

\begin{figure*}[t]
  \centering
  \resizebox{17cm}{!}{
    \includegraphics{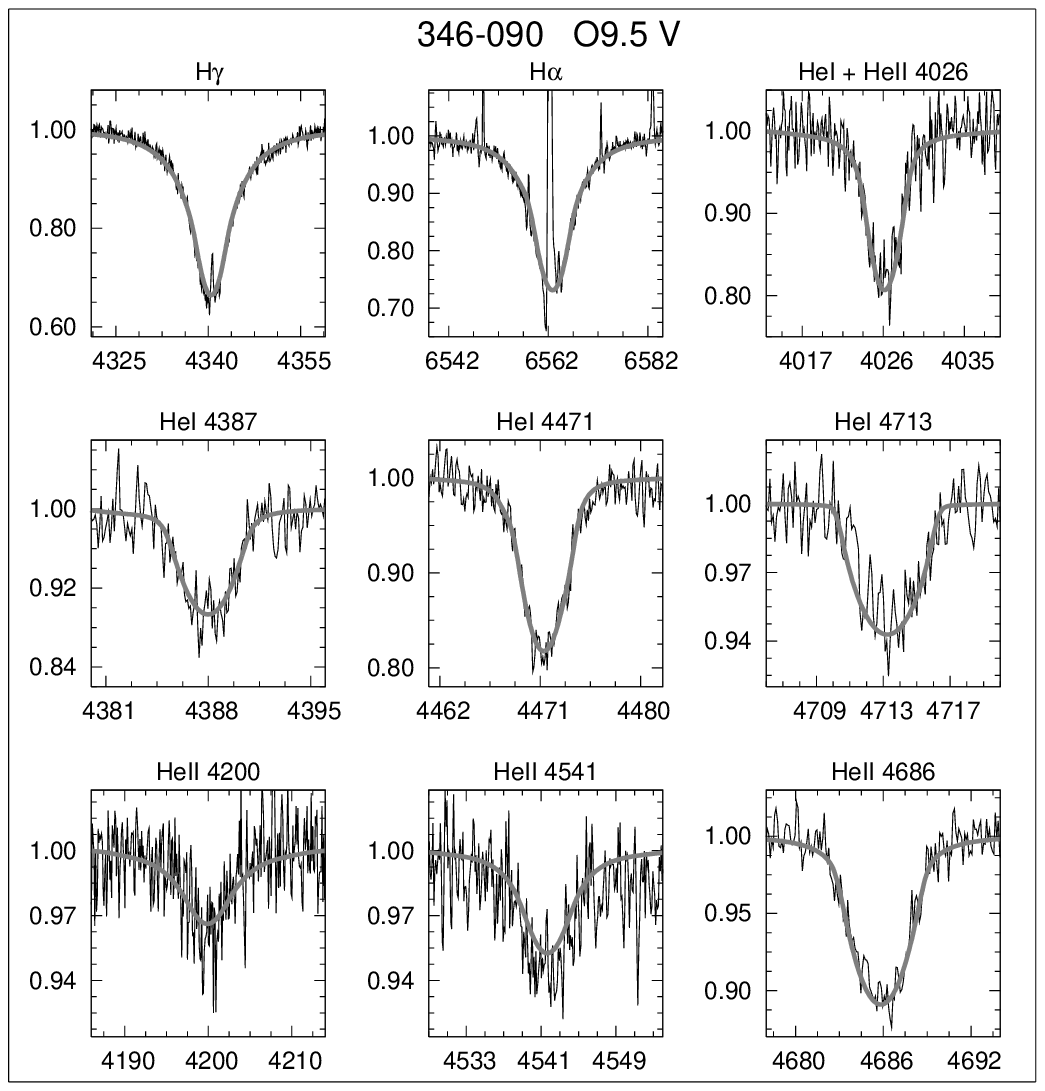}
    \includegraphics{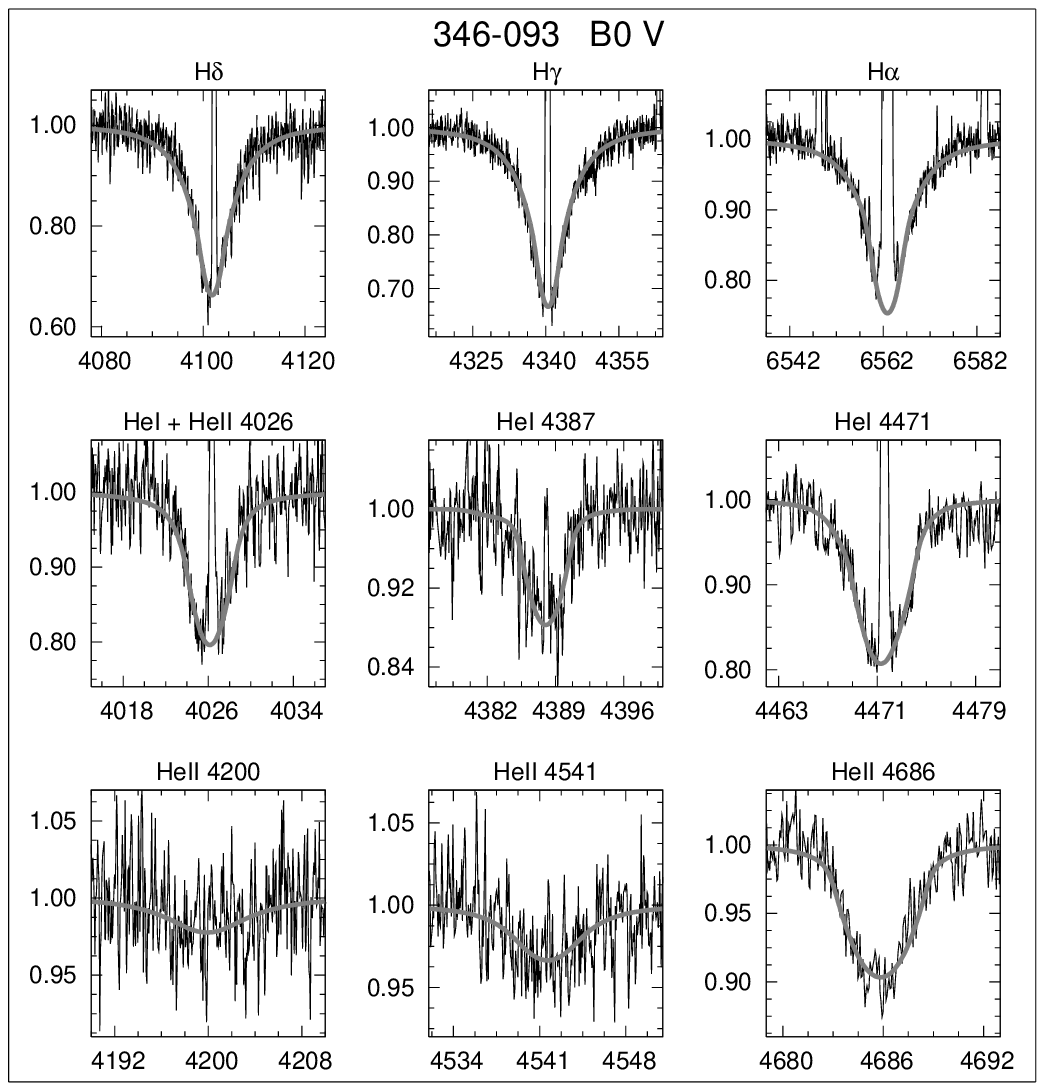}
  }
  \resizebox{1cm}{!}{ }

  \resizebox{17cm}{!}{
    \includegraphics{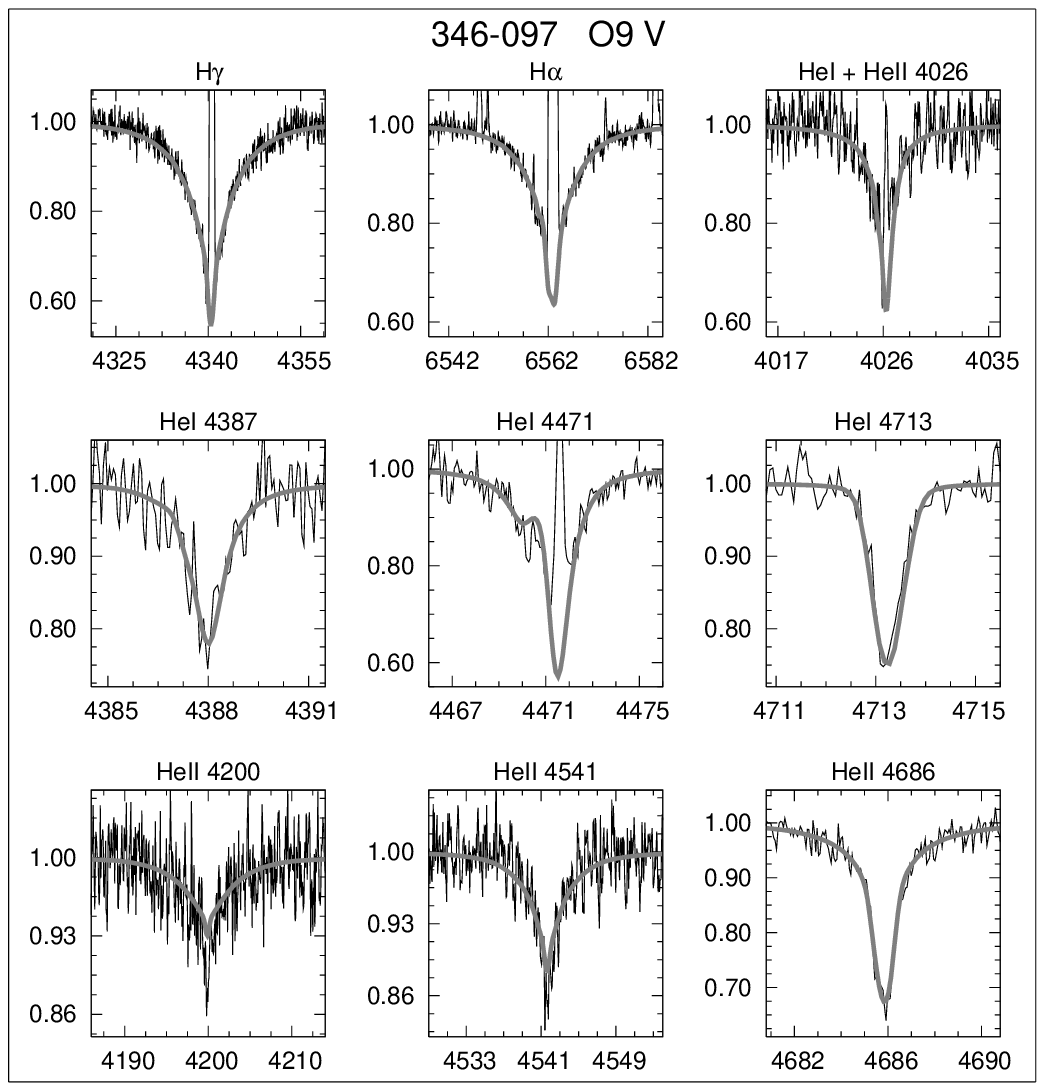}
    \includegraphics{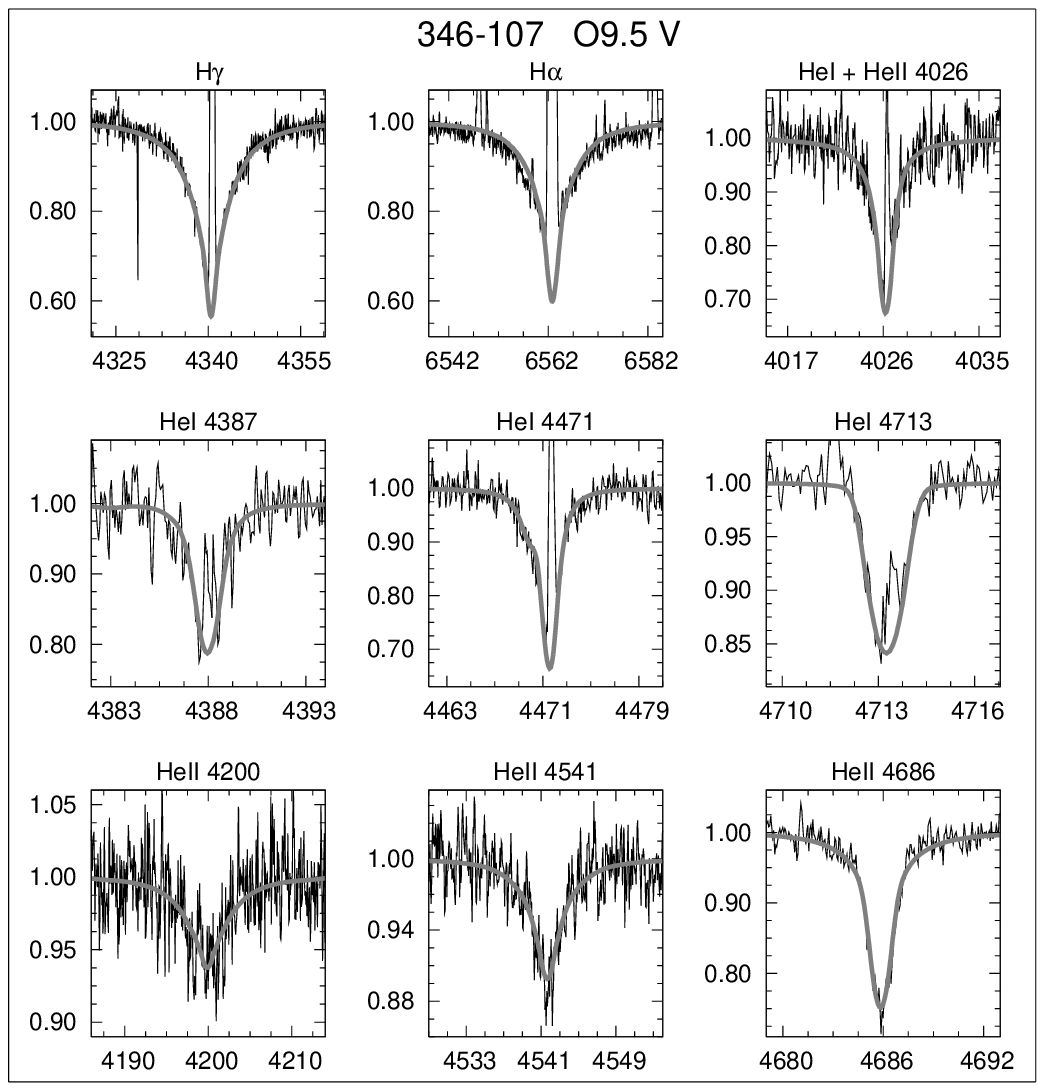}
  }

  \caption{Same as Fig.~\ref{fig:fits_1}, however, for \ngc346-090,
  -093, -097 and -107.}
  \label{fig:fits_5}
\end{figure*}

\begin{figure*}[t]
  \centering
  \resizebox{17cm}{!}{
    \includegraphics{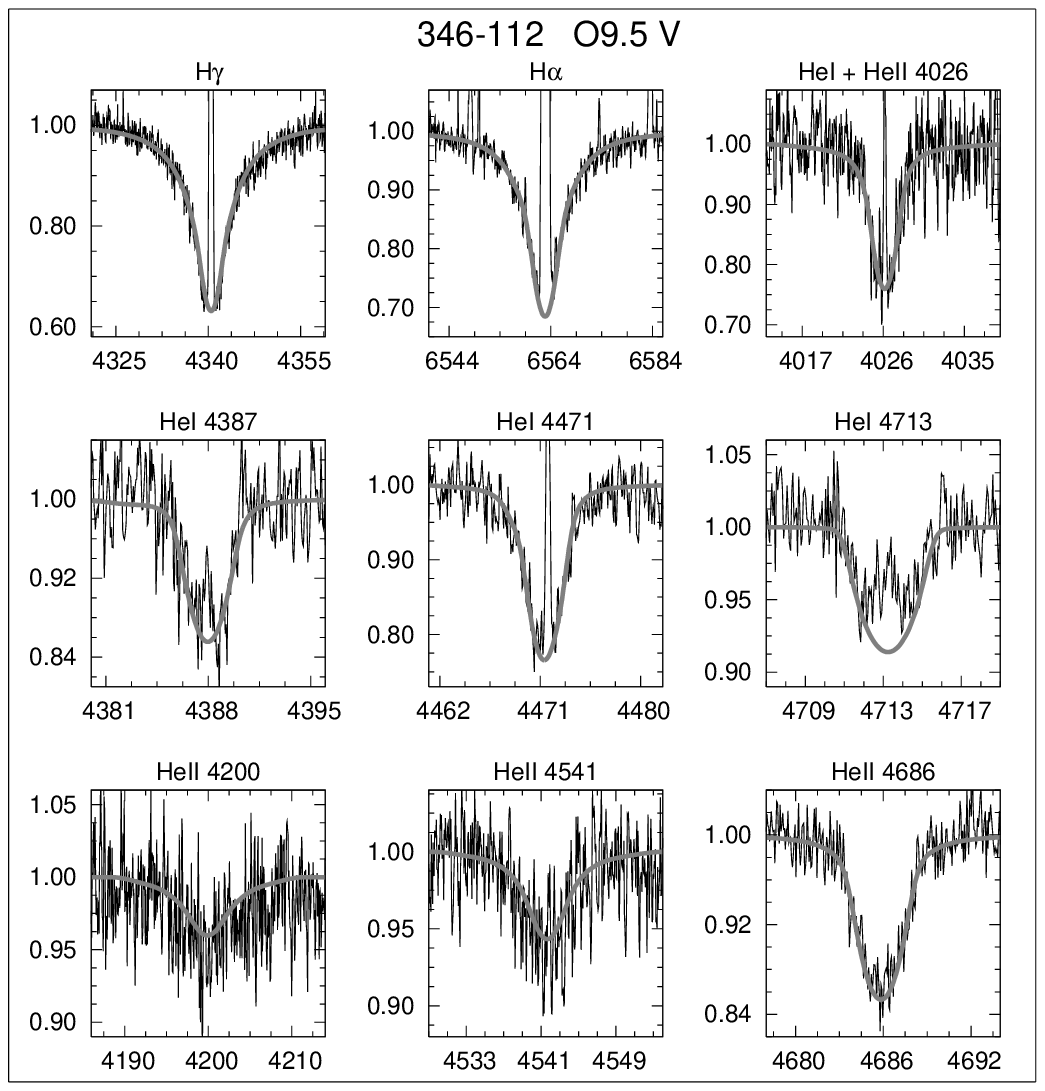}
    \includegraphics{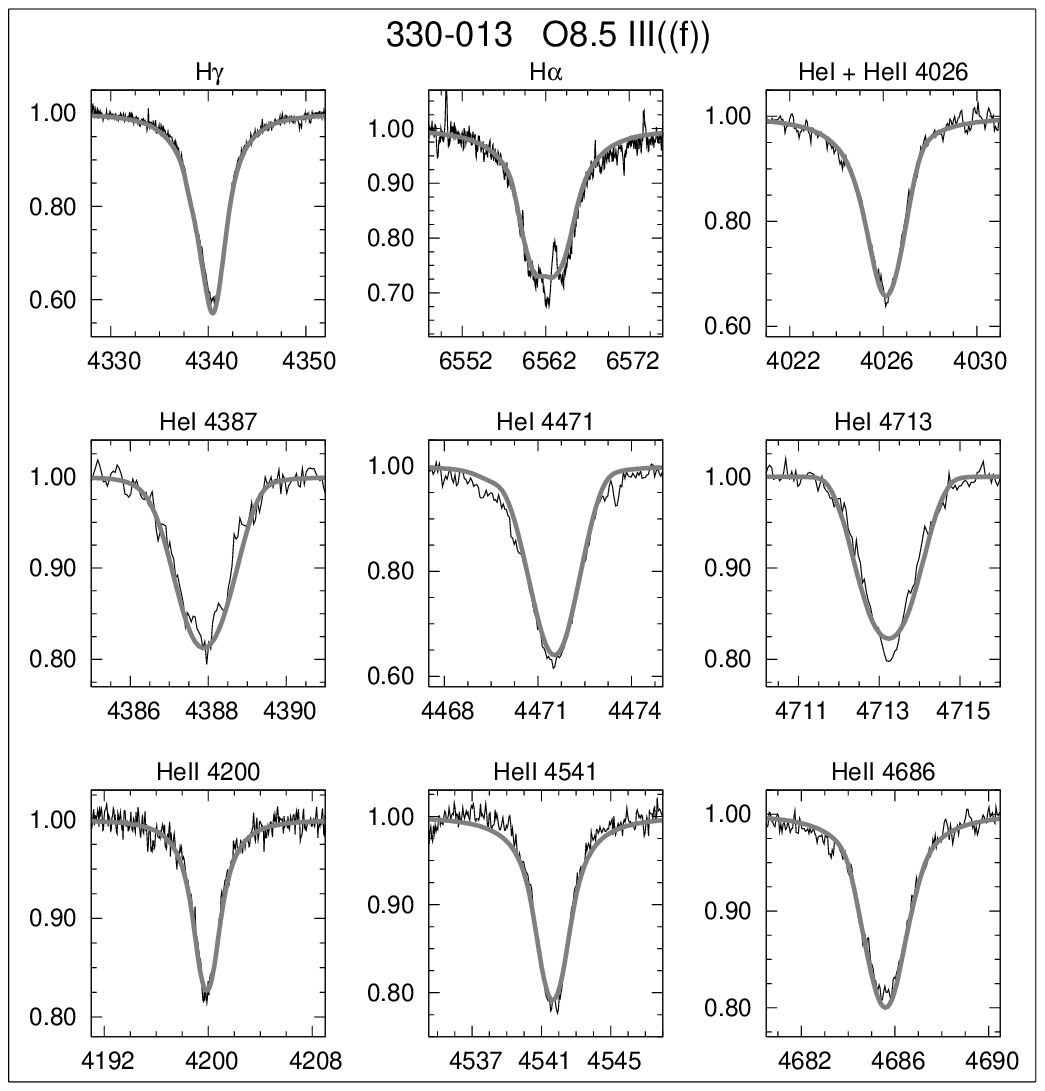}
  }
  \resizebox{1cm}{!}{ }

  \resizebox{8.5cm}{!}{
    \includegraphics{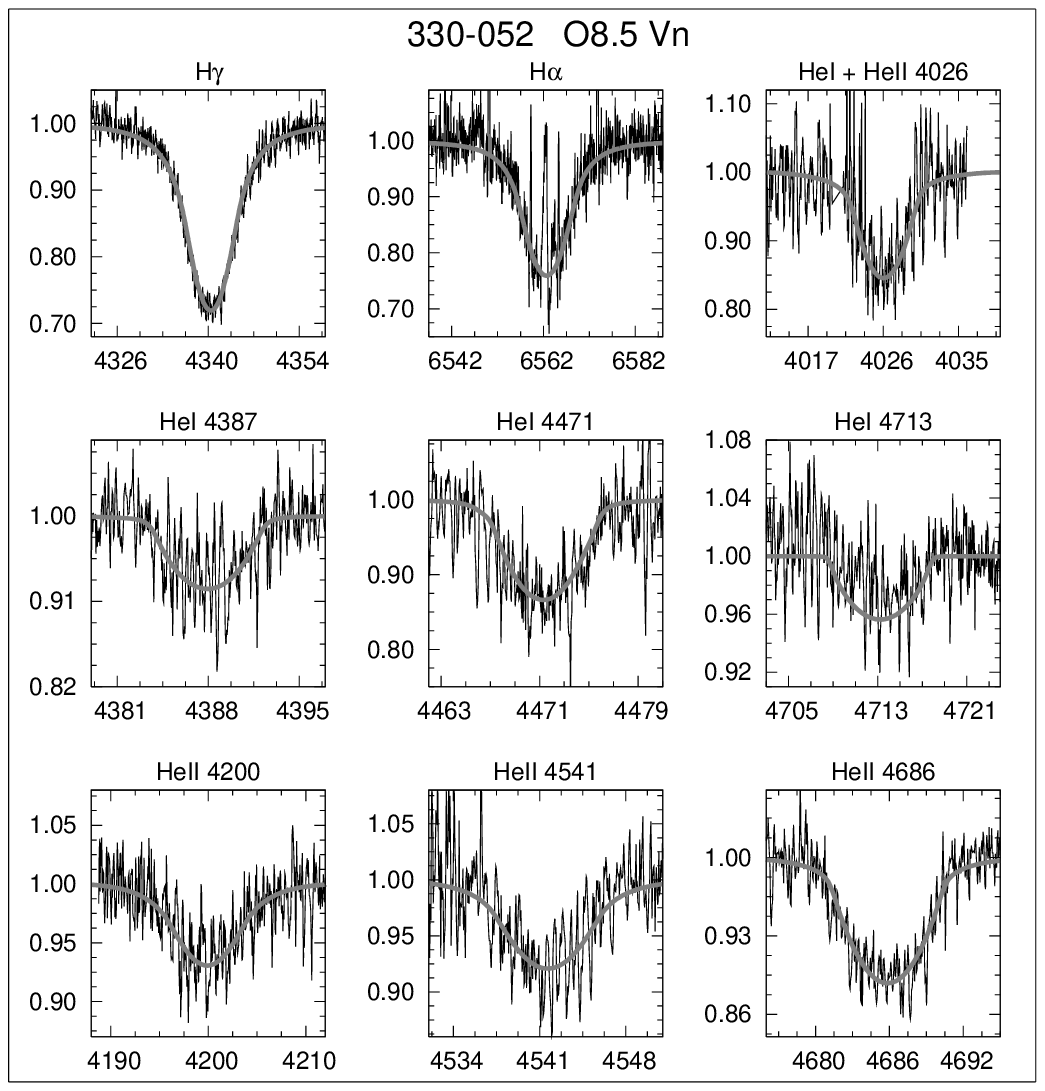}
  }
  \caption{Same as Fig.~\ref{fig:fits_1}, however, for \ngc346-112,
  \ngc330-031 and -052.}
  \label{fig:fits_6}
\end{figure*}

\begin{figure*}[t]
  \centering
  \resizebox{17cm}{!}{
    \includegraphics{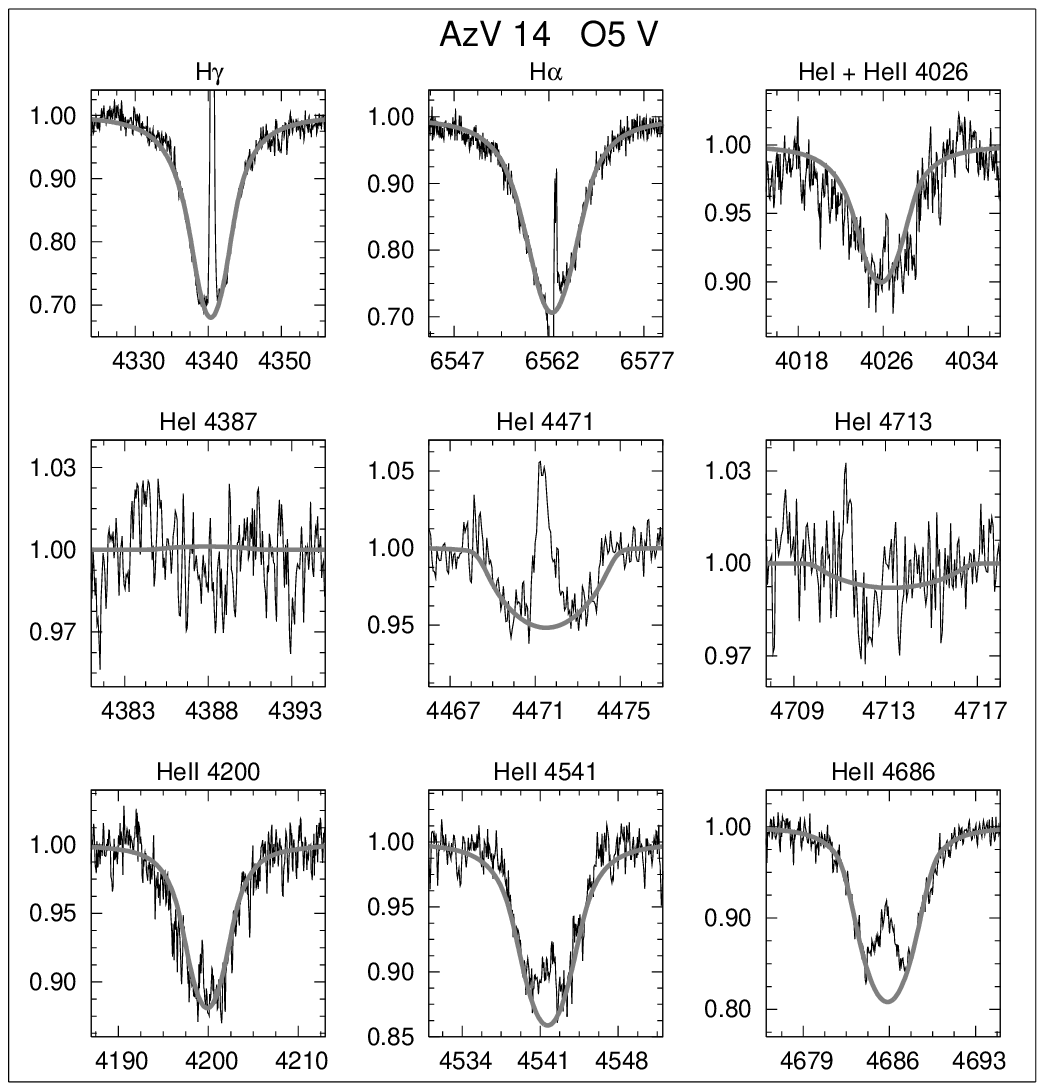}
    \includegraphics{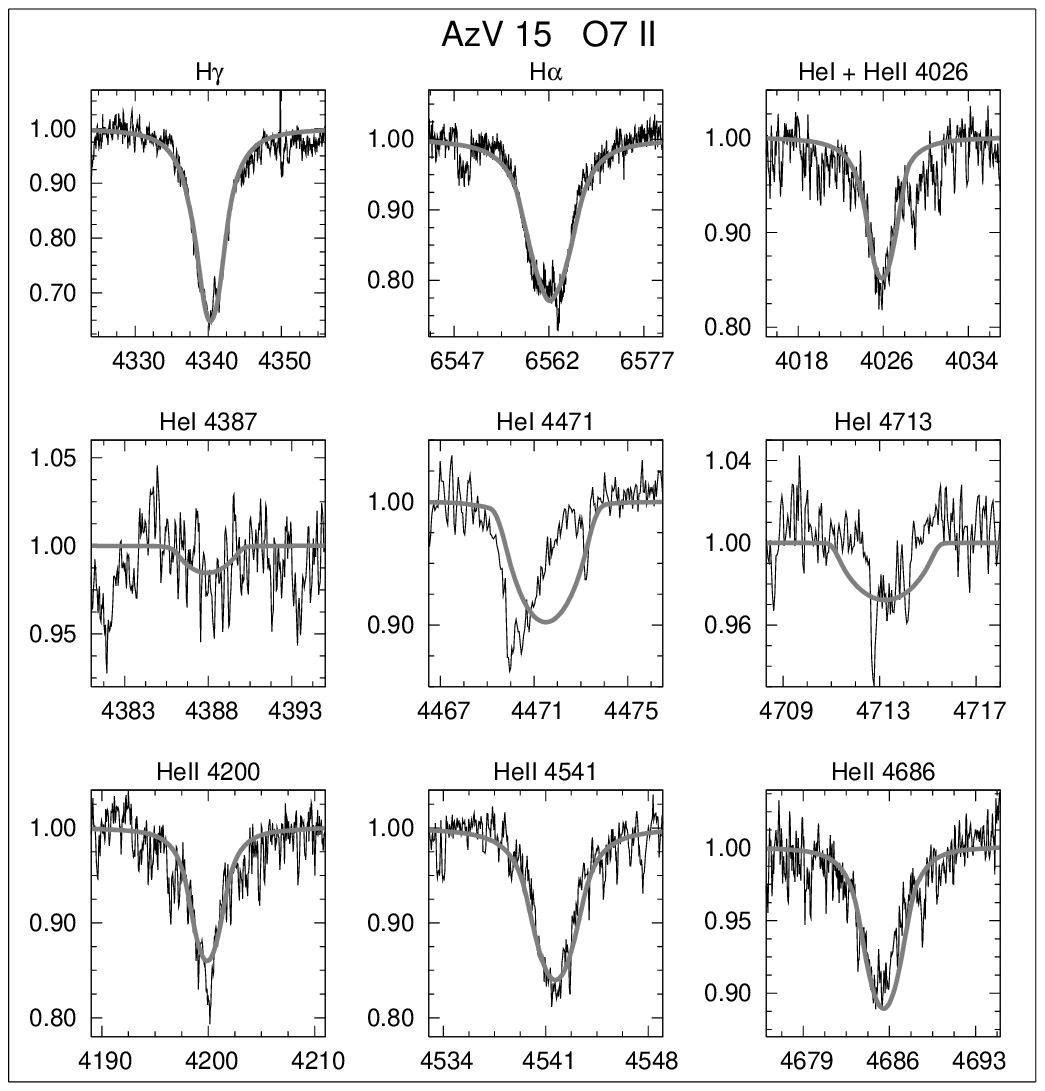}
  }

  \resizebox{1cm}{!}{ }

  \resizebox{17cm}{!}{
    \includegraphics{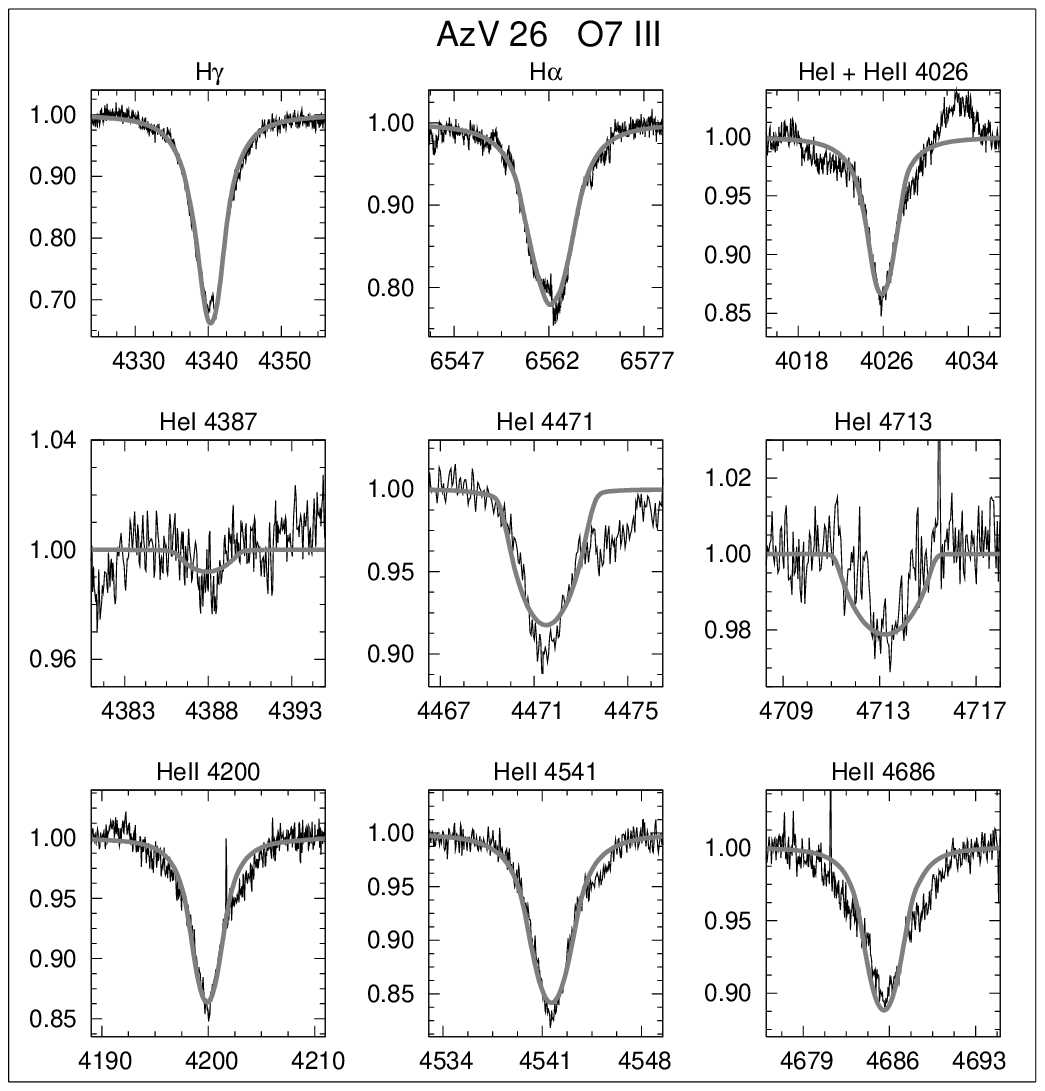}
    \includegraphics{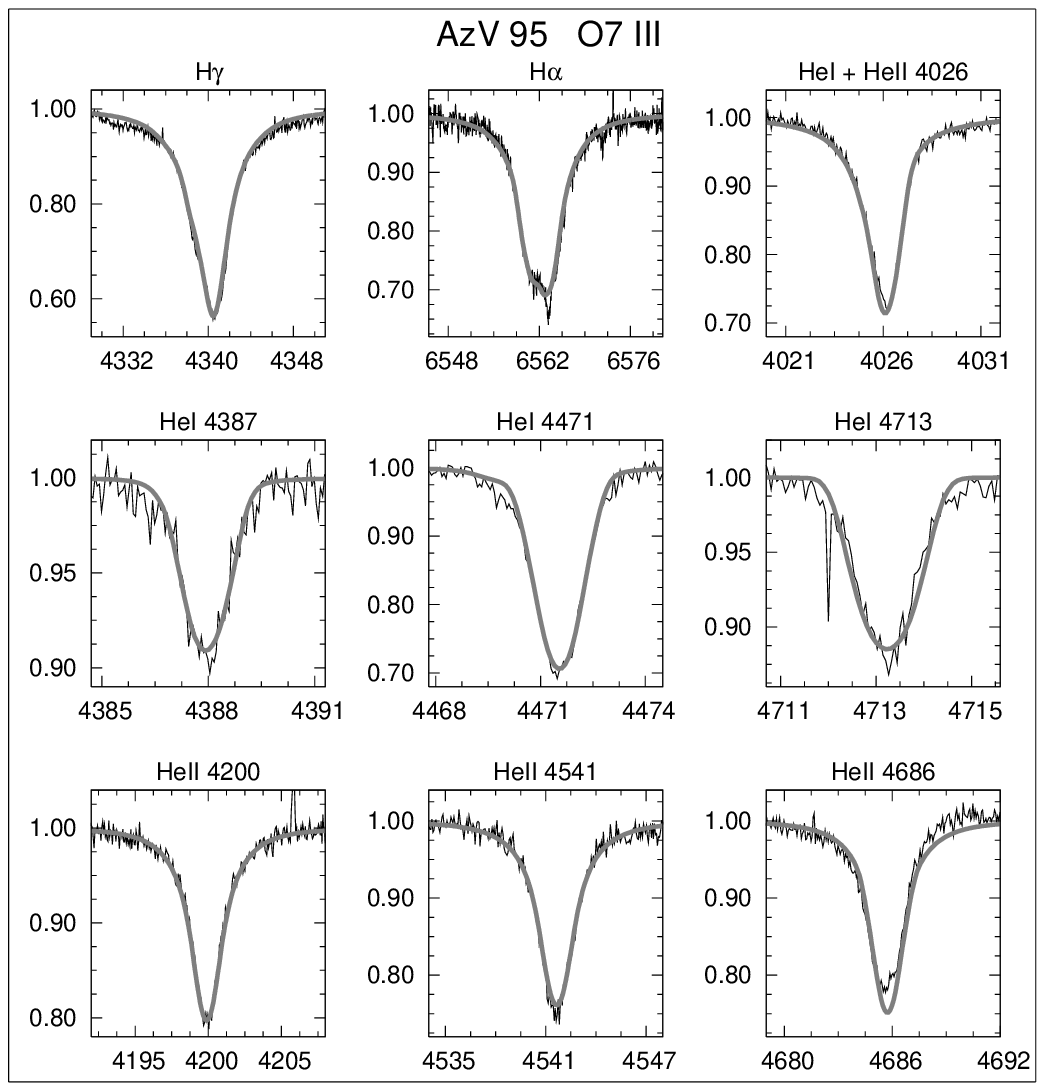}
  }
  \caption{Same as Fig.~\ref{fig:fits_1}, however, for \azv14, 15, 26
  and 95.}
  \label{fig:fits_7}
\end{figure*}

\begin{figure*}[t]
  \centering
  \resizebox{17cm}{!}{
    \includegraphics{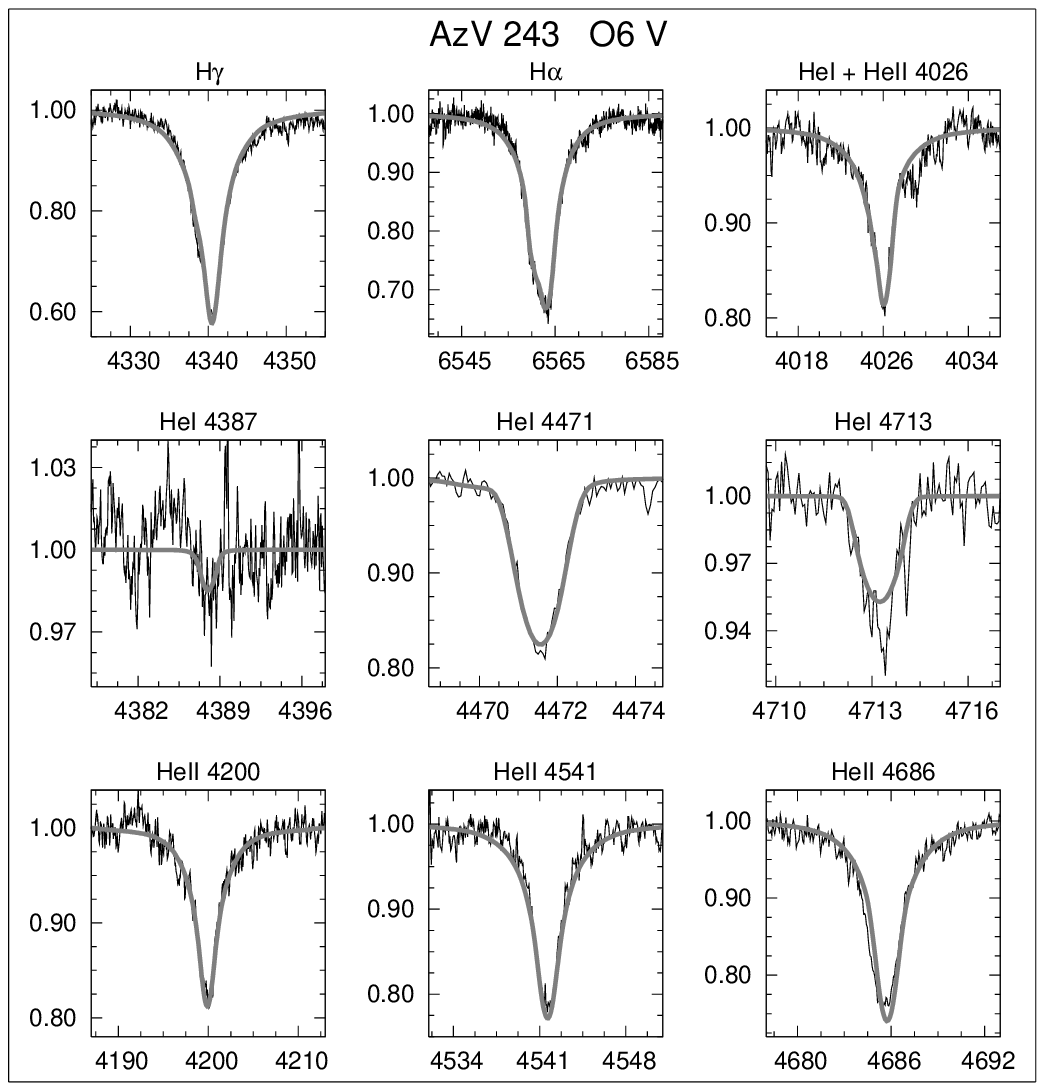}
    \includegraphics{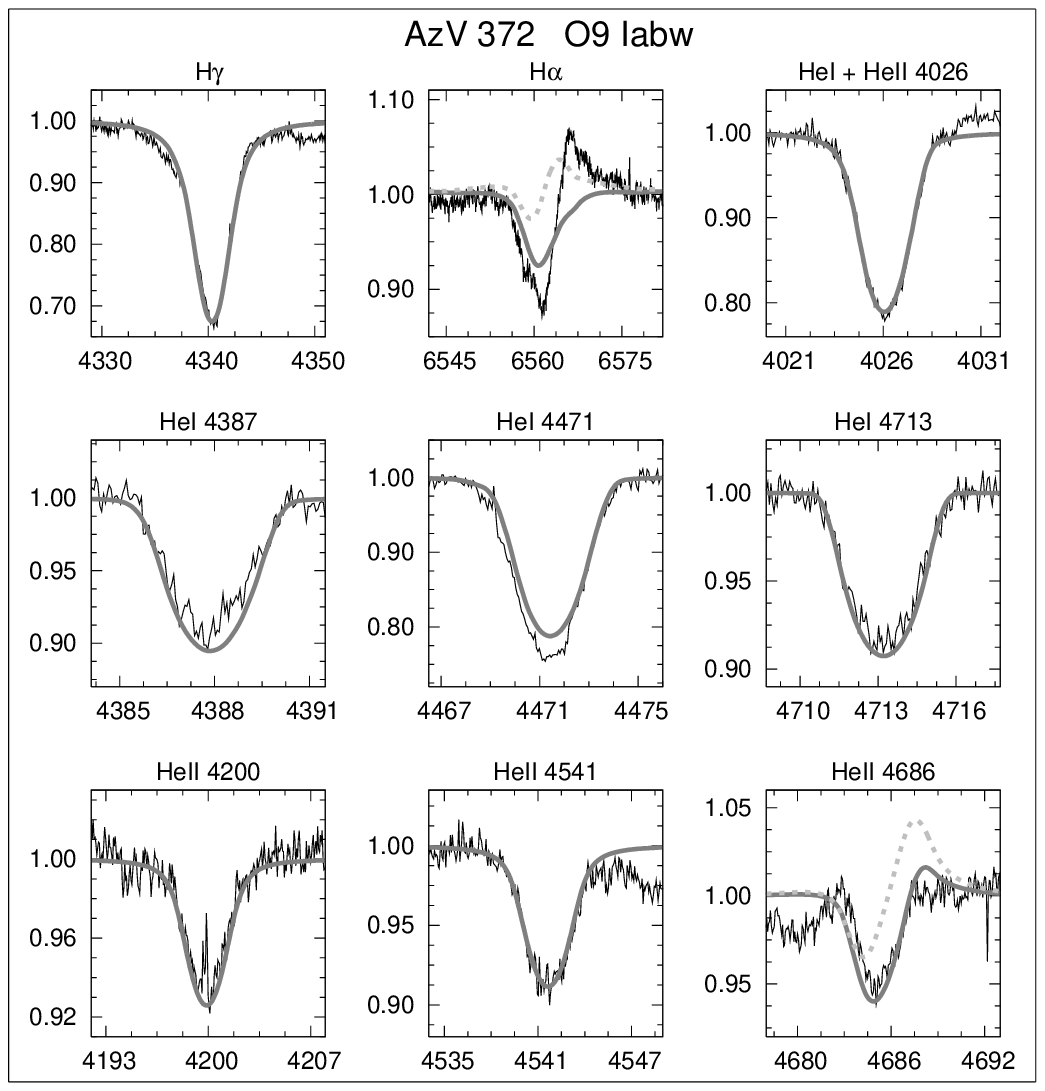}
  }

  \resizebox{1cm}{!}{ }

  \resizebox{17cm}{!}{
    \includegraphics{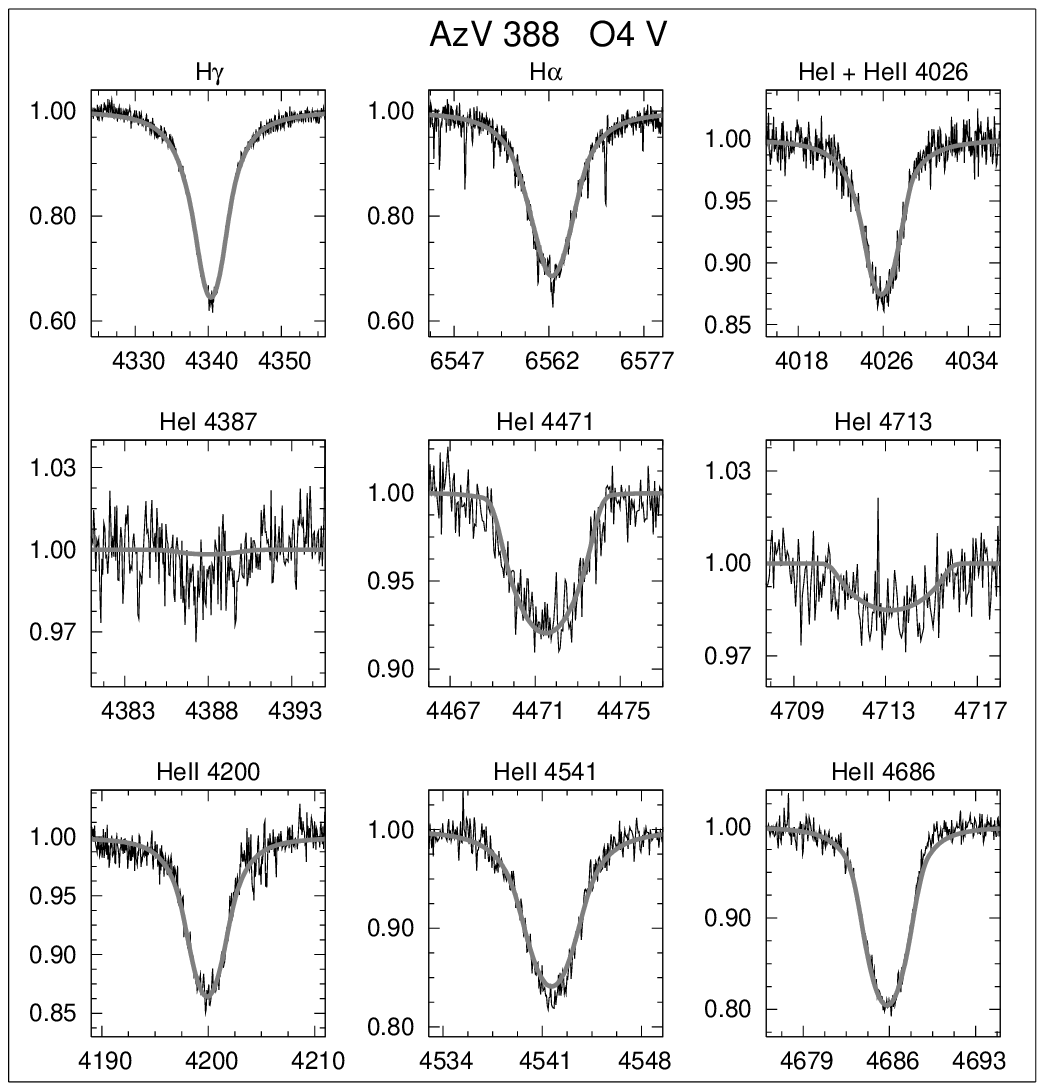}
    \includegraphics{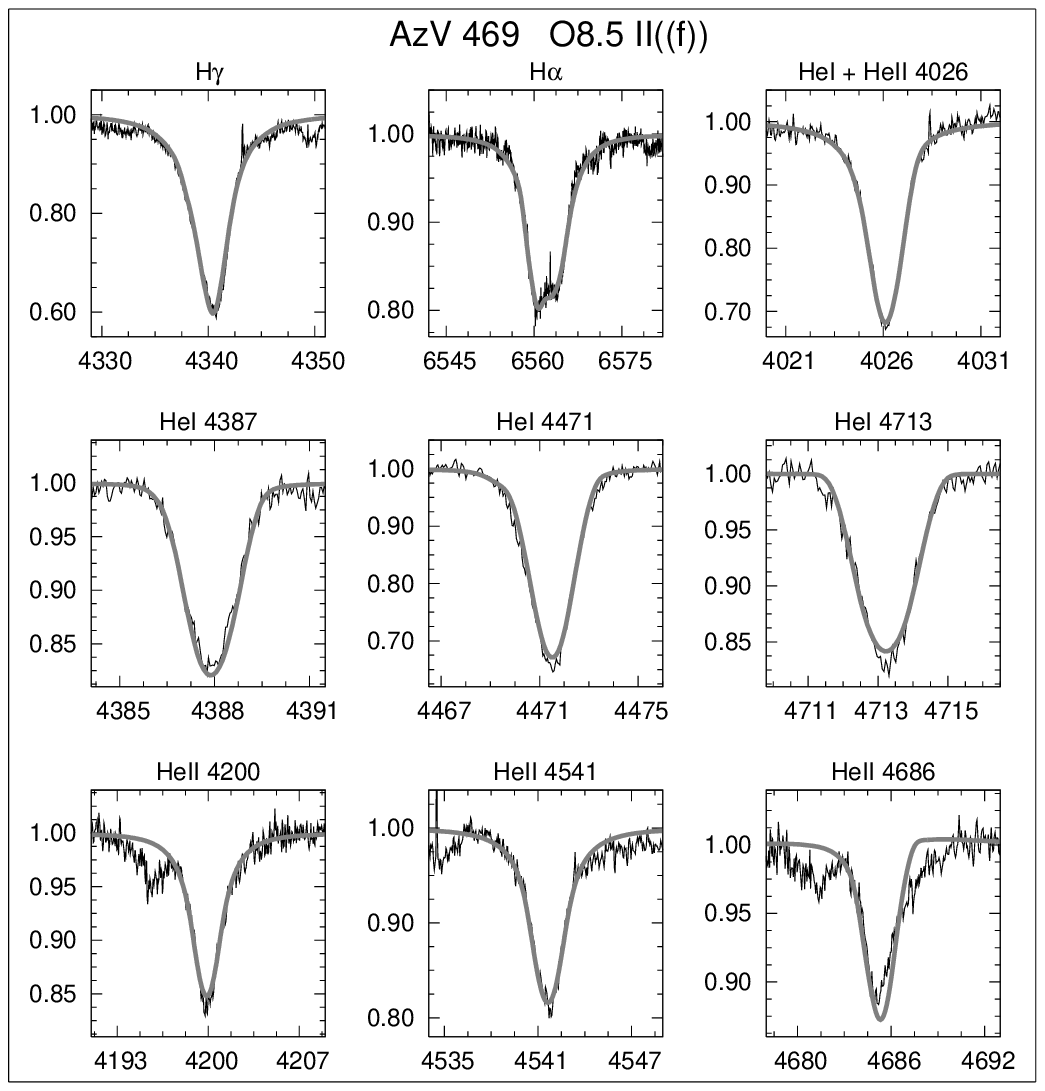}
  }

  \caption{Same as Fig.~\ref{fig:fits_1}, however, for \azv243, 372,
  388 and 469. Shown with a dotted line for \azv372 is the effect of
  an increase of the mass loss rate by 0.1~dex on the line profiles of
  \ha\ and \heii~$\lambda$4686.}
  \label{fig:fits_8}
\end{figure*}

\paragraph{\ngc346-001}
The fit of the spectrum of this O7 supergiant is presented in
Fig.\ref{fig:fits_1}. All lines except \hei~$\lambda$4387 are
reproduced well. Note that the nebular contamination on top of the
\ha\ line core was disregarded in the fit. The inaccurate reproduction
of \hei~$\lambda$4387 is the result of the low weight given to this
line in our weighting scheme (see \citetalias{mokiem05}). For this low
weight the quantitative fit quality of \hei~$\lambda$4387 is
comparable to that of \hei~$\lambda$4026. Therefore, the discrepancy
is not significant and merely reflects the uncertainty with which we
believe that the stellar atmosphere code can reproduce this line.

This object was previously studied by \cite{crowther02} using the
alternative wind code \cmfgen\ \citep{hillier98}. We find good
agreement between the photospheric parameters. Note that the helium
abundance $\yhe = 0.2$ adopted by \citeauthor{crowther02} is in good
agreement with the $\yhe=0.24$ self consistently determined here. The
largest difference is found for the effective temperature (2~kK). This
difference can be explained by the higher surface gravity obtained
from our fit, which requires a higher \teff\ to obtain the same helium
ionisation equilibrium. This occurs in many of the fits discussed
here, and we attribute this to a better multidimensional parameter
optimisation by the automated method.  This assures that the global
optimum in parameter space is found avoiding \logg\ and \teff\ values
which could correspond to a local optimum (see Sect.~5.2 of
\citetalias{mokiem05}).

The mass loss rate obtained from our fit of the optical spectrum is
within the error bars in agreement with the value that
\citeauthor{crowther02} derived from the UV spectrum. With respect to
the acceleration of the wind we find a large difference. From the
optical spectrum we derive $\beta = 1.15$, whereas
\citeauthor{crowther02} find $\beta=1.65$ based on the UV resonance
lines. They do note that from their modelling of \ha\ they obtain a
smaller value for this parameter, indicating a possible systematic
discrepancy between wind accelerations determined from optical and UV
spectra \citep[also see][]{puls96}.

\paragraph{\ngc346-007}
With exception of the \hei\ line at 4471~\AA, which suffers from very
strong nebular contamination, a good fit was obtained for this object.
Note that the fit shown in Fig.~\ref{fig:fits_1} does not include the
\hei~4713 line. The nebular contamination in this line proved to be
too severe to obtain a reliable fit.

Within the uncertainties the photospheric parameters \teff\ and \logg\
are in good agreement with those of \cite{bouret03} who used \cmfgen\
for their analysis. These authors find $\teff = 41.5$~kK and $\logg =
4.0$. The automated method was not able to self consistently derive
$\beta$. Consequently, a fixed value of $\beta=0.8$ was used. With
this fixed value a mass loss rate of $2.3\times 10^{-7}\,\msunyr$ is
found, which is also in good agreement with the rate derived by
\citeauthor{bouret03} from a fit to the wind sensitive UV lines.

The effective temperature of $\teff = 42.8$~kK obtained here is in
poor agreement with the recent analysis of \cite{heap06}. Using the
plane-parallel atmosphere model \tlusty\ these authors derive a
relatively low $\teff$ of 40~kK. The reason for this discrepancy is
unknown, but can probably be attributed to a different treatment of
line blanketing in \tlusty\ and \fastwind, or the coarsely sampled
grid of models used by \citeauthor{heap06}

\paragraph{\ngc346-010}
is a fast rotator. To obtain the fit displayed in
Fig.~\ref{fig:fits_1} a projected rotational velocity of 313~\kmsec\
was needed. Even though all lines seem to be fitted correctly we note
a slight over prediction of the strength of the wings of the wind
sensitive \heii~$\lambda$4686 and \ha\ lines. The shape of the former
line could indicate the existence of denser equatorial material or
possible the presence of a disk.

\paragraph{\ngc346-012}
As can be seen in Fig.~\ref{fig:fits_1} the low projected rotational
velocity of this B-supergiant allowed for a nearly perfect fit. Note
that given the relative scaling of the plots in Fig.~\ref{fig:fits_1}
the fit quality of the \heii\ lines is similar to that of the neutral
helium lines, i.e.\ they are fitted well.

The good fit quality obtained resulted in relatively small error
estimates (cf.\ Tab.~\ref{tab:errors}). Most interestingly, the
uncertainties on the wind parameters, considering the very tenuous
wind of this object, are modest as well. This is mostly because \ha\
is not affected by nebular contamination. As a result of this the
strength and the shape of this wind sensitive line could be matched
very accurately with little room for error.

\paragraph{\ngc346-018}
The line profiles of this object suffer from strong contamination,
making it problematic to obtain a good fit. We believe that this
emission is not only nebular in origin, but, given the breadth in the
hydrogen lines, is possibly also emitted from a circumstellar
disk. The contamination in the \ha\ Balmer line proved so severe that
we did not include it in the fitting process. Our best fitting model
is presented in Fig.~\ref{fig:fits_2}.

\paragraph{\ngc346-022}
A good quality fit was obtained for this object. The nebular emission
in the core of the \hei~4471 line did not hamper an accurate
determination of the effective temperature, hence the small error
given in Tab.~\ref{tab:errors}. This \teff\ is considerably larger
than the effective temperature one would expect for a Galactic O9
dwarf \citep[see][]{martins05a}. However, it is in line with the
expectation of a reduced line blanketing effect as a result of a low Z
environment, and is in agreement with the values obtained for the
other O9~V stars in our sample (see Sect.~\ref{sec:teff}).

\cite{heap06} determined a relatively low \teff\ of 32.5~kK for this
object, which is a consequence of the low surface gravity of $\logg =
3.75$ determined by these authors. In contrast the automated method
obtained $\logg = 4.2$ requiring a higher \teff\ to obtain the correct
helium ionisation equilibrium.

\paragraph{\ngc346-025}
The parameters obtained for this object are very similar to the
parameters used to fit the previously discussed object. The only
difference is in the projected rotational velocity, which is higher by
more than a factor of two.

\paragraph{\ngc346-026}
All lines, including the weak \heii\ lines, are fitted well for this
object. In the Fig.~\ref{fig:fits_2} the final fit is presented. The
luminosity classification IV ascribed to this object is reflected in
the gravity of $\logg = 3.76$. This value is still higher by 0.16~dex
compared to the value determined by \cite{bouret03} and
\cite{heap06}. Accordingly, the \teff\ determined by the automated
method is also higher by 1.1~kK.

\paragraph{\ngc346-028}
This object is classified as an OC6~Vz star, indicating that its
spectrum shows signatures of a relatively low nitrogen abundance and
evidence for an early evolutionary state. Evidence for the latter is
provided by the large relative strength of the \heii~$\lambda$4686
line (also see Sect.~\ref{sec:zams_stars}). In Fig.~\ref{fig:fits_3}
the best fit is presented. Note how well the \heii\ blends in the
hydrogen lines are reproduced. The effective temperature and
luminosity place this object at a position close to the ZAMS in the
HR-diagram shown in Fig.~\ref{fig:hrd_age}, corroborating its young
age.

Compared to the analyses of \cite{bouret03} and \cite{heap06} we find
an effective temperature higher by almost 3~kK. This is likely related
to the large helium abundance we found. In contrast,
\citeauthor{bouret03} and \citeauthor{heap06} adopted the Solar value
for this parameter. In terms of the wind parameters we were only able
to determine an upper limit of $10^{-7}\,\msunyr$ which is at least
consistent with the $3 \times 10^{-9}\,\msunyr$ derived by
\citeauthor{bouret03} from the UV spectrum.

\paragraph{\ngc346-031} is also classified as a Vz type object. The fit
to its spectrum is shown in Fig.~\ref{fig:fits_3}. In contrast to
\ngc346-028 the location of this object in the HR-diagram does not
corroborate its youthful spectral classification. Instead the
isochrones suggest an age of 1--2~Myr.

\paragraph{\ngc346-033}
The fit presented in Fig.~\ref{fig:fits_3} shows that despite the
strong nebular contamination in the neutral helium lines, a good fit
could be obtained for this O8 dwarf. The best fit mass loss rate of
$7.4\times 10 ^{-7}\,\msunyr$ seems rather high for its
luminosity. The reason for the large \mdot\ is likely the anomalously
large wind velocity ($\sim$4100~\kmsec), which resulted from a scaling
with the escape velocity at the stellar surface (see
Sect.~\ref{sec:wind_param}).

\paragraph{\ngc346-046}
The shape of the line profiles of this object are indicative for fast
stellar rotation. This is confirmed by the fit shown in
Fig.~\ref{fig:fits_3}, for which a \vsini\ of 340~\kmsec\ was needed,
classifying 346-046 as a very fast rotator. The helium abundance is
slightly enhanced, which could be related to the rapid rotation of
this object. However, one has to keep in mind that the error estimate
on this parameter is of the same order of the enrichment, when
considering this statement.

\paragraph{\ngc346-050}
To fit the line profiles, displayed in Fig.~\ref{fig:fits_4}, a
\vsini\ of 357~\kmsec\ was needed. Interestingly, similar to the other
fast rotator 346-046 a small helium enhancement was found.

\paragraph{\ngc346-051}
is the final Vz object in our sample. Its spectrum suffers from strong
nebular contamination. Nevertheless a good fit could be obtained,
which is presented in Fig.~\ref{fig:fits_4}. The location of
\ngc346-051 in the HR-diagram presented in Fig.~\ref{fig:hrd_age}
points to an early evolutionary phase for this object.

\paragraph{\ngc346-066}
The line profiles of this object are matched well by the synthetic
profiles. In the final fit, shown in Fig.~\ref{fig:fits_4}, only the
\hei\ line at 4713~\AA\ shows a slight mismatch due to an apparent
over prediction of the width of this line. Given the signal-to-noise
ratio of the spectrum, we do not believe this to be significant.

\paragraph{\ngc346-077}
A good fit was obtained despite the strong nebular contamination
affecting the hydrogen and neutral helium lines. The final fit is
displayed in Fig.~\ref{fig:fits_4}.

\paragraph{\ngc346-090}
Like 346-066 this object is of spectral type O9~V. The fit to its
spectrum is presented in Fig.~\ref{fig:fits_5}. Compared to 346-066 a
\teff\ lower by $\sim$800~K was obtained. This is the result of the
weaker \heii\ lines.

\paragraph{\ngc346-093}
The fit to the spectrum of the B0 dwarf \ngc346-093 is shown in
Fig.~\ref{fig:fits_5}. This fit shows that the very weak \heii\ lines
at 4200 and 4541~\AA\ could still be fitted using the automated
method.

\paragraph{\ngc346-097}
The final fit to the spectrum of the O9 dwarf 346-097 is presented in
Fig.~\ref{fig:fits_5}. Even though the nebular emission is quite
severe in the hydrogen lines and in some of the helium lines a good
fit was obtained. Its parameters are in agreement with other O9~V
objects in our sample. Note that the gravity of $\logg = 4.5$ is
remarkably high.

\paragraph{\ngc346-107}
The nebular contamination, as shown in Fig.~\ref{fig:fits_5}, is
severe in the spectrum of this object. All hydrogen and neutral helium
lines suffer from nebular emission. This did not prevent the automated
fitting method from obtaining a reliable fit. Most importantly, the
effective temperature, determined without information about the cores
of the \hei\ lines, is in good agreement with the other O9.5~V objects
studied here.

\paragraph{\ngc346-112} The spectrum of 346-112 is rather noisy, which
is the result of its low luminosity of $\log L = 4.36~\lsun$, making
it the intrinsically faintest object in our sample. Also strong
nebular contamination is present in the line cores. The fit to the
spectrum is presented in Fig.~\ref{fig:fits_6}. Despite the low
quality of the data the fits of the line profiles are good. The low
S/N value does, however, result in large error bars (cf.\
Tab.~\ref{tab:errors}).

\paragraph{\ngc330-013}
As can be seen in Fig.~\ref{fig:fits_6} all lines of this O8.5 giant
could be fitted accurately. In the HR-diagram in Fig.~\ref{fig:hrd}
this object is found to lie close to the region in which rotating
evolutionary models predict a helium surface enrichment. The
relatively high helium abundance of $\yhe = 0.18$, which was needed to
obtain this fit, seems to be in agreement with rotationally enhanced
mixing effects (see Sect.~\ref{sec:yhe}).

\paragraph{\ngc330-052}
Presented in Fig.~\ref{fig:fits_6} is the spectrum of \ngc330-052. The
\hd\ Balmer line was not fitted as the spectrum around in that region
was of a too low quality. Despite this missing line and the low
signal-to-noise ratio, a good fit could be obtained. The low quality
of the spectrum does result in relatively large error estimates (cf.\
Tab.~\ref{tab:errors}).

\paragraph{\azv14}
For this object and the remaining objects the hydrogen Balmer line
\hb\ was also fitted. The spectrum of the O5~V star suffers from
strong nebular contamination in both the \hei\ and \heii\
lines. However, the fit shown in Fig.~\ref{fig:fits_7}, does seem to
reproduce the observed features correctly. A comparison with the
parameters determined by \cite{massey04}, who also used \fastwind,
confirm this. Within the error bars the parameter values are
comparable.

\paragraph{\azv15}
The \hei~$\lambda$4471 line in the spectrum shown in
Fig.~\ref{fig:fits_7} is strongly affected by nebular emission. This
hampers an accurate determination of the effective temperature and the
surface gravity, which is reflected in relatively large error
estimates listed in Tab.~\ref{tab:errors}. Nebular emission can also
introduce a systematic error in the determination of these
parameters. However, the behaviour of this object in the \teff\ vs.\
spectral type diagram and log \teff -- \loggc\ plane does not indicate
a problem. Moreover, a ``normal'' helium abundance of $\yhe=0.10$ is
recovered from the spectrum, indicating that the \hei\ line profiles
contain enough information to avoid a degeneracy effect between \teff\
and \yhe\ (also see \citetalias{mokiem05}). All in all, we conclude
that despite the strong nebular contamination the parameters of \azv15
are robust. This conclusion is strengthened by the good agreement
between the projected rotational velocity we determine from the
optical (135~\kmsec) and the value determined by \cite{penny04}. These
authors used cross-correlation techniques to determine
$\vsini=128~\kmsec$ from UV spectra.

\cite{heap06} derive a considerably lower \teff\ of 37~kK. This is
connected to the \logg\ value obtained by these authors that was lower
by $\sim$0.2~dex.

\paragraph{\azv26}
A good fit was obtained for \azv26, though in the final fit presented
in Fig.~\ref{fig:fits_7} there is a small under prediction of the line
wings of \heii~$\lambda$4686. A possible explanation of the latter may
be an incorrect projected rotational velocity. However, the value
$\vsini = 132\,\kmsec$ obtained from the automated fit is in perfect
agreement with the value of $\vsini = 127\,\kmsec$ \cite{penny04}
derived from the UV spectrum. Therefore, this small discrepancy
remains unexplained.

Compared to the analysis by \cite{massey04} the largest differences
are found for \teff\ and \logg, for which the automated method
determined values higher by $\sim$2~kK and 0.25~dex, respectively, and
\mdot, for which a value lower by 0.16~dex was obtained. A small part
of the difference in surface gravity can be explained by the larger
\vsini\ of 150~\kmsec\ adopted by these authors. The largest part of
the difference in \logg\ and \teff\ we believe is the result of a
better multidimensional parameter optimisation by the automated method
(see above). The reduced mass loss rate can be explained by the larger
value of $\beta$ found in this study. \citeauthor{massey04} assumed a
fixed value of $\beta=0.8$, whereas we find $\beta=1.17$.

\paragraph{\azv95}
The best fit obtained for this O7 giant is presented in
Fig.~\ref{fig:fits_7}. The accurate reproduction of all lines and the
modest error estimates require no further comments. We note that
\cite{heap06} in their analysis obtained a \teff\ value that is lower
by $\sim$3~kK. We ascribe this difference to the gravity determination
of these authors, which was lower by approximately 0.3~dex.

\paragraph{\azv243}
In Fig.~\ref{fig:fits_8} the spectrum of \azv243 is presented. With
the automated method we were able to obtain a very good fit for all
lines. The final parameters can be compared to the work of
\cite{haser98}. With exception of the effective temperature that
\citeauthor{haser98} adopted from the analysis of \cite{puls96} and
$\beta$, there is good agreement. Our effective temperature is lower
by $\sim$1000~K, which we attribute to the inclusion of line
blanketing in our analysis. The acceleration of the wind was
determined by \citeauthor{haser98} to be $\beta=0.7$, based on UV
resonance lines. Our value of $\beta=1.4$ was derived from the optical
spectrum. The reason for this discrepancy is unclear. However, it
could be related to the fact that the wind is very weak (order
$10^{-7}$~\msunyr), introducing a large error in the determination of
this parameter.

Compared to the study of \cite{penny04} we again find good agreement
between our \vsini\ determined from the optical (59~\kmsec) and their
\vsini\ derived from a UV analysis (62~\kmsec).

\paragraph{\azv372}
Based on its peculiar \ha\ profile and the fact that they were unable
to simultaneously fit the \hei\ and \heii\ lines \cite{massey04}
suggest that this object is a spectroscopic binary. However, the fit
presented in Fig.~\ref{fig:fits_8} shows that a simultaneous fit of
both the \hei\ and \heii\ lines for a single set of parameters is
possible. Therefore, we do not concur with the findings of
\citeauthor{massey04} and treat \azv372 as a single star. Note that we
did encounter a mild form of the ``generalised dilution'' effect in
the \hei~$\lambda$4471 line.

Using our fitting method, reproduction of the \ha\ line still remains
difficult. We are not able to satisfactorily match the absorption and
the weak emission simultaneously. To fit the emission an increase of
\mdot\ of the same order as the error estimate of this parameter is
necessary. In Fig.~\ref{fig:fits_8} the effect of an increase of the
mass loss rate with 0.1~dex on the line profiles of \ha\ and
\heii~$\lambda$4686 is shown using a dotted line. Consequently, given
the modest increase in \mdot\ required we regard our determination of
the mass loss rate as correct within the given error estimates.

\azv372 was also analysed by \cite{evans04b}. Compared to this study
we find differences for \teff, \yhe, \mdot\ and $\beta$. The effective
temperature from the automated method is larger by approximately
3~kK. This is probably linked to the reduced helium abundance we find
and the improved fit of the \heii\ lines (e.g.\ see Fig.~3 of
\citeauthor{evans04b}). Our mass loss rate is approximately a factor
of two larger than the \mdot\ determined by \citeauthor{evans04b}. We
can explain this difference by the large $\beta$ these authors
derive. To match the \ha\ profile \citeauthor{evans04b} needed
$\beta=2.25$, whereas we obtained $\beta=1.27$ using the automated
method. As a larger value for $\beta$ results in a slower accelerating
wind and, consequently, in a higher density in the \ha\ line forming
region, the required mass loss rate to reproduce the \ha\ line profile
is decreased. Note that our wind parameters, in contrast to
\citeauthor{evans04b}, can reproduce the \heii~$\lambda$4686 line
profile correctly. However, one should realise that in our fitting
method this line was given a relatively high weight compared to the
fit of \cite{evans04b}, who gave priority to \ha. Consequently, the
differences in wind parameters for this object also reflect the
particularly peculiar shape of the wind lines and the difficulties
involved in reproducing them.

\paragraph{\azv388}
For the O4 dwarf \azv388 a good fit was obtained. As is shown in
Fig.~\ref{fig:fits_8} all line profiles are reproduced accurately. The
projected rotational velocity of 163~\kmsec\ required to fit the
profiles is in good agreement with the UV determination by
\cite{penny04}. They find $\vsini = 179\,\kmsec$.

\paragraph{\azv469}
The fit to the spectrum of \azv469 is shown in
Fig.~\ref{fig:fits_8}. For all lines a good fit was obtained. Our fit
parameters compare well with the results of \cite{evans04b} who
studied this object using \cmfgen. In contrast we find large
differences for \logg, \teff\ and \mdot\ when comparing our results to
the analysis of \cite{massey04} who also analysed \azv469 using
\fastwind. Note, however, that the differences are likely the result
of a not well constrained fit by \citeauthor{massey04}, as these
authors had difficulties matching the \ha\ core emission and severely
under predicted the strength of the \hei~$\lambda$4471 line.

\end{document}